%
\overfullrule=0pt

\magnification=1400
\hsize=12.2cm
\vsize=16.0cm
\hoffset=-0.4cm
\voffset=.8cm
\baselineskip 16 true pt

\font \smcap=cmcsc10
\font \smcaps=cmbx10 at 12 pt


\font \fauteur=cmbx10 scaled 1167

\font \gcaps=cmbx10 scaled 1333
\font \smcaps=cmcsc10 
\font \pecaps=cmcsc10 at 9 pt

\newtoks \hautpagegauche  \hautpagegauche={\hfil}
\newtoks \hautpagedroite  \hautpagedroite={\hfil}
\newtoks \titregauche     \titregauche={\hfil}
\newtoks \titredroite     \titredroite={\hfil}
\newif \iftoppage         \toppagefalse   
\newif \ifbotpage         \botpagefalse    
\titregauche={\pecaps      Fran\c{c}ois Dubois }
\titredroite={\pecaps    Conservation Laws Invariant for Galileo Group }
\hautpagegauche = { \hfill \the \titregauche  \hfill  }
\hautpagedroite = { \hfill \the \titredroite  \hfill  }
\headline={ \vbox  { \line {  
\iftoppage    \ifodd  \pageno \the \hautpagedroite  \else \the
\hautpagegauche \fi \fi }     \bigskip  \bigskip  }}
\footline={ \vbox  {   \bigskip  \bigskip \line {  \ifbotpage  
\hfil {\oldstyle \folio} \hfil  \fi }}}

\font\smcap=cmcsc10

\def\sqr#1#2{{\vcenter{\vbox{\hrule height.#2pt \hbox{\vrule width .#2pt
height#1pt
\kern#1pt \vrule width.#2pt} \hrule height.#2pt}}}}
\def\square{\mathchoice\sqr64\sqr64\sqr{4.2}3\sqr33}


\def\R{{\rm I}\! {\rm R}}
\def\Z{{\rm Z}\!\! {\rm Z}}

\def\ib#1{_{_{\scriptstyle{\!#1}}}}

\def\abs#1{\mid \! #1 \! \mid }

\def\mod#1{\setbox1=\hbox{\kern 3pt{#1}\kern 3pt}%
\dimen1=\ht1 \advance\dimen1 by 0.1pt \dimen2=\dp1 \advance\dimen2 by 0.1pt
\setbox1=\hbox{\vrule height\dimen1 depth\dimen2\box1\vrule}%
\advance\dimen1 by .1pt \ht1=\dimen1
\advance \dimen2 by .01pt \dp1=\dimen2 \box1 \relax}

\def\nor#1{\setbox1=\hbox{\kern 3pt{#1}\kern 3pt}%
\dimen1=\ht1 \advance\dimen1 by 0.1pt \dimen2=\dp1 \advance\dimen2 by 0.1pt
\setbox1=\hbox{\kern 1pt  \vrule \kern 2pt \vrule height\dimen1
depth\dimen2\box1
\vrule
\kern 2pt \vrule \kern 1pt  }%
\advance\dimen1 by .1pt \ht1=\dimen1
\advance \dimen2 by .01pt \dp1=\dimen2 \box1 \relax}

\def \union {\mathop {\hbox{$\bigcup$}} \limits}

\def\ib#1{_{_{\scriptstyle{#1}}}}

\def\sqr#1#2{{\vcenter{\vbox{\hrule height.#2pt \hbox{\vrule width .#2pt height#1pt 
\kern#1pt \vrule width.#2pt} \hrule height.#2pt}}}}
\def\square{\mathchoice\sqr64\sqr64\sqr{4.2}3\sqr33} 
 


\def\br {\break}

  \def \page #1{\unskip\leaders\hbox to 1.3 mm {\hss.\hss}\hfill   {  $\,\,$ {\oldstyle #1}}}

\rm
\hyphenation {com-pres-sible com-pres-si-bles  me-cha-nics sprin-ger
con-ve-na-ble
si-mu-la-tion si-tua-tion uni-di-men-sio-nal }


~ 

\bigskip
\bigskip
\centerline{\gcaps    Conservation Laws Invariant for Galileo Group;}
\medskip 
\centerline{\gcaps  Cemracs Preliminary resuls}
\bigskip 
\bigskip

\centerline { \fauteur  Fran\c{c}ois Dubois }

\smallskip
\smallskip
\centerline { Conservatoire National des Arts et M\'etiers }
\centerline {  15 rue Marat, F-78 210 $\,$ Saint Cyr
l'Ecole,  Union Europ\'eenne.  } 

\smallskip
\centerline {  and }
\centerline { Centre National de la Recherche Scientifique,}
\centerline {   laboratoire ASCI,  b\^at. 506, BP
167,  F-91~403 Orsay  Cedex, U.E.  }

\bigskip
\bigskip
\centerline {  December  2000  
\footnote{$ ^{^{\scriptstyle  \square}}$}  {\rm 
Con\c cu initialement  au   Centre d'Et\'e de Math\'ematiques et  
de  Recherche Avanc\'ee en Calcul Scientifique 
(Marseille,  Centre International de Rencontres Math\'ematiques) 
en  ao\^ut 1999. 
Publi\'e dans {\it ESAIM: Pro\-ceedings}, volume 10, ``CEMRACS 1999'', 
 Editeurs  Fr\'ed\'eric Coquel et  St\'ephane Cordier,  
pages 233-266,   2001. 
Edition du  11 janvier 2011.  }}

\bigskip
\bigskip \noindent 
{\bf Abstract. }
We propose a notion of hyperbolic system of conservation laws  invariant for the
Galileo  group  of transformations. We show that with natural physical and
mathematical hypotheses, such a system conducts to the gas dynamics
equations or to
exotic systems that are detailed  in this contribution to Cemracs 99.

 

\vfill \eject 
\bigskip 
\bigskip \noindent {\smcaps Contents}   \smallskip  

1)  General framework \page {$\,\,$2}

2)  Conservation laws \page {$\,\,$3} 

3)  Invariance  \page {$\,\,$6}  

4)  Null-velocity manifold   \page {$\,\,$9}  

5)  The case $\, m=1$ \page {$\,\,$12}   

6)   Galilean invariance for systems  of two conservation laws \page {$\,\,$12}  

7)   Galilean invariance for systems  of three conservation laws \page {$\,\,$21}   

8)    The Cemracs System  \page {$\,\,$37}    

9)     Acknowledgements  \page {$\,\,$41}    

10)    References  \page {$\,\,$41}

\bigskip 
\bigskip

\centerline{ \smcap 1. \quad General framework} \smallskip

Let $\, t \,$ and $\, x \,$ two real variables. We denote by $\,\, \R^{2} \,\,$
the set of line vectors~: $\,\,   (t,\, x) \,\in \, \R^2 \,\,  $  and
$\,\, \R^{2,
\, {\rm t}} \,\equiv \, \R \times^{\rm \! t} \R \,   \,$ the set of column
vectors that
are the transposed from line vectors~:

\smallskip \noindent   (1.1)  $ \qquad   \displaystyle 
\pmatrix {t \cr x \cr }\,\, \in \,\, \R^{2,\, {\rm t}} \,.    $ 

\smallskip \noindent
For $\,\, v \in \R \,,$ we denote by $\,\, y\ib{\displaystyle \, v} \,$ the {\bf
special Galileo transform} defined by the relations

\smallskip \noindent   (1.2)  $ \qquad    \displaystyle 
y\ib{\displaystyle \, v} \, \pmatrix{t \cr x \cr} \,=\, 
\pmatrix {t \cr x - v t \cr }
\,=\, \pmatrix { 1 & 0 \cr -v & 1 \cr } \,  \pmatrix{t \cr x \cr}  $ 

\smallskip \noindent
and we introduce also the {\bf  space symmetry} $\, q \,$ defined by the
conditions~:

\smallskip \noindent   (1.3)  $ \qquad   \displaystyle  
q \, \, \pmatrix{t \cr x \cr} \,=\, \pmatrix{t \cr -x \cr} 
\,=\, \pmatrix { 1
& 0 \cr 0 & -1 \cr } \, \, \pmatrix{t \cr x \cr} \,.   $

\bigskip \noindent  {\bf Proposition 1. \quad Relation between the generators.}

\noindent
We have the following relations between the operators $\,
y\ib{\displaystyle \, v} \,$
and $\, q~:$

\smallskip \noindent   (1.4)  $ \qquad    \displaystyle 
\forall \, v ,\,w \in \R \,,\qquad   y\ib{\displaystyle \,  v} \, {\scriptstyle
\bullet} \,   y\ib{\displaystyle \,  w} \,=\, y\ib{\displaystyle \, v+w} \,  $

\smallskip \noindent   (1.5)  $ \qquad    \displaystyle 
q^2 \,=\, {\rm Id} $

\smallskip \noindent   (1.6)  $ \qquad    \displaystyle  
\forall \, v \in \R \,,\qquad  y\ib{\displaystyle \, v} 
\, {\scriptstyle \bullet} \,  q
\, {\scriptstyle \bullet} \,   y\ib{\displaystyle \, v} 
\,=\, q \, $

\smallskip \noindent
where $\,\, {\rm Id } \,\,$ denotes the identity matrix.

\toppagetrue  
\botpagetrue    
%
%
%
\bigskip \noindent  {\bf Definition 1. \quad Galileo group. }

\noindent
We denote by  $\, GL\ib{2}(\R) \,$ the group of  two by two invertible
matrices with
real coefficients.  By definition, the Galileo group $\, G \,$ is the
subgroup of $\,
GL\ib{2}(\R) \,$  generated by the matrices $\, y\ib{\displaystyle \, v}
\,$ for $\, v
\in \R ,\,$ the matrix $\, q \,$ and the relations (1.4), (1.5) and (1.6).

\bigskip 
\noindent  {\bf Definition 2. \quad Space of states.}

\noindent
Let $\, m \,$ be a positive integer. The space of states $\, \Omega  \,$ is an
open convex cone included in $\, \R^{m, \, {\rm t}} ~:\,$

\smallskip \noindent   (1.7)  $ \qquad   \displaystyle 
\Omega \subset \R^{m, \, {\rm t}} \qquad  $ and $ \qquad
\forall \,\, W \in \Omega\,, \quad \forall \, \lambda \, {\rm >}\,0\, ,
\quad \lambda \, W \in \Omega\,. $

\bigskip \noindent  {\bf Hypothesis 1. 

 \noindent 
The Galileo group operates linearly on the space of states. }

\noindent
For $\, g \in G \,$ and $\, W \in \Omega \,$ the action $\,\, g \, {\scriptstyle
\bullet} \, W \,$  of $\, g \,$ on the state $\, W \,$ is well defined as a {\bf
linear}  operation of the group $\, G \,$ on the set $\, \Omega ~:\,$

\smallskip \noindent   (1.8)  $ \qquad    \displaystyle 
\forall \, g \in G ,\quad \forall \, W \in \Omega ,\qquad  g \, {\scriptstyle
\bullet} \, W \in \Omega \, $

\smallskip \noindent   (1.9)  $ \qquad    \displaystyle 
\forall \, g ,\, g' \in G ,\quad \forall \, W \in \Omega ,\qquad  (g \, g') \,
{\scriptstyle \bullet} \, W \,=\, g \, {\scriptstyle \bullet} \, (g' \,
{\scriptstyle \bullet} \,W) \,.  \,$

\smallskip \noindent
From an algebraic point of view, there exists an $\, m \,$ by $\, m \,$ set
of matrices  $\, Y(v) \,$ and a matrix $\, R \,$ with $\,m \,$ lines and
$\, m \,$
columns such that 

\smallskip \noindent   (1.10)  $ \qquad  \displaystyle   
\forall \, v \in \R,\quad \forall \, W \in \Omega ,\qquad Y(v) \, {\scriptstyle
\bullet} \, W \in \Omega  \,$
 
\smallskip \noindent   (1.11)  $ \qquad  \displaystyle 
\forall \, W \in \Omega ,\qquad R \, {\scriptstyle \bullet} \, W \in \Omega
\,. \,$

\smallskip  \noindent
Then the relations (1.4), (1.5) and (1.6) between the generators and the
hypotheses
(1.8) and (1.9) can be written in the vocabulary of the $\,m \,$ by $\, m
\,$ matrices
$\,\, Y(v) \,\,$ and $\,\, R \, : \,$

\smallskip \noindent   (1.12)  $ \qquad  \displaystyle 
\forall \, v,\,w \in \R, \qquad Y(v)  \, {\scriptstyle \bullet} \, Y(w) \,=\,
Y(v+w) \,$
 
\smallskip \noindent   (1.13)  $ \qquad  \displaystyle 
R^2 \,=\, {\rm Id} \,   $   
 
\smallskip \noindent   (1.14)  $ \qquad  \displaystyle  
\forall \, v \in \R, \qquad Y(v)  \, {\scriptstyle \bullet} \, R  \,
{\scriptstyle \bullet} \, Y(v) \,=\, R \, . $ 

\bigskip \bigskip

\centerline{ \smcap 2. \quad Conservation laws} \smallskip

We introduce a regular flux function  $\,\, \Omega \ni W  \longmapsto f(W) \in
\R^{m,\, { \rm t}} \,\,$ and we consider the associated  {\bf system of
conservation laws} in one space dimension~:

\smallskip \noindent   (2.1)  $ \qquad  \displaystyle 
{{\partial }\over{\partial t}} W(t,x)\, + \, {{\partial}\over{\partial
x}}\, f(W(t,x)) \,\,=\,\,0\,$

\smallskip \noindent  where the unknown function $\,\, \,[0,\,+\infty[
\times \R \ni
(t,x) \longmapsto W(t,x) \in \Omega \,\, \,$ takes its values inside the
convex open
cone $\, \Omega .\,$ We suppose that the system (2.1) of conservation laws
admits a
{\bf mathematical entropy} (Friedrichs and Lax [FL71]) $\, \eta\,$  : $\,\,
\Omega \ni
W \longmapsto \eta(W) \in \R\,\,$ which is here supposed to be a  regular
(of $\,
{\cal C}^{2}\,$  class), {\bf strictly convex} function admitting an associated
entropy flux $\, \xi({\scriptstyle \bullet}) .\,$  Recall that a pair of
(mathematical
entropy$\,,\,$ entropy flux) allows  any regular solution of system (2.1)
to satisfy
an additional conservation law (Godunov [Go61])~:

\smallskip \noindent   (2.2)  $ \qquad \displaystyle 
{{\partial }\over{\partial t}} \, \eta(W(t,x))\, + \,
{{\partial}\over{\partial x}}\,
\bigl[\, \xi(W(t,x))  \, \bigr]  \,=\,0 \, $

\smallskip \noindent
and that a weak entropy solution of system (2.1) is by definition (see
e.g. Godlewski-Raviart [GR96]) a weak solution  that satisfies also the
inequality

\smallskip \noindent   (2.3)  $ \qquad \displaystyle  
{{\partial }\over{\partial t}} \, \eta(W(t,x))\, + \,
{{\partial}\over{\partial x}}\,
\bigl[\, \xi(W(t,x))  \, \bigr]  \,\le\,0\, $

\smallskip \noindent
in the sense of distributions.

\bigskip \noindent $\bullet \quad$
We introduce the Frechet derivative $\,{\rm d}\eta(W) \,$ of the entropy and its
associated partial derivatives $\,\, \Omega \ni W  \longmapsto \varphi(W)
\in \Phi
\,\,$ that takes its values in some set $\,\, \Phi\, \subset \, \R^m \,\,  $ and
defines the so-called {\bf entropy variables}~: 

\smallskip \noindent   (2.4)  $ \qquad \displaystyle  
\forall \, W \in \Omega\,,\quad \forall r \, \in \R^{m,\,{ \rm t}}
\,,\qquad  {\rm d}\eta(W) \,{\scriptstyle \bullet} \, r \,\, = \,\,
\varphi(W) \,
{\scriptstyle \bullet}\,  r \,. \,     $

\smallskip \noindent
We consider also  the dual function of the entropy~:  $\,\, \Phi \ni \varphi
\longmapsto \eta^{\star}(\varphi)  \in \R \,\, $  in the sense proposed by
Moreau
[Mo66], that is 

\smallskip \noindent   (2.5)  $ \qquad \displaystyle  
\eta^{\star}(\varphi) \,\,= \,\, \sup_{W\in \, \Omega} \, \bigl( \, \varphi
\,{\scriptstyle \bullet} \, W \,-\, \eta(W) \, \bigr) \,.\, $

\smallskip \noindent
This dual function satisfies 

\smallskip \noindent   (2.6)  $ \qquad \displaystyle 
{\rm d}\eta^{\star}(\varphi) \,\, = \,\, {\rm d}\varphi  \,{\scriptstyle
\bullet} \,
W(\varphi)\,\,, \quad {\rm where} \,\,\, \varphi \, = \, {\rm
d}\eta(W(\varphi)) \,. $

\bigskip 
 \noindent  {\bf Hypothesis 2. \quad  Entropy flux and velocity. }

\noindent There exists a regular  (of $\, {\cal C}^{1}\,$  class) function
$u~: \,\, \Omega \ni W \longmapsto u(W) \in \R\,\,  $  which allows to write the
entropy flux $\, \xi({\scriptstyle \bullet})\,$ associated to the strictly
convex
entropy $\, \eta({\scriptstyle \bullet})\,$ under the form

\smallskip \noindent   (2.7)  $ \qquad \displaystyle  
\forall \, W \in \Omega\,,\qquad  \xi(W) \,\, = \,\, \eta(W) \,\, u(W)\,.
\,  $

\bigskip \noindent $\bullet \quad$
For the study of gas dynamics in Lagrangian coordinates, Despr\'es [De98] has
developed  the general case where the function $  \quad \Omega \ni W \longmapsto u(W)
\in \R\,\,  $ is identically null. Hypothesis~2 is very little restrictive
because
we can define $\,\, u(W)\,\,$ by the relation $\,\, u(W) \,\,=\,\,
{{\xi(W)}\over{\eta(W)}}\,\,\,  $ as soon as the state $W$ is such that the
entropy
$\, \eta(W) \,$ is not null. We observe also that field $\,u({\scriptstyle
\bullet})\,$ is homogeneous  to the ratio of space variable $x$ over time
variable
$t$. For this reason, we have introduced in [Du99] the following

\bigskip \noindent  {\bf Definition 3. \quad   Velocity of the state $W$. }

\noindent When hypothesis 2 is satisfied, we call velocity of the state $W$
the scalar
 $\,u(W)\,$ defined according to  the relation (2.7).

\bigskip \noindent  {\bf Definition 4. \quad  Thermodynamic flux. }

\noindent When the hypothesis 2 is satisfied, the thermodynamic flux $j$~:
$\,\,
\Omega
\ni W  \longmapsto j(W) \in \R^{m, \, {\rm t}} \,\,$ is defined according to the
relation

\smallskip \noindent   (2.8)  $ \qquad \displaystyle 
j(W) \,\,\equiv \,\, f(W) \,-\, u(W)\,W\,.\, $


\noindent
The hypothesis 2 is satisfied if and only if we have the following relation
between
the  thermodynamic flux $\,j({\scriptstyle \bullet})\,$ and  the velocity
$\,u({\scriptstyle \bullet})\,$~: 

\smallskip \noindent   (2.9)  $ \qquad \displaystyle 
\forall \, W \in \Omega \,, \qquad  \varphi(W) \,{\scriptstyle \bullet}\,
{\rm d}j(W)
\,\,+ \,\,\eta^{\star}(\varphi(W)) \,\, {\rm d}u(W) \,\,=\,\,0  \,. \, $

\bigskip  \noindent {\bf Proof of Proposition 2.}

\noindent
It is sufficient  to express the classical compatibility relation (see e.g.
[GR96])
between  the entropy flux, the derivative of entropy and the derivative of
the flux
function, i.e. $\,\,  {\rm d}(\eta \, u) \,\equiv$ $   \, \varphi \,
{\scriptstyle
\bullet} \, {\rm d}f \,;\,\, $  this relation  is what is usefull to deduce the
relation (2.2) from the original system (2.1) of conservation laws.  We
derive the
relation (2.8). Then we get    $ \quad  \displaystyle  
{\rm d}f(W) \,=\,  ({\rm d}u(W)) \, W \,+ u(W) \, {\rm d}W \,+\, {\rm
d}j(W) \,   $ \quad  and

\smallskip \noindent     $  \displaystyle  
\varphi \,{\scriptstyle \bullet}\,  {\rm d}f \,-\, {\rm d}(\eta \, u) \,=\, $

\smallskip \noindent     $ \quad  \displaystyle   \,=\,
 ({\rm d}u(W)) \, \varphi \, {\scriptstyle \bullet}
\, W \,+\, u(W) \,  \varphi  \,{\scriptstyle \bullet}\, {\rm d}W \,+\,
\varphi  \,{\scriptstyle \bullet}\,  {\rm d}j(W) \,- \, (\varphi \,u(W)
\,+\, \eta \, {\rm d}u(W))   $

\smallskip \noindent     $ \quad  \displaystyle   \,=\, 
{\rm d}u(W)\,\bigl( \,\varphi \, {\scriptstyle
\bullet}\, W \,-\,\eta(W) \,\bigr) \,+\, \varphi \,{\scriptstyle \bullet}\, {\rm
d}j(W) \, \hfill $ 
because $\,\, {\rm d}W \,= \,{\rm Id} \,(\R^{m, \, {\rm t}}) $

\smallskip \noindent     $ \quad  \displaystyle   \,=\, 
\eta^{\star}(\varphi) \, {\rm d}u(W) \,\,+\,\, \varphi \,{\scriptstyle
\bullet}\, {\rm d}j(W) $

\smallskip \noindent
and the desired result is established.  $ \hfill \square \kern0.1mm    $

\bigskip \noindent  {\bf Example. \quad Gas dynamics (i). }

\noindent For the Euler equations of gas dynamics, we have $\,m=3,$

\smallskip \noindent   (2.10)  $ \qquad \displaystyle 
\Omega \,\,=\,\, \biggl\{ \, (\rho \,,\,  q  \,,\,  \epsilon
\,)^{\displaystyle \rm t}
\in \R^{3, \, {\rm t}} \,, \quad \rho \, {\rm >} \, 0\,, \quad \epsilon \,
{\rm >} \,
{{q^{2}}\over{2 \rho}} \, \biggr\}\,.\,$

\smallskip \noindent
Classically, the velocity $\,u\,$ and the internal energy $\,e\,$ are
introduced thanks
to the relations 

\smallskip \noindent   (2.11)  $ \qquad \displaystyle 
q \,\,=\,\, \rho \, u \,\, , \quad \epsilon \,\, = \,\, \rho \, e \,\, + \,\, {1
\over 2}\,\rho \, u^{2}\, \, \,$

\smallskip \noindent
and because the pair  $\,\,$ (mathematical entropy , entropy flux) $\,\,$
is of the
form

\smallskip \noindent   (2.12)  $ \qquad \displaystyle  
(\eta\,,\, \xi) \,\,=\,\, (\, -\rho \, s \,,\, -\rho \, s \, u \,)  \, $

\smallskip \noindent
where the entropy $s$ is the specific  thermostatic entropy (see e.g.
[Du90]), the
relation (2.7) is satisfied and  the definition 3 is perfectly compatible  with
this classical physical model for gas dynamics.  Moreover, the classical
relation between the massic internal energy $\, e \,$, the massic entropy
$\,s\,$ and
the specific volume $\,\, \tau \, \equiv \, {1 \over \rho}\,:\,$

\smallskip \noindent   (2.13)  $ \qquad \displaystyle   
{\rm d}e \,\,= \,\, T\,{\rm d}s \,-\, p \, {\rm d}\tau \,$

\smallskip \noindent
allows us to define the temperature $\,T\,$ and the thermodynamic static
pressure
$\,p\,$ (see e.g. Callen [Ca85]). Then the Gibbs-Duhem relation  associated
to extensive  fields shows
that the specific chemical potential $\,\mu\,$ can be defined as a function of
temperature and of static pressure according to the relation

\smallskip \noindent   (2.14)  $ \qquad \displaystyle  
\mu \,\,= \,\, \mu(T,\, p) \,\, \equiv \,\, e \,- \, T \, s \, 
+ \, p \, \tau\,. \, $

\smallskip \noindent
After some lines of easy algebra, it is possible to express the entropy
variables for
the system of Euler equations for gas dynamics~:

\smallskip \noindent   (2.15)  $ \qquad \displaystyle  
\varphi \,\,= \,\, {1 \over T} \, \bigl( \, \mu \,-\, {1 \over 2}\,u^{2}
\,{\bf ,}\,
u \,{\bf ,}\,-1 \, \bigr) \,.\, $

\smallskip \noindent
It is also clear that the thermodynamic flux $\,j({\scriptstyle
\bullet})\,$ admits
the following simple form~:

\smallskip \noindent   (2.16)  $ \qquad \displaystyle  
j(W) \,\,= \,\,\bigl( \, 0 \,\,,\,\,  p \,\,,\,\, p \,u  \,\,  \bigr)^{ \rm
t}\,. \, $

\smallskip \noindent
The explicit evaluation of matrix $\,Y(v)\,$ is  simple ; we have

\smallskip \noindent   (2.17)  $ \qquad \displaystyle   
Y(v) \, \, =\,\,\,   \pmatrix {1 & 0 & 0
\cr -v & 1 & 0 \cr {1 \over 2}\,v^{2} & -v & 1  \cr }\, \, $

\smallskip \noindent
and the orbit of the particular state $\,\, \displaystyle
\,W\,=\,\bigl(\,\rho\,,\,
\rho\,u\,,\,\rho\,e \,+\, {1 \over 2}\,\rho \, u^{2}\,
\bigr)^{\displaystyle \rm t}\,\,
\in \Omega \,\,\,$ is a parabola that is  expressed by

\smallskip \noindent   (2.18)  $ \qquad \displaystyle    
G \,{\scriptstyle \bullet}\, W\,\,=\,\, \Bigl\{  \bigl(\,\rho\,,\,
\rho\,(u-v)\,,\,\rho\,e \,+\, {1 \over 2}\,\rho \, (u-v)^{2}\,
\bigr)^{\displaystyle
\rm t} \,,\quad v \in \R\, \Bigr\} \,.\,$

\bigskip \bigskip
\centerline{ \smcap 3. \quad Invariance}

\smallskip \noindent  {\bf Hypothesis 3. 

\noindent The mathematical entropy is invariant
under the Galileo group. } 

\smallskip \noindent   (3.1)  $ \qquad \displaystyle    
\forall \, g \in G ,\quad \forall \, W \in \Omega ,\qquad \eta(g \,
{\scriptstyle
\bullet} \, W) \,=\, \eta(W)  \, $

\smallskip \noindent
and this hypothesis implies in particular that

\smallskip \noindent   (3.2)  $ \qquad \displaystyle  
\forall \, v \in \R ,\quad \forall \, W \in \Omega ,\qquad \eta( Y(v) \,
{\scriptstyle
\bullet} \, W) \,=\, \eta(W)  \, $

\smallskip \noindent   (3.3)  $ \qquad \displaystyle     
\forall \, W \in \Omega ,\qquad \eta( R \,  {\scriptstyle \bullet} \, W)
\,=\, \eta(W) .\, $

\bigskip \noindent  {\bf Definition 5. 

\noindent Systems of conservation laws invariant
for the  Galileo group. }

\noindent
The system of conservation laws (2.1) associated with a mathematical entropy
$\,\eta\,$ and an entropy flux satisfying hypothesis 2 (relation (2.7)) is
said to be
invariant for the Galileo group $\,\, G\,\,$  of transformations if for each
{\bf regular} solution $\,\, \R^2 \, \supset \Xi^{\displaystyle \rm t} \ni
(t,\,x)
\longmapsto W(t,x) \in \Omega \,\,$ of the conservation law (2.1) and each
Galilean
transformation $\,\, g \in G ,\,$ the function $\,\, V(\theta,\, \xi) \,\,$
defined by the relation 

\smallskip \noindent   (3.4)  $ \qquad \displaystyle  
\R^2  \,\,  \supset \,\,  g(\Xi^{\displaystyle \rm t}) \,\, \ni (\theta,\, \xi)
\longmapsto V(\theta,\, \xi) \,=\, (g\,  {\scriptstyle \bullet} \, W)
(g^{-1}(\theta,\,
\xi))\,  \in \Omega \, $

\smallskip \noindent
is also a regular solution of the conservation law (2.1). It is the case in
particular
when $\,\, g \,=\, y\ib{\displaystyle \, v} \,\,$ is a special Galilean
transform~:

\setbox21=\hbox{$\displaystyle   
{{\partial}\over{\partial \theta}} \bigl[ Y(v) \, {\scriptstyle \bullet} \,
W(\theta,\, \xi+v \theta) \bigr]  \,+ \,    $}
\setbox22=\hbox{$\displaystyle \quad   
+\,  {{\partial}\over{\partial \xi}} \Bigl[ u  ( Y(v) \, {\scriptstyle
\bullet} \,  W(\theta,\, \xi+v \theta)   \, (Y(v) \, {\scriptstyle \bullet} \,
W(\theta,\, \xi+v \theta))   $}
\setbox23=\hbox{$\displaystyle \qquad    \quad   
\,+\,  j(Y(v) \, {\scriptstyle \bullet} \, W(\theta,\, \xi+v
\theta)) \Bigr] \,\,=\,\, 0  \,   $ }
\setbox30= \vbox {\halign{#\cr \box21 \cr   \box22 \cr  \box23 \cr    }}
\setbox31= \hbox{ $\vcenter {\box30} $}
\setbox44=\hbox{\noindent  (3.5) $\displaystyle \, \,   \left\{ \box31 \right. $}  
\smallskip \noindent $ \box44 $

\setbox21=\hbox{$\displaystyle  
{{\partial}\over{\partial \theta}} \,\eta \Bigl(  Y(v)  {\scriptstyle
\bullet}  W(\theta,\,\xi+v \theta) \Bigr) \, + \,  $}
\setbox22=\hbox{$\displaystyle \qquad    \, + \, 
{{\partial}\over{\partial \xi}}
\Bigl[ u  ( Y(v)  {\scriptstyle \bullet} W(\theta,\, \xi+v \theta))
\,\eta \bigl( Y(v)  {\scriptstyle \bullet}  
W(\theta,\, \xi+v \theta) \bigr)  \Bigr] \,=\, 0 \,   $ }
\setbox30= \vbox {\halign{#\cr \box21 \cr   \box22 \cr     }}
\setbox31= \hbox{ $\vcenter {\box30} $}
\setbox44=\hbox{\noindent  (3.6) $\displaystyle \, \,   \left\{ \box31 \right. $}  
\smallskip \noindent $ \box44 $

\smallskip \noindent
and when $\,\, g \,=\, q \,\,$ is the reflection operator :

\setbox21=\hbox{$\displaystyle  
{{\partial}\over{\partial \theta}}  \bigl( R \, {\scriptstyle \bullet} \,
W(\theta,\, -\xi) \bigr) \, + \,   $}
\setbox22=\hbox{$\displaystyle \quad    \, + \, 
{{\partial}\over{\partial \xi}} \Bigl[ u  (
 R \, {\scriptstyle \bullet} \, W(\theta,\, -\xi)) \,\, R \, {\scriptstyle
\bullet} \,
W(\theta,\, -\xi)  \,+\, j \bigl(  R \, {\scriptstyle \bullet} \,
W(\theta,\, -\xi) \bigr) \Bigr]  \,=\, 0  \,  $ }
\setbox30= \vbox {\halign{#\cr \box21 \cr   \box22 \cr     }}
\setbox31= \hbox{ $\vcenter {\box30} $}
\setbox44=\hbox{\noindent  (3.7) $\displaystyle \, \,   \left\{ \box31 \right. $}  
\smallskip \noindent $ \box44 $

\smallskip \noindent   (3.8)  $ \qquad \displaystyle   
{{\partial}\over{\partial \theta}} \,\eta \Bigl( R \, {\scriptstyle \bullet} \,
W(\theta,\, -\xi) \Bigr) \, + \, {{\partial}\over{\partial \xi}}
\Bigl[ u  (  R \, {\scriptstyle \bullet} \, W(\theta,\, -\xi)) \,\, \eta \bigl(
R \, {\scriptstyle \bullet} \, W(\theta,\, -\xi) \bigr)  \Bigr] \,=\, 0 \,.\, $

 \bigskip  \noindent  {\bf Proposition 3. \quad Transformation of the
velocity
field.}

\noindent
We have the following properties for the  velocity field associated with a
special Galilean transformation and with the space reflection when the
hypotheses 1,
2  and 3 are satisfied~:

\smallskip \noindent   (3.9)  $ \qquad \displaystyle   
\forall \, v \in \R ,\quad \forall \, W \in \Omega ,\qquad  u \bigl( Y(v) \,
{\scriptstyle \bullet} \,W \bigr) \,\,=\,\, u(W) - v \,  $ 

\smallskip \noindent   (3.10)  $ \qquad \displaystyle  
\forall \, W \in \Omega ,\qquad  u \bigl( R \,
{\scriptstyle \bullet} \,W \bigr) \,\,=\,\, -u(W) \,.\,\, $

\bigskip  \noindent {\bf Proof of Proposition 3.}

\noindent
We first consider the elementary calculus that  express the partial
derivatives on each
side of the Galilean transformation~: $\,\, {{\partial}\over{\partial \theta}}
\,=\, {{\partial}\over{\partial t}} \,+\, v\,  {{\partial}\over{\partial x}}
\,\,;\,\,  {{\partial}\over{\partial \xi}} \,\,=\,\,
{{\partial}\over{\partial x}}
\,\, $ and we remark that the relation (3.2) implies

\smallskip \noindent   (3.11)  $ \qquad \displaystyle   
\eta \bigl( Y(v) \,  {\scriptstyle \bullet} \,W (\theta,\, \xi+v \theta)
\bigr) \,=\, \eta \bigl(W(\theta,\, \xi+v \theta) \bigr)\,.\,\, $

\smallskip \noindent
We develop the left hand side of relation (3.6). We get

\smallskip \noindent  $  \displaystyle   
\varphi \, \Bigl[  {{\partial W}\over{\partial t}} \,+\, v\,  {{\partial
W }\over{\partial x}} \Bigr] \,+\, \, \, {{\partial}\over{\partial \xi}}
\Bigl[ \, u  ( Y(v)  {\scriptstyle \bullet} W(\theta,\, \xi+v \theta))   \,
\, \eta \bigl( Y(v)  {\scriptstyle \bullet}  
W(\theta,\, \xi+v \theta) \bigr) \,\Bigr]  \,\,= $

\smallskip \noindent  $  \displaystyle     \qquad = \quad
-\,  {{\partial}\over{\partial x}} \bigl[ u(W(\theta, \xi+ v \theta)) \,\, \eta
(W(\theta, \xi+ v \theta)) \bigr]  $

\smallskip \noindent  $  \displaystyle     \qquad = \quad
 \,+\, v \, {\rm d}\eta (W(\theta, \xi+ v
\theta)) \,  {\scriptstyle \bullet} \, {{\partial W }\over{\partial x}}
(\theta, \xi+ v \theta) \,+\, \,$

\smallskip \noindent  $\displaystyle   \qquad \qquad + \,
{{\partial }\over{\partial \xi}} \bigl[ u(Y(v) \, \,  {\scriptstyle
\bullet} \, W
(\theta, \xi+ v \theta)) \,\, \eta (W (\theta, \xi+ v \theta)) \bigr] \,\,
\hfill $  due to (3.11)

\smallskip \noindent $\displaystyle  \qquad = \quad
{{\partial }\over{\partial x}} \Bigl\{ \, \Bigl[ \, -u(W) \,+\, v \,+\,
u(Y(v) \,
{\scriptstyle \bullet} \, W ) \,\Bigr] \,\, \eta \bigl( W(\theta,\, \xi+ v
\theta)
\bigr) \, \Bigr\} \,\,=\,\, 0 \,.\, $
\smallskip \noindent

This last expression is identically null  for any regular solution $\,
W(t,x)\,$. In
consequence the coefficient in front of $\,\, \eta(W(\theta,\, \xi+v
\theta)) \,\,$ is
null  (see e.g. Serre [Se82]) and this fact is  exactly expressed by the
relation
(3.9). In order to prove the relation (3.10), we develop the left hand side of
the relation (3.8). We obtain

\smallskip \noindent  $  \displaystyle   
{\rm d}\eta(W (\theta,\,-\xi)) \, {{\partial W}\over{\partial t}}(\theta,\,-\xi)
\,+\,  {{\partial }\over{\partial \xi}} \Bigl[ \, u(R \, {\scriptstyle
\bullet} \, W
(\theta,\,-\xi)) \,\,\, \eta(W (\theta,\,-\xi)) \, \Bigr] \,\,=\,     $

\smallskip \noindent  $  \displaystyle \quad = \,  
- {{\partial }\over{\partial x}} \Bigl[ \,u(W (\theta,\,-\xi)) \,\,  \eta(W
(\theta,\,-\xi)) \, \Bigr] \, +\,  {{\partial }\over{\partial \xi}} \Bigl[
\, u(R \, {\scriptstyle \bullet} \, W (\theta,\,-\xi)) \, \,\, \eta(W
(\theta,\,-\xi)) \, \Bigr] \,\,$
 
\smallskip \noindent  $  \displaystyle \quad = \,  
{{\partial }\over{\partial \xi}} \Bigl\{ \, \Bigl[ \, u(W (\theta,\,-\xi))
\,+\, u(R \, {\scriptstyle \bullet} \, W (\theta,\,-\xi)) \,  \Bigr] \,\, \eta (W
(\theta,\,-\xi)) \, \Bigr\} \,\, = \,\, 0 \,.\, \,$

\smallskip \noindent
Then the bracket is null as above and the relation (3.10) is established.
$ \hfill \square \kern0.1mm    $

\bigskip \noindent  {\bf Proposition 4. \quad Transformation of the
thermodynamic
flux.}

\noindent
Let (2.1) a system of conservation laws satisfying the hypotheses 1 to 3 and
invariant for the Galileo group. Then for a special Galilean transformation $\,
y\ib{\, \displaystyle v} \,$  and the space reflection $ \, q \,$ we have 

\smallskip \noindent   (3.12)  $ \qquad \displaystyle  
\forall \, v \in \R ,\quad \forall \, W \in \Omega ,\qquad  j \bigl( Y(v) \,
{\scriptstyle \bullet} \,W \bigr) \,\,=\,\, Y(v) \,  {\scriptstyle \bullet}
\, j(W)  \, $ 

\smallskip \noindent   (3.13)  $ \qquad \displaystyle   
\forall \, W \in \Omega ,\qquad  j \bigl( R \, {\scriptstyle \bullet} \,W \bigr)
\,+\, R \,   {\scriptstyle \bullet} \, j(W) \,\,= \,\, 0 \,. \, $

\bigskip  \noindent {\bf Proof of Proposition 4.}

\noindent
For a special Galilean transformation, we have from the relation (3.5)~:

\smallskip \noindent  $ \displaystyle  
Y(v) \, {\scriptstyle \bullet} \, \Big[ {{\partial W}\over{\partial t}} + v
{{\partial W}\over{\partial x}} \Bigr] \,+\, {{\partial}\over{\partial \xi}}
\Bigl\{ \, u(Y(v) \,  {\scriptstyle \bullet} \,W)) \, (Y(v) \,
{\scriptstyle \bullet}
\,W) \,+\, j(Y(v) \,  {\scriptstyle \bullet} \,W) \, \Bigr\} \,\,= \,   $

\smallskip \noindent  $ \displaystyle    \quad = \, 
- {{\partial }\over{\partial x}} \Bigl\{ \, Y(v) \,  {\scriptstyle \bullet}
\, [ u(W) \, W \, +\, j(W) ] \, \Bigr\} 
\,+\,  {{\partial }\over{\partial x}} \Bigl( \, v \,
Y(v)  \,  {\scriptstyle \bullet} \, W \, \Bigr) \,\,+\, $

\smallskip \noindent $\displaystyle   \qquad \qquad + \,
 {{\partial}\over{\partial x}} \Bigl\{ \, u(Y(v) \,  {\scriptstyle \bullet}
\,W)) \, (Y(v) \,  {\scriptstyle \bullet} \,W) 
\,+\, j(Y(v) \,  {\scriptstyle \bullet} \,W) \, \Bigr\} \,\, \,$
 
\smallskip \noindent  $ \displaystyle    \quad = \, 
{{\partial }\over{\partial x}} \Bigl\{ \, \bigl[ -u(W) \,+\,v \,+\, u(Y(v) \,
{\scriptstyle \bullet} \,W) \bigr] \, (Y(v) \,  {\scriptstyle \bullet} \,W) \, + \,$

\smallskip \noindent $\displaystyle   \qquad \qquad  
\,+\, \bigl[ - Y(v) \,  {\scriptstyle \bullet} \, j(W) 
\,+\, j(Y(v) \,  {\scriptstyle \bullet} \,W) \bigr] \, \Bigr\} \,$

\smallskip \noindent  $ \displaystyle    \quad = \,  
{{\partial }\over{\partial x}} \Bigl\{ \, \bigl[ - Y(v) \,  {\scriptstyle
\bullet} \, j(W) \,+\, j(Y(v) \,  {\scriptstyle
\bullet} \,W) \bigr] \, \Bigr\} \, \hfill $  due to (3.9)

\smallskip \noindent  $ \displaystyle    \quad = \,    0 \, \hfill $
according to the relation (3.5).

\smallskip \noindent
Then  the relation (3.12) is established~; the end of the proof is obtained
by writing
that the conservation law (3.7) is satisfied for $\,R\,{\scriptstyle
\bullet}\,W \,$~:

\smallskip \noindent  $ \displaystyle   
{{\partial }\over{\partial \theta}}  \,\bigl( R\,{\scriptstyle \bullet}\,W
\bigr)
\,+ \,  {{\partial}\over{\partial \xi}} \,\bigl(f( R\,{\scriptstyle \bullet}\,W)
\bigr) \,\, = \,\,  \,$

\smallskip \noindent  $ \displaystyle    \quad = \,  
 {{\partial  }\over{\partial t}} \, \bigl(  R\,{\scriptstyle
\bullet}\,W \bigr) \,-\, {{\partial}\over{\partial x}} \Bigl[ \, u(
R\,{\scriptstyle \bullet}\,W)\, (R\,{\scriptstyle \bullet}\,W ) 
\,+\, j( R\,{\scriptstyle \bullet}\,W ) \, \Bigr] \,   \,$

\smallskip \noindent  $ \displaystyle    \quad = \,  
R \,{\scriptstyle \bullet}\,  {{\partial W }\over{\partial t}} \,-\,
 {{\partial}\over{\partial x}} \Bigl[ \, -u(W)\,\,(R\,{\scriptstyle
\bullet}\, W) \,+\, j(R \,{\scriptstyle \bullet}\, W) \, \Bigr]  $

\smallskip \noindent  $ \displaystyle    \quad = \,   
 - R \,{\scriptstyle \bullet}\, {{\partial}\over{\partial x}}  \Bigl[ \,
u(W) \, W \,+\, j(W) \, \Bigr] 
\,+\,  {{\partial}\over{\partial x}}  \Bigl[ u(W) \,R
\,{\scriptstyle \bullet}\, W \,-\, j(R\,{\scriptstyle \bullet}\,W) \Bigr]   $

\smallskip \noindent  $ \displaystyle    \quad = \,    
- {{\partial}\over{\partial x}}
\Bigl[ \, R \,{\scriptstyle \bullet}\, j(W) \,\,+\,\, j(R\,{\scriptstyle
\bullet}\,W)  \,  \Bigr]   $

\smallskip \noindent  $ \displaystyle    \quad = \,    
0 \,.\,   $

\smallskip  \noindent
Then the  relation (3.13) is established and the proposition 4 is proven.
$ \hfill \square \kern0.1mm    $

\bigskip \bigskip
\centerline{ \smcap 4. \quad Null-velocity manifold}

\smallskip \noindent  {\bf Definition 6. $\quad$  Null-velocity manifold. }

\noindent
We denote by $\,\Omega_{0}\,$ the null-velocity manifold, {\it i.e.} the
set of all
states $\,  W \in \Omega \,$ whose associated velocity $\,u(W)\,$ is equal
to zero~:

\smallskip \noindent   (4.1)   $ \qquad \displaystyle    
W \in \Omega_0 \quad $ if and only if $ \quad u(W)\,=\,0. $

\smallskip \noindent
We remark that the denonination of manifold is appropriate because the
function $\,
\Omega \ni W \longmapsto u(W) \in \R \,$ is regular.

\bigskip \noindent  {\bf Proposition 5. \quad  Decomposition of the space of
states.}

\noindent
The open cone $\,\Omega\,$ is decomposed under the following form

\smallskip \noindent   (4.2)   $ \qquad \displaystyle     
\Omega \,\,=\,\, \union_{\displaystyle v \in \R} \, Y(v) \, {\scriptstyle
\bullet} \, \Omega_0 \,$

\smallskip \noindent
and the sets $\,Y(v)  \, {\scriptstyle \bullet} \, \Omega_0 \,$ have no two
by two
intersection : 

\smallskip \noindent   (4.3)   $ \qquad \displaystyle    
\forall \, v,\,w \in \R\,,\qquad v \ne w \,\,\,\,  \Longrightarrow \,\,\,\,
\bigl(
Y(v) \,{\scriptstyle \bullet}\, \Omega_0 \bigr) \,\, \cap \,\, \bigl(   Y(w)
\,{\scriptstyle \bullet}\, \Omega_0 \bigr) \,\,= \,\, $\O$  \,.\, $

\smallskip  \noindent
In other words, the cone $\,\Omega\,$ is a boundle space with basis
$\,\Omega_0, \,$
fiber $\,G  \,{\scriptstyle \bullet}\, \Omega_0 \,$ over the manifold $\,
\Omega_0\,$ and projection $\,\Pi\,$ defined by

\smallskip \noindent   (4.4)   $ \qquad \displaystyle  
\Omega \ni W \longmapsto \Pi(W) \,=\, Y(u(W)) \,{\scriptstyle \bullet}\, W
\,\, \in \Omega_0\,.  \,  $ 

\bigskip \noindent  {\bf Proof of proposition 5.}
\smallskip \noindent
The relation (4.2) is a consequence of the following remark :

\smallskip \noindent   (4.5)   $ \qquad \displaystyle  
Y \bigl( u(W) \bigr)  \,{\scriptstyle \bullet}\, W  \,\,  \in \,\Omega_0
\,  $

\smallskip \noindent
that takes into account the relation (3.9) and the definition 6 of $\,
\Omega_0 .$ In a
similar way, $\,Y(v)  \,{\scriptstyle \bullet}\, \Omega_0 \,$ is also the
set of all
the states having a velocity exactly equal to $\,-v\,$; then assertion (4.3) is clear.
The end of the proposition consists  simply in using the vocabulary of
topologists. We
refer the reader for example to the book of Godbillon [Go71].   $ \hfill \square
\kern0.1mm $

\bigskip \noindent  {\bf Proposition 6. \quad  The null velocity manifold is of
co-dimension $\,1 .\,$ } 

\smallskip \noindent   (4.6)   $ \qquad \displaystyle  
\forall \,\,  W \in \Omega_0 \,,\qquad {\rm dim} \,\,T\ib{W}\Omega_0  \,=\,
m-1 \,.\, $  

\bigskip  \noindent {\bf Proof of Proposition 6.}

\noindent
We start from the relation (3.9) : $ \,\, u \bigl( Y(v) \,  {\scriptstyle
\bullet} \,W
\bigr) \,=\, u(W) - v \,\,$ and we derive this expression relatively to the
variable
$\, v .\,$ We obtain $\,\,\,  {\rm d}u \bigl(  Y(v) \,  {\scriptstyle
\bullet} \,W \bigr)
\,  {\scriptstyle \bullet} \, \bigl(  {\rm d}Y(v) \,  {\scriptstyle
\bullet} \,W \bigr)
\,=\, -1 \,,\,\, $ then we consider  the particular value  $\, v= 0 .\,$ It
comes : 

\smallskip \noindent   (4.7)   $ \qquad \displaystyle  
\forall \, W \in \Omega \,,\qquad  {\rm d}u ( W ) \,  {\scriptstyle
\bullet} \, \bigl(
{\rm d}Y(0)  \,  {\scriptstyle \bullet} \,W \bigr) \,\,= \,\, -1 \,.\, $

\smallskip \noindent
For $\,\,  W \in \Omega \,\,$ and $\,\,  r \in \R^{m,\,  { \rm t}} \,,\,$
we have

\smallskip \noindent $ \displaystyle   
{\rm d}u(W) \, {\scriptstyle \bullet} \, \bigl\{ \, r \,+\, ({\rm d}u(W) \,
{\scriptstyle  \bullet} \,r ) \,\, ({\rm d}Y(0) \, {\scriptstyle \bullet}
\, W ) \, \bigr\} \,\,=\,\,  $ 

\smallskip \noindent $ \displaystyle  \qquad = \, 
({\rm d}u(W) \, {\scriptstyle \bullet} \,r) \,+\,  ({\rm
d}u(W) \, {\scriptstyle \bullet} \,r) \,(-1) \,\,= \,\, 0 \,  \,$

\noindent
Then the vector $\,\, r \in \R^{m,\,  { \rm t}} \,\,$ can be decomposed
under the form \br 
$\, r \,=\, \rho \,+\, \theta \, \bigl(  {\rm d}Y(0) \, {\scriptstyle
\bullet} \, W \bigr) \,\, \,$ with $\,\, {\rm d}u(W) \,  {\scriptstyle
\bullet} \,\rho \,=\, 0 \,.\,\, $ If we suppose now that the state $\,W \,$
belongs to
the null-velocity manifold  $\,\, \Omega_0 \,,\,\,$ the
condition  $\,\, {\rm d}u(W) \,  {\scriptstyle \bullet} \,\rho \,=\, 0 \,\,\,$
implies that $\,\,\, \rho \in T\ib{W}\Omega_0 .\, \,$ We deduce from this
point the
following decomposition of space $\, \R^{m,\,  { \rm t}} \,:$

\smallskip \noindent   (4.8)   $ \qquad \displaystyle   
\R^{m,\,  { \rm t}} \,\,\,= \,\,\, T\ib{W}\Omega_0 \,\, \,+\,\,\, \R \, ({\rm
d}Y(0) \, {\scriptstyle \bullet} \, W ) \,,\qquad W \in \Omega_0 \,.\, $
 
\smallskip \noindent
The decomposition (4.8) is in fact a direct sum.  

\noindent If $\,\, r \in \, \bigl(
T\ib{W}\Omega_0 \bigr) \,\, \cap  $ $  \,\,  \bigl( \R \, ({\rm d}Y(0) \, {\scriptstyle
\bullet} \, W )\bigr)  \,,\,$ there exists some scalar $\, \mu \in \R \,$
such that
$\,\, r \, =\, \mu \, {\rm d}Y(0) \, {\scriptstyle \bullet} \, W .\,$ Then, the
property that $\,\, r \in  T\ib{W}\Omega_0 \,$ implies $\,\, {\rm d}u(W) \,
{\scriptstyle \bullet} \,r \,=\, 0.\,$ Due to the relation (4.7), we deduce
that $\,\,
\mu = 0 \,\,$ and the vector $\,\, r\,\, $ is null. Then the property (4.6) is
established.  $ \hfill \square \kern0.1mm    $

\bigskip \vfill \eject \noindent  {\bf Hypothesis 4.

\noindent Null-velocity manifold is invariant by
space reflection. }

\noindent
The null-velocity manifold $\,\Omega_0 \,$  is supposed to be invariant point by
point by space reflection~: 

\smallskip \noindent   (4.9)   $ \qquad \displaystyle   
\forall \, W \in \Omega_0 \,,\qquad R  \,{\scriptstyle \bullet}\,  W \,\,=\,\, W
\,.\, $

\bigskip  \noindent  {\bf Definition 7.\quad  Decomposition of space.}

\noindent
We introduce the two eigenspaces associated with the reflection operator
$\, R :\,$ 

\smallskip \noindent   (4.10)   $ \qquad \displaystyle   
\Lambda\ib{1} \,\,=\,\, \bigl\{ \, W \in \R^{m,\,  { \rm t}} \,,\,\, R
\,  {\scriptstyle \bullet} \, W \,=\, W \, \bigr\} \, \,  $

\smallskip \noindent   (4.11)   $ \qquad \displaystyle     
\Lambda\ib{-1} \,\,=\,\, \bigl\{ \, W \in \R^{m,\,  { \rm t}} \,,\,\, R
\,  {\scriptstyle \bullet} \, W \,=\, -W \, \bigr\} \, .\,  $
 
\smallskip \noindent
An immediate consequence of the property (1.13) is the decomposition

\smallskip \noindent   (4.12)   $ \qquad \displaystyle   
\R^{m,\,  { \rm t}} \,\,= \,\,\Lambda\ib{1} \, \oplus \,
\Lambda\ib{-1} \,. \, $

\bigskip \noindent  {\bf Proposition 7. \quad  Constraint for the
thermodynamic flux.}

\smallskip \noindent   (4.13)   $ \qquad \displaystyle    
\forall \, W \in \Omega_0 \,,\qquad j(W) \in \Lambda\ib{-1} \,. \, $

\bigskip  \noindent {\bf Proof of Proposition 7.}

\noindent
It is an immediate consequence of the relation (3.13) : $\,\, \, j(R\,
{\scriptstyle
\bullet} \, W) \,+\, j(W) \,=\, 0 \,\, \,$ and of the hypothesis 4 : $\,\, R \,
{\scriptstyle \bullet} \, W \,=\, W \,\,$ when $\,\,  W \in \Omega_0 \,.\,$
$ \hfill \square \kern0.1mm $

\bigskip \noindent  {\bf Remark 1. \quad Linear geometry. }

\noindent The hypothesis 4 can also be written as

\smallskip \noindent   (4.14)   $ \qquad \displaystyle  
\Omega_0 \, \subset \, \Lambda\ib{1} \, \, $

\smallskip  \noindent
and the null-velocity manifold is flat.

\bigskip \noindent  {\bf Example. \quad Gas dynamics (ii). }

\noindent
In the case of the Euler equations of gas dynamics, we have

\smallskip \noindent   (4.15)   $ \qquad \displaystyle   
\Lambda_{1} \,\,=\,\, \bigl\{ \, (\rho \,,\, 0 \,,\,
\epsilon)^{\displaystyle \rm t}
\,,\, \rho \in \R \,, \,\, \epsilon \in \R \, \bigr\}$
  
\smallskip \noindent   (4.16)   $ \qquad \displaystyle   
\Lambda_{-1} \,\,=\,\, \bigl\{ \, (0 \,,\, \sigma \,,\, 0 ) ^{\displaystyle
\rm t} \,,\, \sigma \in \R \, \bigr\} \,. $

\smallskip \noindent
We remark also that all the matrices $\, Y(v)\,$ have a common eigenvector that
generates a linear space $\, \Gamma_1 \,$ of dimension $1$ included in $\,
\Lambda_{1} \,$: 

\smallskip \noindent   (4.17)   $ \qquad \displaystyle   
\Gamma_1 \,\,=\,\, \{ \, (0 \,,\, 0 \,,\, \epsilon )^{\displaystyle \rm t}
\,,\, \epsilon \in \R \, \} \,\,\, \subset \,\,\, \Lambda_{1}. $

\smallskip \noindent
Moreover, the half manifold
$\, \, \Gamma_1 ^{+} \,=\, \{  \, (0 \,,\, 0 \,,\, \epsilon)
^{\displaystyle \rm t} \,,\, \epsilon  \, {\rm >} \, 0  \, \} \,$ is a part
of the
boundary of $\, \Omega_0 \,\, \,$ which is composed by (unphysical ?)
states without any matter $\,( \rho \,= \, 0 ),  $
undefined velocity $\displaystyle \, \Bigl( u = {{\rho\,u}\over{\rho}} =
{0\over0}
\Bigr) \, $  and full of energy  $\,( \epsilon  \, {\rm >} \, 0 ) \,$ !

\bigskip \bigskip

\centerline{ \smcap 5. \quad The case $\, m=1$ }

\smallskip \noindent  {\bf Proposition 8. \quad   }

\noindent
It does not exist any hyperbolic equation ($m=1$) invariant for the Galileo
group.

\bigskip  \noindent {\bf Proof of Proposition 8.}

\noindent
The operator $\, R \,$ is linear $\, \R \longrightarrow \R \,$ and
satisfies $\, R^2
\,=\, {\rm Id} .\,$ If $\,\, {\rm dim} \, \Lambda\ib{1} = 1 ,\,\,$ then
$\,\, u(R \,W)
\,+$ $+\, R \,u(W) \,= \, 0 \,\,$ implies that $\,\, \, u(W) \equiv 0 \,\,$
for all
$\,\, W \in \Omega .\,$ Then  $\,\, T\ib{W}\Omega_0 \,=\, \R \,\,$ and this
property
is in contradiction with the proposition 6. If $\,\, {\rm dim}
\,\Lambda\ib{-1} = 1
,\,\,$ {\it i.e.} $\,\, R \, W \, = \, - W ,\,\,$ we have from (1.12) and the
derivability of the function $\,\, \R \ni v \longmapsto Y(v) \,$ : $\,\,
Y(v) \,=\,
a^v \,\,$ for some $\, a \geq 0 .\, $ On the other hand, from relation
(1.14) : $\,\,
Y(v) \, R \, Y(v) \,=\, R \,$ we deduce that $\,\, \forall \, v \in \R,\,\,
Y(v)^2
\equiv 1 .\,\,\, $ Then $\, a = 0 \,$ and $\,\, Y(v) \equiv 1 .\,\,$ But
this property is in
contradiction with the property (3.9) : $\,\,  u ( Y(v) \,W ) \,=\, u(W) -
v \,.\,\, $
The proposition is established.  $ \hfill \square \kern0.1mm $

\bigskip \bigskip
\centerline{ \smcap 6. \quad Galilean invariance for systems
of two conservation laws}

\smallskip \noindent  {\bf Theorem 1. \quad   }

\noindent
When $\,\, \Omega \ \subset \ \, ]0,\, +\infty[ \,\times^{\rm \! t} \,\R \
\subset \ \R^{2, \, { \rm t}} \,,\,$ a system of two conservation
laws invariant for the Galileo group  is parameterized  by the scalars
$\,\, \alpha
\!>\! 0 \,,\,\, \beta \!>\! 0 \, \,$ and by a derivable  strictly convex function $
\,\, ]0,\, +\infty[ \, \, \ni \xi \longmapsto \sigma(\xi) \in \R \,.\,$ It
takes one of
the following forms~:
\smallskip \noindent {\bf (i) \quad  Hyperbolic Galileo}.

\noindent
We have  for this first case

\smallskip \noindent   (6.1)   $ \qquad \displaystyle 
 \sigma'(\xi) < 0  \, $

\smallskip \noindent
and the space of states $\,\, \Omega \,\,$  is included in  the following one : 

\smallskip \noindent   (6.2)   $ \qquad \displaystyle 
\Omega_{+} \,\,=\,\, \biggl\{ \, W \,=\, \pmatrix{\theta\cr \zeta \cr} \in
\R^{2,\,{\rm t}} \,, \quad   \theta > 0 \,, \quad  \abs{\zeta} \, < \,
\sqrt{{{\alpha}\over{\beta}}} \,  \abs{\theta} \,  \biggr\}  \,. \, $

\smallskip \noindent
The  system of conservation laws takes the algebraic form

\smallskip \noindent   (6.3)   $ \qquad \displaystyle  
{{\partial}\over{\partial t}} \pmatrix{\theta \cr \zeta \cr} \,+\,
{{\partial}\over{\partial x}} \, \Biggl\{ \, u(W) \,  \pmatrix{\theta \cr
\zeta \cr}
\,+\, {{\Pi\bigl( \sqrt{\theta^2 - \beta \, \zeta^2 / \alpha \, } \,
\bigr)}\over{\sqrt{\theta^2 - \beta \, \zeta^2 / \alpha \,}}} \,\,
\pmatrix{ \beta \, \zeta / \alpha \cr \theta \cr} \, \Biggr\} \,\,= \,\, 0  \, $ 

\smallskip \noindent
with a velocity $\,\, u({\scriptstyle \bullet} ) \,\,$  given by the relation 

\smallskip \noindent   (6.4)   $ \qquad \displaystyle 
u(W) \,\,=\,\, {{1}\over{\sqrt{\alpha \, \beta}}} \,\,  {\rm argth} \biggl(
\sqrt{{{\beta}\over{\alpha}}} \, {{\zeta}\over{\theta}} \biggr) \, $
 
\smallskip \noindent
and a function $\,\, \Pi({\scriptstyle \bullet})\,\,$ named here the {\bf
mechanical
pressure} and  satisfying the relation 

\smallskip \noindent   (6.5)   $ \qquad \displaystyle 
\Pi(\xi) \,\,=\,\, -{{1}\over{\beta}}
\,{{\sigma^*(\sigma'(\xi))}\over{\sigma'(\xi)}} \,  \,\, $

\smallskip \noindent
if we  denote by $\,\, \sigma^* ( {\scriptstyle \bullet}) \,\,$ the dual
function of
$\, \sigma ({\scriptstyle \bullet}) .\,$  Moreover, the function  $\,\,
\eta({\scriptstyle \bullet})\,\,$ defined by

\smallskip \noindent   (6.6)   $ \qquad \displaystyle  
\eta(\theta,\, \zeta) \,=\, \sigma \biggl( \sqrt{\theta^2 - {{\beta \,
\zeta^2}\over{\alpha}}} \, \biggr) \,\,$

\smallskip \noindent
is a mathematical entropy associated with the hyperbolic system (6.3).

\smallskip \noindent {\bf (ii) \quad  Elliptic Galileo}.

\noindent
In this second case, we have

\smallskip \noindent   (6.7)   $ \qquad \displaystyle  
 \sigma'(\xi) > 0 \,. \, $

\smallskip \noindent
The  elliptic Galileo system of conservation laws admits the expression 

\smallskip \noindent   (6.8)   $ \quad \displaystyle  
{{\partial}\over{\partial t}} \pmatrix{\theta \cr \zeta \cr} \,+\,
{{\partial}\over{\partial x}} \, \Biggl\{ \, u(W) \,  \pmatrix{\theta \cr
\zeta \cr}
\,+\, {{\Pi\bigl( \sqrt{\theta^2 + \beta \, \zeta^2 / \alpha \,}\,
\bigr)} \over {\sqrt{\theta^2 + \beta \, \zeta^2 / \alpha \,}}} \,\,
\pmatrix{ -\beta \, \zeta / \alpha \cr \theta \cr} \, \Biggr\} \,=\, 0 \, $
 
\smallskip \noindent
with a velocity $\,\, u({\scriptstyle \bullet}),\,\,$  a mechanical pressure
$\,\, \Pi ({\scriptstyle \bullet}) \,\,$ and a mathematical entropy $\,\, \eta
({\scriptstyle \bullet}) \,\,$ defined by the relations

\smallskip \noindent   (6.9)   $ \qquad \displaystyle  
u(W) \,\,=\,\, {{1}\over{\sqrt{\alpha \, \beta}}} \,\,  {\rm arctg} \biggl(
\sqrt{{{\beta}\over{\alpha}}} \, {{\zeta}\over{\theta}} \biggr) \,$

\smallskip \noindent   (6.10)   $ \qquad \displaystyle   
\Pi(\xi) \,\,=\,\, {{1}\over{\beta}}
\,{{\sigma^*(\sigma'(\xi))}\over{\sigma'(\xi)}} \,$

\smallskip \noindent   (6.11)   $ \qquad \displaystyle    
\eta(\theta,\, \zeta) \,=\, \sigma \biggl( \sqrt{\theta^2 + {{\beta \,
\zeta^2}\over{\alpha}}} \, \biggr) \,.\, \, $

\bigskip  \noindent {\bf Proof of Theorem 1.}

\smallskip \noindent $\bullet \quad$
We have $\,\, R^2 \,=\, {\rm Id} \,\,$ in the linear space $\,\, \R^{2,\,
{\rm t}}
.\,\,$ If $\, \, {\rm dim}\, \Lambda\ib{1}  = 2 ,\,$ then $\,\, R \,=\, {\rm Id}
,\,\,\,\, \Lambda\ib{-1} \,=\, \{0\} \,\,$ and due to the relation  (4.13),  $\,\,
j(W_0) \,\,$ belongs to $\,\, \Lambda\ib{-1} \,\,$ when $\, W_0 \,$ lies in $\,
\Omega_0.\,$ Then $\,\,  j(W_0) \,  =\, 0 \,$ if $\, W_0 \in \Omega_0.\,$
We deduce from (3.12) that  $\,\,
j\bigl( Y(v) \,{\scriptstyle \bullet}\, W_0 \bigr) \,=\, Y(v) \,{\scriptstyle
\bullet}\,j(W_0) \,\,$ for each $\, W_0 \in \Omega_0 .\,$ Then, according
to (4.2) and
the preceding point, we have  $\,\, j(W) \,=\, 0 \,\, $  for each $\, W \in
\Omega .\,$
We deduce from the proposition  2 and  the property (2.9) that $\,\, {\rm
d}u \,\equiv
\, 0 \,\,$ and this fact contradicts the relation (4.7). Then $\, \, {\rm dim}\,
\Lambda\ib{1}  \le 1 .\, \,$ Moreover the unidimensional (due to (4.6))
flat manifold
$\,\, T\ib{W}\Omega_0 \,\,$ is included in $\, \Lambda\ib{1} . \,$ Then $\,
\, {\rm
dim}\, \Lambda\ib{1}  \ge 1 \,\,$ and  $\, \, {\rm dim}\, \Lambda\ib{1}
\,=\, {\rm
dim}\, \Lambda\ib{-1}  \,=\, 1 .\, \,$

\smallskip \noindent $\bullet \quad$
We differentiate the relation (1.14) relatively to the variable $\, v \,:\,$

\smallskip \noindent  $  \displaystyle     
{\rm d} Y(v) \,{\scriptstyle \bullet}\, R \,{\scriptstyle \bullet}\,Y(v)
\,+\, Y(v)  \,{\scriptstyle \bullet}\, 
R \,{\scriptstyle \bullet}\, {\rm d} Y(v) \,\,=\,\, 0 \, $ \quad  
and we take $\, v=0 \,: \,$

\smallskip \noindent   (6.12)   $ \quad \displaystyle  
{\rm d} Y(0) \,{\scriptstyle \bullet}\, R \,\,+\,  R \,{\scriptstyle
\bullet}\, {\rm d} Y(0) \,\,=\,\, 0 \,.\, $

\smallskip \noindent
For $\,\, r \in \Lambda\ib{1} ,\,$ we have $\,\,
{\rm d} Y(0) \,{\scriptstyle \bullet}\, r \,\,+\,  R \,{\scriptstyle
\bullet}\, {\rm
d} Y(0)  \,{\scriptstyle \bullet}\, r \,=\, 0 \,\,$ and

\smallskip \noindent   (6.13)   $ \quad \displaystyle   
\forall \, r \in \Lambda\ib{1} \,,\qquad {\rm d} Y(0) \,{\scriptstyle
\bullet}\, r \in \Lambda\ib{-1} \,.\,$

\smallskip \noindent
In a similar way, for $\,\, r \in \Lambda\ib{-1} ,\,$ we have $\,\,  -{\rm
d} Y(0)
\,{\scriptstyle \bullet}\, r \,\,+\,  R \,{\scriptstyle \bullet}\, {\rm d}
Y(0)  \,
{\scriptstyle \bullet}\, r \,=\, 0 \,\,$ and this implies

\smallskip \noindent   (6.14)   $ \quad \displaystyle   
\forall \, r \in \Lambda\ib{-1} \,,\qquad {\rm d} Y(0) \,{\scriptstyle
\bullet}\, r \in \Lambda\ib{1} \,.\,$

\smallskip \noindent $\bullet \quad$
Let $\,\, (r_+ ,\, r_-) \,\,$ be a basis of the linear space  $\,\,
\R^{2,\, {\rm t}}
\,\,$ composed by a non null vector  $\,\, r_+ \,\,$  of $\, \Lambda\ib{1}
\,$ and
a non null vector  $\,\, r_- \,\,$ of $\, \Lambda\ib{-1} .\,$  We define
$\,\, \alpha
\geq 0 \, \,$ by the condition $\,\, {\rm d}Y(0)  \,{\scriptstyle
\bullet}\, r_+ \,=\,
- \alpha \, r_- \,\,$  after an eventual change of the  sign of $\, r_- . $
  Due to (6.14),  the vector 
$\,\, {\rm d}Y(0)  \,{\scriptstyle \bullet}\, r_- \,\,$
belongs to the linear space $\,\, \Lambda\ib{1} \,\,$ 
and can be written under the form : $\,\,
{\rm d}Y(0)  \, {\scriptstyle \bullet}\, r_- \,=\, 
\beta \, r_+ \,.\,\,$  If the scalar $\,\, \alpha \,\,$ is  null, 
we can express the matrix $\, Y(v) \,$ as

\smallskip \noindent   (6.15)   $ \quad \displaystyle   
Y(v) \,\,= \,\, {\rm e}^{\displaystyle v \, {\rm d}Y(0)} \,\,= \,\,  {\rm
exp} \, \biggl[ \, v \,\pmatrix{0 & \beta \cr 0 & 0 \cr} \, \biggr]  
\,\,= \,\, \pmatrix {1 & \beta  \, v \cr  0   & 1 \cr} \,;\,$

\smallskip \noindent
and for $\,\, W_0 \,=\, (\theta_0 ,\, 0)^{\rm t} \,\in \Omega_0 \,,\,$  we
have $\,\,
Y(v) \, {\scriptstyle \bullet}\, W_0  \, =\,   (\theta_0 ,\, 0)^{\rm t}
.\,\,$  Due
to the relation (4.2), this property implies that $\,\, \Omega \,\,$ is
included inside
the subspace $\,\, \Lambda\ib{1},\,\,$ that contradicts the definition~2
that claims
that $\,\, \Omega \,\, $ is an open set of $\,\, \R^{2,\, {\rm t}} .\,\,$
Then $ \,\,
\alpha \!>\! 0 \,.\,\,$

\smallskip \noindent $\bullet \quad$
If the scalar $\,\, \beta \,\,$ is  null, we can express the matrix $\,
Y(v) \,$ as 

\smallskip \noindent   (6.16)   $ \quad \displaystyle   
Y(v) \,\,= \,\, {\rm e}^{\displaystyle v \, {\rm d}Y(0)} \,\,= \,\,  {\rm
exp} \,
\biggl[ \, v \,\pmatrix{0 & 0 \cr -\alpha & 0 \cr} \, \biggr] \,\,= \,\, \pmatrix {1 &
0 \cr -\alpha \, v  & 1 \cr} \,.\,$

\smallskip \noindent
Then for $\,\, W_0 \,=\, (\theta_0 ,\, 0)^{\rm t} \,\in \Omega_0 \,,\,$  we
have $\,\,
Y(v) \, {\scriptstyle \bullet}\, W_0  \, =\,  (\theta_0 ,\,-v \,
\theta_0)^{\rm t}
\,\,$ with $\,\, u( Y(v) \, {\scriptstyle \bullet}\, W_0 )\,=\, -v \,\,$
due to the
relation (3.9). We deduce the expression $\,\, u(W) \,=\, \zeta / \theta  \,$
for the velocity field. Moreover due to the invariance (3.2) of the mathematical
entropy, we have the following calculus~:

\smallskip \noindent  $  \displaystyle    
\eta (\theta_0 ,\,-v \, \theta_0) \,=\, \eta(W) \, =\,  \eta \bigl( Y(v) \,
{\scriptstyle \bullet}\, W_0 \bigr) \,=\, \eta(W_0) \,=\, \eta(\theta_0,\,
0) \,  $

\smallskip \noindent
and the mathematical entropy is function of the {\bf unique} variable $\,
\theta .\,$
In consequence the function $\,\, \eta({\scriptstyle \bullet},\, {\scriptstyle
\bullet}) \,\,$  can not be a {\bf strictly} convex function of the  pair $\,
(\theta,\, \zeta) .\,$ Due to the general choices done in the section 2,  this
case must be excluded and  the matrix of the operator $\, {\rm d}Y(0) \,$ has
relatively to this basis one among  the two  following expressions~:

\smallskip \noindent   (6.17)   $ \quad \displaystyle   
{\rm d}Y(0) \,\,= \,\, \pmatrix { 0 & -\beta  \cr -\alpha & 0 \cr }
\,,\qquad \alpha >0 \,,\quad \beta > 0 \,, \,\,$

\smallskip \noindent   (6.18)   $ \quad \displaystyle    
{\rm d}Y(0) \,\,= \,\, \pmatrix { 0 & \beta  \cr -\alpha & 0 \cr }
\,,\qquad \alpha >0 \,,\quad \beta > 0 \,. \,\,$

\smallskip \noindent $\bullet \quad$ {\bf Case (i).} \quad
When the operator is defined in the basis $\,\, (r_+,\, r_-) \in \Lambda\ib{1}
\times \Lambda\ib{-1} \, \,$ with the matrix (6.17), the end of the
construction of
the system of conservation laws can be done as follows. We first remark that

\smallskip \noindent   (6.19)   $ \quad \displaystyle   
\bigl( {\rm d}Y(0) \bigr)^2 \,\,= \,\, \alpha \, \beta \,\,\,  {\rm Id} \,;
\, $

\smallskip \noindent
 we have the  expansion

\smallskip \noindent  $  \displaystyle    
Y(v) \,\,= \,\,  {\rm e}^{\displaystyle v \, {\rm d}Y(0)} \,\,= \,\,{\rm
Id} \,+\, v
\, {\rm d}Y(0) \,+\, {{v^2}\over{2 !}} \,  \alpha \, \beta \, {\rm Id} \,+\,
{{v^3}\over{3 !}} \,  \alpha \, \beta \, {\rm d}Y(0) \,+\,\cdots \,\, $ 
 
\smallskip \noindent
and the sum of the previous series is equal to

\smallskip \noindent   (6.20)   $ \quad \displaystyle    
Y(v) \,\,= \,\, \pmatrix { {\rm ch} \bigl( v \, \sqrt{ \alpha \, \beta \, }\,
\bigr) &  -\sqrt{\beta / \alpha \,} \, \, {\rm sh} \bigl( v \, \sqrt{\alpha
\, \beta} \,\bigr) \cr 
-\sqrt{\alpha / \beta \,} \, {\rm sh} \bigl( v \, \sqrt{\alpha \, \beta}
\,\bigr) &  {\rm ch} \bigl( v \, \sqrt{ \alpha \, \beta \, } \,\bigr) \cr
}\,.\, $
 
\smallskip \noindent
For a state $\,\, W_0 \,=\, (\theta_0,\, 0)^{\rm t} \in \Omega_0 \,,\,$ we
have due to the expression  (6.20),

\smallskip \noindent  $  \displaystyle   
W \,=\, Y(v) \, {\scriptstyle \bullet}\, W_0 \,\,=\,\, \pmatrix {
\theta_0 \, {\rm ch} \bigl( v \, \sqrt{\alpha \, \beta \, } \, \bigr) \cr
-\sqrt{\alpha / \beta \,} \,\,\theta_0 \,\,\, {\rm sh} \bigl( v \,
\sqrt{\alpha \, \beta} \,\bigr) \cr } \,   $

\smallskip \noindent
with $\,\, u(W) \,=\, -v. \,\,$ We deduce the relation (6.4) with the adding
condition 

\smallskip \noindent   (6.21)   $ \quad \displaystyle   
\theta_0 \,\,=\,\, {{\theta}\over{{\rm ch} \,\bigl(  \, u(W) \,
\sqrt{\alpha \, \beta
\, } \,\bigr)}} \,\,=\,\, \sqrt{\theta^2 - {{\beta \,
\zeta^2}\over{\alpha}}\,} \qquad
{\rm if} \,\,\, W \,=\, (\theta ,\, \zeta)^{\rm t} \in \Omega_+ \,.\, $

\smallskip \noindent $\bullet \quad$
We focus now on the mathematical entropy. We first note $\,\, \sigma
({\scriptstyle
\bullet})  \,\,$ the restriction of the mathematical entropy $\,\, \eta
({\scriptstyle
\bullet}) \,\,$ to the subset  $\,\, \Omega_0 \,=\, \Omega \, \cap \,
\bigl( \, ]0,\,
+\infty [ \times^{\rm \! t} \{ 0 \} \, \bigr)  .\,\,$ With the preceding notations, we
have necessarily from the hypothesis (3.2) $\, \,\eta(W) \,$ $=\, \eta(W_0)
\,$ $=\,
\sigma(\theta_0) \, \,$ that establishes exactly the relation (6.6). A natural
question is to verify that the mathematical entropy $\, \eta({\scriptstyle
\bullet})
\,$ is a  strictly convex function, when $\, \sigma ({\scriptstyle \bullet}) \,$
satisfies the same property. We set

\smallskip \noindent   (6.22)   $ \quad \displaystyle   
\xi \,\,=\,\, \sqrt{\theta^2 - {{\beta \, \zeta^2}\over{\alpha}}} \, $
 
\smallskip \noindent
and we have from the relation (6.6) the following calculus :

\smallskip \noindent $  \displaystyle  
\xi \, {\rm d} \xi \,\,= \,\, \theta \,  {\rm d}\theta \,-\,
{{\beta}\over{\alpha}}
\, \zeta \,  {\rm d}\zeta \, \,,\qquad
{{\partial \eta}\over{\partial \theta}} \,=\, {{\theta}\over{\xi}} \, \sigma'
\,,\qquad   {{\partial \eta}\over{\partial \zeta}} \,=\,
-{{\beta}\over{\alpha}} \,
{{\zeta}\over{\xi}} \, \sigma'  \,, \,  $

\smallskip \noindent $  \displaystyle   
 {{\partial^2 \eta}\over{\partial \theta^2}} \,=\, -{{\beta}\over{\alpha}} \,
{{\zeta^2}\over{\xi^3}} \,\sigma'  \,+\,   {{\theta^2}\over{\xi^2}} \, \sigma ''
\,,\quad  {{\partial^2 \eta}\over{\partial  \theta \, \partial \zeta}}
\,=\, {{\beta}\over{\alpha}} \,   {{\theta \, \zeta}\over{\xi^3}} \,\sigma'
\,-\,
{{\beta}\over{\alpha}} \,   {{\theta \, \zeta}\over{\xi^2}} \, \sigma ''
\,,\quad $

\smallskip \noindent $  \displaystyle  
{{\partial^2 \eta}\over{\partial \zeta^2}} \,=\, -{{\beta}\over{\alpha}} \,
{{\theta^2}\over{\xi^3}} \, \sigma'  \,+\, {{\beta^2}\over{\alpha^2}} \,
{{\zeta^2}\over{\xi^2}} \, \sigma'' .  \,  $ \quad  
Then

\smallskip \noindent $  \displaystyle  
{\rm det} \bigl( {\rm d}^2\eta \bigr) \,\,=
\,\,{{\partial^2 \eta}\over{\partial \theta^2}} \,\,  {{\partial^2
\eta}\over{\partial
\zeta^2}} \,-\, \Bigl( {{\partial^2 \eta}\over{\partial  \theta \, \partial
\zeta}}
\Bigr)^2 \,\,= \, $

\smallskip \noindent $  \displaystyle  \quad = \,  
\Bigl( -{{\beta}\over{\alpha}} \,  {{\zeta^2}\over{\xi^3}} \,\sigma'  \,+\,
{{\theta^2}\over{\xi^2}} \, \sigma '' \Bigr) \,\, \Bigl(
-{{\beta}\over{\alpha}} \,
{{\theta^2}\over{\xi^3}} \, \sigma'  \,+\, {{\beta^2}\over{\alpha^2}} \,
{{\zeta^2}\over{\xi^2}} \, \sigma'' \Bigr) \,-\, \Bigl(
{{\beta}\over{\alpha}} \,
{{\theta \, \zeta}\over{\xi^3}} \,\sigma' \,-\, {{\beta}\over{\alpha}} \,
{{\theta
\, \zeta}\over{\xi^2}} \, \sigma '' \Bigr)^2   \,$

\smallskip \noindent $  \displaystyle  \quad = \,   
{{\beta}\over{\alpha}} \, {{\sigma' \,\, \sigma''}\over{\xi^5}} \,\, \Bigl( \,
-\theta^4 \,-\, {{\beta^2}\over{\alpha^2}} \, \zeta^4 \,+\, 2 \,
{{\beta}\over{\alpha}}
\, \theta^2 \,\zeta^2 \, \Bigr) \,\,=\,\, -{{\beta}\over{\alpha}} {{\sigma' \,\,
\sigma''}\over{\xi}} \,\quad > \,\,0 \,$

\smallskip \noindent
when the condition (6.1) is satisfied. Then, joined to 
the hypothesis $\,\, \sigma'' \!>\! 0 \,\,$ 
the function $\,\, \eta({\scriptstyle \bullet}) \,\,$ is scritly
convex.

\smallskip \noindent $\bullet \quad$
For $\,\,W_0 = (\theta_0 ,\,0)^{\rm t} \in \Omega_0 \,,\,$ we have from
(4.13) :
$\,\, j(W_0) \in \Lambda\ib{-1} \,,\,$ then we can write it under the form :

\smallskip \noindent   (6.23)   $ \quad \displaystyle
\forall \,\,  W_0 \, =\,  (\theta_0 ,\,0)^{\rm t} \,  \in \Omega_0
\,,\qquad  j(W_0)
\,\,=\,\, \bigl( \,  0 \,,\,  \Pi(\theta_0) \, \bigr) ^{\rm t} \,\, \in
\Lambda\ib{-1} \,,\, $

\smallskip \noindent
where the function $\,\, ]0,\, +\infty[ \, \, \ni \theta_0 \longmapsto
\Pi(\theta_0)
\in \R \,\,$ remains to be determined.  In order to find the algebraic
expression of
the thermodynamic flux $\, j({\scriptstyle \bullet}), \,$ we  derive now
the relation
(3.12) relatively to the variable $\, v .\,$ It comes : 

\smallskip \noindent  $  \displaystyle
 {\rm d}j(Y(v) \,
{\scriptstyle \bullet}\, W) \, {\scriptstyle \bullet}\, {\rm d}Y(v) \,
{\scriptstyle \bullet}\,W \,=\, {\rm d}Y(v)
\, {\scriptstyle \bullet}\,j(W) \,\,$ 

\smallskip \noindent
and taking the particular value $\, v=0 \,:\,$ 

\smallskip \noindent   (6.24)   $ \quad \displaystyle
\forall \, W \in \Omega \,,\qquad  {\rm d}j(W) \, {\scriptstyle \bullet} \, {\rm
d}Y(0) \, {\scriptstyle \bullet}\, W \,\,= \,\,  {\rm d}Y(0) \, {\scriptstyle
\bullet}\,j(W) \,.\,   $ 

\smallskip \noindent
We apply the compatibility condition (2.9) to the vector $\,\, {\rm d}Y(0)  \,
{\scriptstyle \bullet}\,W_0 \,\,$ with $\,\, W_0 \in \Omega_0.\,$ Taking
into account
the relation (4.7), we get

\smallskip \noindent  $  \displaystyle 
\varphi \, {\scriptstyle \bullet}\,  {\rm d}Y(0) \, {\scriptstyle
\bullet}\, j(W_0)
\,\,=\,\, \varphi \, {\scriptstyle \bullet}\,  {\rm d} j(W_0) \, {\scriptstyle
\bullet}\,  {\rm d}Y(0) \, {\scriptstyle \bullet}\,W_0 \,\,=\,\, $ 

\smallskip \noindent  $  \displaystyle \qquad  = \,\,   
-\eta^*(\varphi) \,
{\rm d}u (W_0) \, {\scriptstyle \bullet}\, {\rm d}Y(0) \, {\scriptstyle
\bullet}\,W_0 \,\,=\,\,\eta^*(\varphi) \, $ \quad  
{\it id est}

\smallskip \noindent   (6.25)   $ \quad \displaystyle 
\forall \, W_0 \in \Omega_0 \,,\qquad  \varphi\, {\scriptstyle \bullet}\,
 {\rm d}Y(0) \, {\scriptstyle \bullet}\,j(W_0) \,\,= \,\, \eta^*(\varphi)
\qquad {\rm
with} \,\,\, \varphi \,=\, {\rm d} \eta(W_0) \,. \,  $ 

\smallskip \noindent
We have also from the relations (6.6) and (6.22) :

\smallskip \noindent   (6.26)   $ \quad \displaystyle  
\forall \, W \,=\, (\theta,\, \zeta)^{\rm t} \in \Omega \,,\qquad  \varphi
\,\,=\,\, {{\sigma'}\over{\xi}} 
\,\, \Bigl( \, \theta \,,\, -{{\beta}\over{\alpha}} \, \zeta \, \Bigr)  \,   $ 

\smallskip \noindent
and if $\,\, W = W_0 = (\theta_0 ,\,0)^{\rm t} \in \Omega_0 \,,\,$ we have
$\,\, \varphi \,=\, (\sigma',\,0) .\,$ We have also the following calculus~:

\smallskip \noindent  $  \displaystyle  
\varphi \,  {\scriptstyle \bullet}\, {\rm d}Y(0)  \,  {\scriptstyle
\bullet}\, j(W_0)
\,=\, \bigl( \, \sigma' \, ,\,0 \, \bigr) \,  \pmatrix { 0 & -\beta \cr
-\alpha & 0 \cr} \, \pmatrix {0 \cr \Pi \cr}  $

\smallskip \noindent  $  \displaystyle \qquad  = \,\,
\Bigl( 0 \,,\,\,  - \beta \,
\sigma'  \Bigr) \,\, \pmatrix {0 \cr \Pi \cr} \,= \, -\beta \,
\sigma'(\theta_0) \, \Pi(\theta_0) \,. \,    $

\smallskip \noindent
We have also $\,\, \eta^*(\sigma'(\xi),\,0) \,=\, \xi \, \sigma'(\xi) \,-\,
\eta(\xi)
\,=\, \xi \sigma'(\xi) \,-\, \sigma(\xi) \,=\, \sigma^*\bigl(\sigma'(\xi)
\bigr) \,\,$
and due to the relation (6.25) and the preceding development,  we have
necessarily 

\smallskip \noindent   (6.27)   $ \quad \displaystyle  
\Pi(\theta_0) \,\,= \,\, -{{\sigma^* \bigl(  \sigma'(\theta_0) \bigr)
}\over{\beta \, \sigma'(\theta_0)}} \,  \,$ 

\smallskip \noindent
in coherence with the relation (6.5). Then the thermodynamic flux function
$\,\, j(W)
\,\, $ can be easily deduced from the relation (3.12). We get, due to the
condition
$\,\, v \,= \, -u(W) \,: \,$

\smallskip \noindent  $  \displaystyle  
j \bigl( Y(v) \, {\scriptstyle \bullet}\, W_0 \bigr) \,\,=\,\,
 \pmatrix { {\rm ch} \bigl( u \, \sqrt{\alpha \, \beta \, } \bigr) &
\sqrt{\beta
 / \alpha \,} \, \, {\rm sh} \bigl( u \, \sqrt{\alpha \, \beta} \bigr) \cr
\sqrt{\alpha / \beta \,} \, {\rm sh} \bigl( u \, \sqrt{\alpha \, \beta}
\bigr) &
{\rm ch} \bigl( u \, \sqrt{\alpha \, \beta \, } \bigr) \cr  } \,\,  \pmatrix {0
\cr \Pi(\theta_0) } \,\,= \,   $

\smallskip \noindent  $  \displaystyle  \qquad = \,\,  
\Pi(\theta_0) \,\,{\rm ch} \bigl( v \, \sqrt{ \alpha \, \beta \, } \bigr) \,\,
\pmatrix { \sqrt{\beta / \alpha \,} \, \sqrt{\beta / \alpha \,} \, \zeta /
\theta \, \cr 1 \, } \,\,\hfill   $ due to (6.4)

\smallskip \noindent  $  \displaystyle  \qquad = \,\,   
\Pi \Bigl( \sqrt{\theta^2 \,-\, \beta \, \zeta^2 / \alpha \, } \, \Bigr) \,\,
{{1}\over{ \sqrt{\theta^2 - {{\beta \, \zeta^2}\over{\alpha}}} }} \,\,
\pmatrix { \beta \, \zeta / \alpha \cr \theta } \, \hfill $  thanks to (6.21)

\smallskip \noindent
and the expression (6.3) of the hyperbolic system is established in this
first case.

\smallskip \noindent $\bullet \quad$
We still have to verify the global coherence of what have been done, {\it
i.e.} that
the function $\, \eta({\scriptstyle \bullet}) \,$ introduced at the
relation (6.6) is
really a mathematical entropy for the system (6.3). We first have 
$\,\,\, {\rm th}(u \, \sqrt{\alpha \, \beta}) 
\,=\, \sqrt{\beta / \alpha \,} \,\, \zeta / \theta \,\,\,\, $ 
then by derivation $ \,\,\,  
{{1}\over{ {\rm ch}^2 (u \, \sqrt{\alpha \, \beta})}} \, \sqrt{\alpha \,
\beta} \,  {\rm d}u \,= \, \sqrt{{{\beta}\over{\alpha}}} \,{{1}\over{\theta^2}}
\,  \bigl( \, \theta \, {\rm d}\zeta \,-\, \zeta \, {\rm d} \theta \, \bigr) \,\,\,   $
and due to the relation (6.21) we have the following expression for the
derivative of velocity :

\smallskip \noindent   (6.28)   $ \quad \displaystyle  
\alpha \, {\rm d}u \,\,= \,\, {{ 1}\over{\theta^2 \,-\, \beta \, \zeta^2 /
\alpha}} \, \, \bigl( \, \theta \, {\rm d}\zeta 
\,-\, \zeta \, {\rm d} \theta \, \bigr) \,.\,$

\smallskip \noindent
Let $\,\, W(t,\,x) \,\equiv\, (\theta(t,\,x) ,\, \zeta(t,\,x) )^{\rm t}
\,\,$ be a
regular solution of the system (6.3). We have, with the notation (6.22) :

\smallskip \noindent  $  \displaystyle   
{{\partial \eta}\over{\partial t}} \,+\,  {{\partial}\over{\partial x}}
\bigl( \eta \,
u \bigr) \,\,= \,\, \Bigl(  {{\partial}\over{\partial t}} \,+\, u \,
{{\partial}\over
{\partial x}} \Bigr) \eta(W) \,+\, \eta \,  {{\partial u}\over{\partial x}} $

\smallskip \noindent  $  \displaystyle   \quad  =  \,
{{{\rm d}\sigma}\over{{\rm d}\xi}} \, \Bigl(  {{\partial}\over{\partial t}}
\,+\, u \, {{\partial}\over{\partial x}} \Bigr) \xi(W)  
\,+\, \eta \,  {{\partial u}\over{\partial x}}    $

\smallskip \noindent  $  \displaystyle   \quad  =  \, 
\sigma'\, \biggl( \, {{\theta}\over{\xi}} \,  \Bigl( {{\partial
\theta}\over{\partial
t}} \,+\, u \,  {{\partial \theta}\over{\partial x}} \Bigr) \,-\,
{{\beta}\over{\alpha}} \, {{\zeta}\over{\xi}} \,  \Bigl(
 {{\partial \zeta}\over{\partial t}} \,+\, u \,  {{\partial
\zeta}\over{\partial x}}
\Bigr) \, \biggr) \,+\, \eta \,  {{\partial u}\over{\partial x}}  $

\smallskip \noindent  $  \displaystyle   \quad  =  \,  
{{\sigma'}\over{\xi}} \, \biggl( \, \theta \Bigl[ -\theta \,  {{\partial
u}\over{\partial x}}  \,- \, {{\partial}\over{\partial x}} \Bigl(
{{\Pi}\over{\xi}} \, {{\beta}\over{\alpha}} \, \zeta \Bigr) \Bigr] \,+\,
{{\beta}\over{\alpha}} \, \zeta \, \Bigl[ \, \zeta \, 
 {{\partial u}\over{\partial x}} \,+\, {{\partial
}\over{\partial x}} \Bigl( {{\Pi}\over{\xi}} \, \theta \Bigr) \, \Bigr] \,
\biggr) \,+\,  \eta \,  {{\partial u}\over{\partial x}} \, $ 

\hfill   due to (6.3)

\smallskip \noindent  $  \displaystyle   \quad  =  \,   
{{\sigma'}\over{\xi}} \, \biggl( \, -\xi^2 \,  {{\partial u}\over{\partial
x}} \, +\,
{{\beta}\over{\alpha}} \,  {{\Pi(\xi)}\over{\xi}} \, \Bigl[ \,\zeta \,
{{\partial
\theta  }\over{\partial x}} \,-\, \theta \,  {{\partial \zeta
}\over{\partial x}}\,
\Bigr] \,  \biggr) \,+  \eta \,  {{\partial u}\over{\partial x}} \,\hfill
$ due to (6.22)

\smallskip \noindent  $  \displaystyle   \quad  =  \,    
\bigl( \, -\xi \, \sigma'(\xi) \,+\, \eta \, \bigr) \,  {{\partial
u}\over{\partial x}}
\,-\, \beta \, \sigma'(\xi) \,\, \Pi(\xi) \, \,  {{\partial
u}\over{\partial x}} \, \hfill  $ due to (6.28)

\smallskip \noindent  $  \displaystyle   \quad  =  \,   
\Bigl( \, -\xi \, \sigma'(\xi) \,+\, \sigma(\xi) \, +\, \sigma^* \bigl(
\sigma'(\xi) \bigr) \, \Bigr) \,\,  {{\partial u}\over{\partial x}} \,\, 
\,\hfill  $ due to (6.5) and (6.6)

\smallskip \noindent  $  \displaystyle   \quad  =  \,   0 \,$

\smallskip \noindent
by definition of the dual of a function in the sense of Moreau. Then any regular
solution of the system (6.3) is also solution of the equation (2.2) of
conservation
of the entropy~; in other terms, the relation (2.9) is identically
satisfied. This
fact ends the first part of the theorem~1 where has been developed the case
of a
jacobian matrix $\,\, {\rm d}Y(0) \,\,$  given by the relation (6.17).

\smallskip \noindent $\bullet \quad$ {\bf Case (ii).} \quad
When the matrix $\,\, {\rm d}Y(0) \,\,$ is given by the relation (6.18), we
have

\smallskip \noindent   (6.29)   $ \quad \displaystyle   
\bigl( {\rm d}Y(0) \bigr)^2 \,\,= \,\, -\alpha \, \beta \,\,\, 
 {\rm Id} \,. \,$

\smallskip \noindent   Then

\smallskip \noindent  $ \displaystyle   
Y(v) \,\,= \,\,  {\rm e}^{\displaystyle v \, {\rm d}Y(0)} $ 

\smallskip \noindent  $ \displaystyle  \qquad \,\,\, = \,  
  {\rm Id} \,+\, v
\, {\rm d}Y(0) \,-\, {{v^2}\over{2 !}} \,  \alpha \, \beta \, {\rm Id} \,-\,
{{v^3}\over{3 !}} \,  \alpha \, \beta \, {\rm d}Y(0) \,+\,{{v^4}\over{4 !}} \, (
\alpha \, \beta)^2 \,\,  {\rm Id} \,+\, \cdots \,    $
 
\smallskip \noindent  $ \displaystyle  \qquad \,\,\, = \,  
\,\, {\rm cos} \bigl( v \, \sqrt{\alpha \, \beta}\, \bigr) \,\, {\rm Id}
\,+\, {{1}\over{\sqrt{\alpha \, \beta}}} \,\,  {\rm sin} \bigl( v \,
\sqrt{\alpha \, \beta}\, \bigr) \,\, {\rm d}Y(0) \,; \,$

\smallskip \noindent   (6.30)   $ \quad \displaystyle  
Y(v) \,\,= \,\, \pmatrix { {\rm cos} \bigl( v \, \sqrt{ \alpha \, \beta \, }\,
\bigr) &  \sqrt{\beta / \alpha \,} \, \, {\rm sin} \bigl( v \, \sqrt{\alpha
\, \beta} \,\bigr) \cr -\sqrt{\alpha / \beta \,} \, {\rm sin} \bigl( v \,
\sqrt{\alpha \, \beta} \,\bigr) &  {\rm cos} 
\bigl( v \, \sqrt{\alpha \, \beta \, } \,\bigr) \cr }\,.\, $

\smallskip  \noindent
For $\,\, W_0 \,=\, (\theta_0,\,0)^{\rm t} \in \Omega_0 \,,\,\,$ we have

\smallskip \noindent  $  \displaystyle   
W \,= \, \pmatrix { \theta \cr \zeta \cr} \,=\, Y(v) \, {\scriptstyle
\bullet} \, W_0
\,=\, \pmatrix { \, \theta_0 \,\, {\rm cos} \bigl( v \, \sqrt{ \alpha \,
\beta \, }\, \bigr)  \cr  
-\sqrt{\alpha / \beta \,} \,  \theta_0 \,   {\rm sin}
\bigl( v \, \sqrt{\alpha \, \beta} \,\bigr)} \,$ \quad 
with $\,  u(W) \,=\, -v   .\,\,  $

\smallskip  \noindent
We deduce that necessarily the relation (6.9) is compatible with the previous
expression  and

\smallskip \noindent   (6.31)   $ \quad \displaystyle 
\theta_0 \,\,=\,\, {{\theta}\over{{\rm cos} \, \bigl(  \, u(W) \,
\sqrt{\alpha \, \beta
\, } \,\bigr)}} \,\,=\,\, \sqrt{\theta^2 + {{\beta \,
\zeta^2}\over{\alpha}}\,} \qquad
{\rm if} \,\,\, W \,=\, (\theta ,\, \zeta)^{\rm t} \in \Omega \,.\, $ 

\smallskip \noindent $\bullet \quad$
As in the hyperbolic case, the condition 
$\,\, u(W) \,=\, 0 \,\,$ implies
that the second component $\, \zeta \,$ of the state $\, W \,$ is null. 
Then the necessary condition\br 
 $ \eta(W) \,=\, \sigma(\theta_0) \,\,$ conducts to fix the
entropy on
the null velocity manifold $\, \Omega_0 \,$ as a strictly convex function
of some
variable $\, \xi > 0 \,; \,$ we set : 

\smallskip \noindent   (6.32)   $ \quad \displaystyle 
\eta(\theta ,\, \zeta) \,=\, \sigma(\xi) \,\,,\qquad $ with \quad $
\displaystyle \xi \,=\, \sqrt{\theta^2 
+ {{\beta \, \zeta^2}\over{\alpha}}\,}\,.\,$

\smallskip \noindent
We have again to determine the condition for the convex function $\,\, \sigma
({\scriptstyle \bullet})  \,\,$ to construct a stricly convex entropy $\,
\eta \,$ from the relation (6.32). We have :

\smallskip \noindent $  \displaystyle  
\xi \, {\rm d} \xi \,\,= \,\, \theta \,  {\rm d}\theta \,+\,
{{\beta}\over{\alpha}}
\, \zeta \,  {\rm d}\zeta \, \,,\qquad
{{\partial \eta}\over{\partial \theta}} \,=\, {{\theta}\over{\xi}} \, \sigma'
\,,\qquad   {{\partial \eta}\over{\partial \zeta}} \,=\,
{{\beta}\over{\alpha}} \,
{{\zeta}\over{\xi}} \, \sigma'  \,, \, $   

\smallskip \noindent $  \displaystyle   
 {{\partial^2 \eta}\over{\partial \theta^2}} \,=\, {{\beta}\over{\alpha}} \,
{{\zeta^2}\over{\xi^3}} \,\sigma'  \,+\,   {{\theta^2}\over{\xi^2}} \, \sigma ''
\,, 
\quad  {{\partial^2 \eta}\over{\partial  \theta \, \partial \zeta}}
\,=\, -{{\beta}\over{\alpha}} \,   {{\theta \, \zeta}\over{\xi^3}}
\,\sigma' \,+\, {{\beta}\over{\alpha}} \, 
  {{\theta \, \zeta}\over{\xi^2}} \, \sigma '' \,, \, $

\smallskip \noindent $  \displaystyle    
 {{\partial^2 \eta}\over{\partial \zeta^2}} \,=\, {{\beta}\over{\alpha}} \,
{{\theta^2}\over{\xi^3}} \, \sigma'  \,+\, {{\beta^2}\over{\alpha^2}} \,
{{\zeta^2}\over{\xi^2}} \, \sigma'' \,  .   \quad    $ 
Then

\smallskip \noindent $  \displaystyle    
 {\rm det} \bigl( {\rm d}^2\eta \bigr) \,\,=
\,\,{{\partial^2 \eta}\over{\partial \theta^2}} \,\,  {{\partial^2
\eta}\over{\partial
\zeta^2}} \,-\, \Bigl( {{\partial^2 \eta}\over{\partial  \theta \, \partial
\zeta}} \Bigr)^2 \,    $

\smallskip \noindent $  \displaystyle  \quad = \,    
\Bigl( {{\beta}\over{\alpha}} \,  {{\zeta^2}\over{\xi^3}} \,\sigma'  \,+\,
{{\theta^2}\over{\xi^2}} \, \sigma '' \Bigr) \,\, \Bigl( {{\beta}\over{\alpha}} \,
{{\theta^2}\over{\xi^3}} \, \sigma'  \,+\, {{\beta^2}\over{\alpha^2}} \,
{{\zeta^2}\over{\xi^2}} \, \sigma'' \Bigr) \,-\, \Bigl(
-{{\beta}\over{\alpha}} \,
{{\theta \, \zeta}\over{\xi^3}} \,\sigma' \,+\, {{\beta}\over{\alpha}} \,
{{\theta
\, \zeta}\over{\xi^2}} \, \sigma '' \Bigr)^2  \,  $

\smallskip \noindent $  \displaystyle  \quad = \,    
{{\beta}\over{\alpha}} \, {{\sigma' \,\, \sigma''}\over{\xi^5}} \,\, \Bigl( \,
\theta^4 \,+\, {{\beta^2}\over{\alpha^2}} \, \zeta^4 \,+\, 2 \,
{{\beta}\over{\alpha}}
\, \theta^2 \,\zeta^2 \, \Bigr) \,\,=\,\, {{\beta}\over{\alpha}} {{\sigma' \,\,
\sigma''}\over{\xi}} \,\quad > \,\,0 \,$

\smallskip \noindent
when the condition $\,\, \sigma'(\xi) > 0 \,\, $ of relation (6.7) is true.
In these
conditions, we have also $\,\, {{\partial^2 \eta}\over{\partial \theta^2}} > 0
\,,\,\,$ and we have established that the function $\, \, \eta({\scriptstyle
\bullet}) \,\,$ is a  scritly  convex function of the pair $\,\, (\theta,\,
\zeta) .\,\,$

\smallskip \noindent $\bullet \quad$
We search now the thermodynamic flux $\,\, j(W_0) \,\,$ on the form (6.23). The
determination of the entropy variables is easy :

\smallskip \noindent   (6.33)   $ \quad \displaystyle  
\forall \, W \,=\, (\theta,\, \zeta)^{\rm t} \in \Omega \,,\qquad  \varphi
\,\,=\,\, {{\sigma'}\over{\xi}} 
\,\, \Bigl( \, \theta \,,\, {{\beta}\over{\alpha}} \,
\zeta \, \Bigr)  \, $

\smallskip \noindent
and $\,\, \varphi (W_0) \,=\, ( \sigma' ,\, 0 ) \,\,$ when $\,\, W_0 \in
\Omega_0 .\,\,$ We have now

\smallskip \noindent  $  \displaystyle  
\varphi \,  {\scriptstyle \bullet}\, {\rm d}Y(0)  \,  {\scriptstyle
\bullet}\, j(W_0)
\,=\, \bigl( \, \sigma' \, ,\,0 \, \bigr) \,  \pmatrix { 0 & \beta \cr
-\alpha & 0 \cr} \, \pmatrix {0 \cr \Pi \cr} \,=\, \Bigl( 0 \,,\,\,  - \beta \,
\sigma'  \Bigr) \,\, \pmatrix {0 \cr \Pi \cr}  \,$

\smallskip \noindent  $  \displaystyle  \qquad  \,= \, 
\beta \, \sigma'(\theta_0) \, \Pi(\theta_0) \, $

\smallskip \noindent
and the relation (6.10) is established due to (6.25) which remains true for this
second case. The end of the determination of the thermodynamic flux $\,\,
j(W) \,\,$ for an arbitrary state $\, W \,$ is easy, taking into account the relation
(3.12) and the expression (6.30) :

\smallskip \noindent  $  \displaystyle  
j(W) \,\,= \,\, Y(-u(W)) \,{\scriptstyle \bullet}\, j \bigl(
Y(u(W))\,{\scriptstyle
\bullet}\, W\bigr) \,\,= \,\, Y(-u(W)) \,{\scriptstyle \bullet}\, j (W_0) $

\smallskip \noindent  $  \displaystyle  \qquad \,\,\,\,  = \,  
\pmatrix { {\rm cos}  \bigl( u \, \sqrt{ \alpha \, \beta \, }\, \bigr) &
-\sqrt{\beta
/ \alpha \,} \, \, {\rm sin} \bigl( u \, \sqrt{\alpha \, \beta} \,\bigr) \cr
\sqrt{\alpha / \beta \,} \, {\rm sin} \bigl( u \, \sqrt{\alpha \, \beta}
\,\bigr) &
{\rm cos} \bigl( u\, \sqrt{\alpha \, \beta \, } \,\bigr) \cr  } \,\,
\pmatrix{ 0 \cr \Pi(\theta_0) \cr} \,$

\smallskip \noindent  $  \displaystyle  \qquad \,\,\,\,  = \,   
\Pi(\theta_0) \,\,  {\rm cos} \bigl( u \, \sqrt{ \alpha \, \beta \, } \,\bigr)
\,\,\, \pmatrix { -\sqrt{ \beta / \alpha \,} \, \sqrt{ \beta / \alpha \,}
\,  \, \zeta \,  / \theta \cr 1 \cr } \,$

\smallskip \noindent  $  \displaystyle  \qquad \,\,\,\,  = \,   
\Pi \Bigl( \sqrt{\theta^2 \,+\, \beta \, \zeta^2 / \alpha \, } \, \Bigr) \,\,
{{1}\over{ \sqrt{\theta^2 + {{\beta \, \zeta^2}\over{\alpha}}} }} \,\,
\pmatrix { -\beta \, \zeta / \alpha \cr \theta } \, 
\hfill $  thanks to (6.31)

\smallskip  \noindent
and the algebraic expression (6.8) of the hyperbolic system is established.

\smallskip \noindent $\bullet \quad$
As in the first case, we must verify that the candidate $\,\, \eta({\scriptstyle
\bullet}) \,\,$ for a mathematical entropy is in fact a correct  one, {\it
i.e.} that
the quantity $\,\, \eta(W) \,\,$ is advected with the velocity $\,\, u(W)
\,\,$ for the
regular solutions of the equation (6.8), with the entropy  flux $\,\, u(W) \,\,
\eta(W) .\,\, $ Taking into account the relation (6.9), we have  
$\,\,\, {\rm tg}(u \, \sqrt{\alpha \, \beta}) \,=\, 
\sqrt{\beta / \alpha \,} \,\, \zeta / \theta \,\,\,\, $
then 
 $  \,\,\,
{{1}\over{ {\rm cos}^2 (u \, \sqrt{\alpha \, \beta})}} \, \sqrt{\alpha \,
\beta} \, 
{\rm d}u \,= \, \sqrt{{{\beta}\over{\alpha}}} \,{{1}\over{\theta^2}}
\,\, \bigl( \, \theta \, {\rm d}\zeta 
\,-\, \zeta \, {\rm d} \theta \, \bigr)    $
and due to the relation (6.31) we have the following expression for the
derivative of
velocity : 

\smallskip \noindent   (6.34)   $ \quad \displaystyle  
\alpha \, {\rm d}u \,\,= \,\, {{ 1}\over{\theta^2 \,+\, \beta \, \zeta^2 /
\alpha}} \, \, \bigl( \, \theta \, {\rm d}\zeta 
\,-\, \zeta \, {\rm d} \theta \, \bigr) \,.\, $

\smallskip \noindent
Let $\,\, W(t,\,x) \,\equiv\, (\theta(t,\,x) ,\, \zeta(t,\,x) )^{\rm t}
\,\,$ be a
regular solution of system (6.8). We have, with the notation (6.32) :

\smallskip \noindent $  \displaystyle   
{{\partial \eta}\over{\partial t}} \,+\,  {{\partial}\over{\partial x}}
\bigl( \eta \,
u \bigr) \,\,= \,\, \Bigl(  {{\partial}\over{\partial t}} \,+\, u \,
{{\partial}\over
{\partial x}} \Bigr) \eta(W) \,+\, \eta \,  {{\partial u}\over{\partial x}} $ 

\smallskip  \noindent $\displaystyle \quad = \,\, 
{{{\rm d}\sigma}\over{{\rm d}\xi}} \, 
\Bigl(  {{\partial}\over{\partial t}} \,+\, u \,
{{\partial}\over{\partial x}} \Bigr) \xi(W)  \,+\, \eta \,  {{\partial
u}\over{\partial x}}  \,     $

\smallskip  \noindent $\displaystyle \quad = \,\,
\sigma'\, \biggl( \, {{\theta}\over{\xi}} \,  \Bigl( {{\partial
\theta}\over{\partial
t}} \,+\, u \,  {{\partial \theta}\over{\partial x}} \Bigr) \,+\,
{{\beta}\over{\alpha}} \, {{\zeta}\over{\xi}} \,  \Bigl(
 {{\partial \zeta}\over{\partial t}} \,+\, u \,  {{\partial
\zeta}\over{\partial x}}
\Bigr) \, \biggr) \,+\, \eta \,  {{\partial u}\over{\partial x}}   \,$

\smallskip  \noindent $\displaystyle \quad = \,\, 
{{\sigma'}\over{\xi}} \, \biggl( \, \theta \Bigl[ -\theta \,  {{\partial
u}\over{\partial x}}  \,+ \, {{\partial }\over{\partial x}} \Bigl(
{{\Pi}\over{\xi}}
\, {{\beta}\over{\alpha}} \, \zeta \Bigr) \Bigr] \,+\,
{{\beta}\over{\alpha}} \, \zeta \, \Bigl[ 
\, -\zeta \,  {{\partial u}\over{\partial x}} \,-\, {{\partial
}\over{\partial x}} \Bigl( {{\Pi}\over{\xi}} \, \theta \Bigr) \, \Bigr] \,
\biggr) \,+\,  \eta \,  {{\partial u}\over{\partial x}} \, $ 

\hfill   due to (6.8)

\smallskip  \noindent $\displaystyle \quad = \,\,  
{{\sigma'}\over{\xi}} \, \biggl( \, -\xi^2 \,  {{\partial u}\over{\partial
x}} \, +\, {{\beta}\over{\alpha}} \,
  {{\Pi(\xi)}\over{\xi}} \, \Bigl[ \,-\zeta \, {{\partial
\theta  }\over{\partial x}} \,+\, \theta \,  {{\partial \zeta
}\over{\partial x}}\, \Bigr] \,  \biggr) 
\,+  \eta \,  {{\partial u}\over{\partial x}} 
\,\hfill $ due to (6.32)

\smallskip  \noindent $\displaystyle \quad = \,\,   
\bigl( \, -\xi \, \sigma'(\xi) \,+\, \eta \, \bigr) \,  {{\partial
u}\over{\partial x}}
\,+\, \beta \, \sigma'(\xi) \,\, \Pi(\xi) \, \,  {{\partial
u}\over{\partial x}} \, \hfill  $ due to (6.34)

\smallskip  \noindent $\displaystyle \quad = \,\,   
\Bigl( \, -\xi \, \sigma'(\xi) \,+\, \sigma(\xi) \, +\, \sigma^* \bigl(
\sigma'(\xi) \bigr) \, \Bigr) \,\, 
 {{\partial u}\over{\partial x}} \,\, \, 
\hfill  $ due to (6.10) and (6.11)

\smallskip  \noindent $\displaystyle \quad = \,\,   0 \,.\,$

\smallskip  \noindent
The proof of Theorem 1 is completed.  $ \hfill \square \kern0.1mm    $

\bigskip \noindent  {\bf Remark 2. \quad About the $\, p-$system. }

\noindent
Following a remark proposed by F. Coquel in march 2000,  we can fix $\,\,
\beta > 0 \,
\,$ and take the limit $\,\, \alpha \longrightarrow +\infty \,\,$ for the
systems
found at the theorem 1. Then  the velocities $\,\, u(W) \,$ defined in (6.4) and
(6.9) tend to zero and both systems of conservation laws (6.3) and (6.8)
admit the
following formal limit

\smallskip \noindent   (6.35)   $ \quad \displaystyle   
{{\partial}\over{\partial t}} \pmatrix{\theta \cr \zeta \cr} \,+\,
{{\partial}\over{\partial x}} \, \, \pmatrix {0 \cr \Pi ( \theta ) \cr }
\,\,= \,\, 0 \,;\, \,\,   $
 
\smallskip  \noindent
we observe here that  this limit system is {\bf not} the $p$-system that takes the
classical form (see {\it e.g.} [GR96]) :

\smallskip \noindent   (6.36)   $ \quad \displaystyle    
{{\partial}\over{\partial t}} \pmatrix{\theta \cr \zeta \cr} \,+\,
{{\partial}\over{\partial x}} \, \, \pmatrix {-\zeta \cr p ( \theta ) \cr }
\,\,= \,\, 0 \,.\, \,\, $

\bigskip  \noindent  {\bf Remark 3. \quad Preliminaries. }

\noindent
In all fairness, the elliptic Galileo version of two by two  Galileo group
preserving
systems of conservation laws  satisfies  the relation (3.9)   $\,\, u
\bigl( Y(v) \, {\scriptstyle \bullet} \,W \bigr) \,= $\br  
$\,=\, u(W) - v \,\,$  only for sufficiently small
velocities $\, v .\,$ It is clear from the relation (6.9) that the velocity
$\,\, u(W) \,\,$ 
should be defined {\it modulo} some additive constant (equal to
$\, \pi /
\sqrt{\alpha \, \beta} \, $) because only the expression $\,\, {\rm tg} \bigl(
\sqrt{\alpha \, \beta} \,  u(W) \bigr) \,\,$ is well defined.  Then the
hypothesis 2
should be adapted to systems of conservation  laws whose velocity belongs
to some
quotient group of the type $\,\, \R / ( \mu \, \Z) \,. \,$ These
developments have not
been realized at this moment [november 2000], and this fact explains the word
``preliminary'' in the  title of our contribution. This remark is also to
be done for the three by three ``elliptic Galileo'' 
 system developed in the next section.

\bigskip \bigskip 
\centerline{ \smcap 7. \quad Galilean invariance for systems of three
conservation laws}

\smallskip \noindent  {\bf Theorem 2. \quad   }

\noindent
When $\,\, \Omega \ \subset \ ]0,\,+\infty[ \,\, \times^{\rm \! t} \,\,
\R^{2, \, {
\rm t}} \ \subset \   \R^{3, \, { \rm t}} \,,\,$ a system of three conservation
laws invariant for the Galileo group has one of the three following types :
hyperbolic
Galileo, elliptic Galileo or nilpotent Galileo. The hyperbolic system  is
parameterized  by the scalars $\,\, a \!>\! 0 \,,\,\, b \!>\! 0 \, \,$ and  by a
derivable  strictly convex function $ \,\, ]0,\, +\infty[ \,  \times \, \R
\,\, \ni
(\alpha,\, \beta)  \longmapsto \sigma (\alpha,\, \beta) \in \R \,; \,$ we
denote by
$\,\, \sigma^* ( {\scriptstyle \bullet}) \,\,$ the dual function of $\, \sigma
({\scriptstyle \bullet}) .\,$

\smallskip \noindent {\bf (i) \quad  Hyperbolic Galileo}.

\noindent
We have  for this first case 

\smallskip \noindent   (7.1)   $ \quad \displaystyle    
\forall \, \alpha > 0 \,,\quad \forall \, \beta \in \R \,,\qquad
{{\partial \sigma}\over{\partial \alpha}}(\alpha,\, \beta) < 0 \, \, $ 

\smallskip \noindent
and the space of states $\,\, \Omega \,\,$  is included in the cone $\,\,
\Omega_+
\,\,$  defined by 

\smallskip \noindent   (7.2)   $ \quad \displaystyle    
\Omega_{+} \,\,=\,\, \biggl\{ \, W \,=\, \bigl( \,  \theta ,\, \zeta ,\, \psi \,
\bigr)^{\rm t} \in \R^{3,\,{\rm t}} \,, \quad \, \theta > 0 \,, \quad
\abs{\zeta} \, <
\, \sqrt{{{a}\over{b}}} \,  \abs{\theta} \,  \biggr\}  \,. \, $

\smallskip \noindent
The  system of conservation laws takes the algebraic form

\smallskip \noindent   (7.3)   $ \,\,  \displaystyle     
{{\partial}\over{\partial t}} \pmatrix{\theta \cr \zeta \cr \psi \cr } \,+\,
{{\partial}\over{\partial x}} \, \Biggl\{ \, u(W) \,  \pmatrix{\theta \cr
\zeta \cr
\psi \cr} \,+\, {{\Pi\bigl( \, \sqrt{\theta^2 - b \, \zeta^2 / a \, } \,,\,
\psi \,  \bigr)}\over{\sqrt{\theta^2 - b \, \zeta^2 / a \,}}} \,\,
\pmatrix{ b \, \zeta / a \cr \theta \cr 0 \cr } \, \Biggr\} \,=\, 0 \, $

\smallskip \noindent
with a velocity $\,\, u({\scriptstyle \bullet} ) \,\,$  given by the relation 

\smallskip \noindent   (7.4)   $ \quad \displaystyle  
u(W) \,\,=\,\, {{1}\over{\sqrt{a \, b}}} \,\,  {\rm argth} \biggl(
\sqrt{{{b}\over{a}}} \, {{\zeta}\over{\theta}} \biggr) \,$

\smallskip \noindent
and a so-called mechanical pressure function  $\,\, \Pi({\scriptstyle
\bullet})\,\,$ satisfying 

\smallskip \noindent   (7.5)   $ \quad \displaystyle  
\Pi(\xi,\, \beta) \,\,=\,\, -{{1}\over{b}} \,{{\sigma^*(A,\,B)} \over{A}}
\,\,,\qquad {\rm with} \,\, (A,\, B) 
\,\,=\,\, {\rm d}\sigma(\xi,\, \beta) \, . \,  $

\smallskip \noindent
Moreover, the function  $\,\, \eta({\scriptstyle \bullet})\,\,$ defined by 

\smallskip \noindent   (7.6)   $ \quad \displaystyle  
\eta(\theta,\, \zeta ,\, \psi ) \,=\, \sigma \biggl( \sqrt{\theta^2 - {{b \,
\zeta^2}\over{a}}} \,,\,\psi  \, \biggr) \,\, $

\smallskip \noindent
is a mathematical entropy associated with the hyperbolic system (7.3).

\smallskip \noindent {\bf (ii) \quad  Elliptic Galileo}.

\noindent
We have  in this second case

\smallskip \noindent   (7.7)   $ \quad \displaystyle   
\forall \, \alpha > 0 \,,\quad \forall \, \beta \in \R \,,\qquad
{{\partial \sigma}\over{\partial \alpha}}(\alpha,\, \beta) > 0 \, . \, $
 
\smallskip \noindent
The  elliptic Galileo system of conservation laws admits the expression 

\smallskip \noindent   (7.8)   $ \,\,  \displaystyle 
{{\partial}\over{\partial t}} \pmatrix{\theta \cr \zeta \cr \psi \cr } +
{{\partial}\over{\partial x}} \, \Biggl\{ \, u(W) \,  \pmatrix{\theta \cr
\zeta \cr
\psi \cr } +  {{\Pi\bigl( \, \sqrt{\theta^2 + b \, \zeta^2 / a \,} ,\,\psi
\,  \bigr)} \over {\sqrt{\theta^2 + b \, \zeta^2 / a \,}}} \,  \pmatrix{
-b \, \zeta / a \cr \theta \cr 0 \cr } \, \Biggr\} \,= \, 0 \, $

\smallskip \noindent
with a velocity $\,\, u({\scriptstyle \bullet}),\,\,$  a function $\,\, \Pi
({\scriptstyle \bullet}) \,\,$ and a mathematical entropy $\,\, \eta
({\scriptstyle
\bullet}) \,\,$ defined by the relations 

\smallskip \noindent   (7.9)   $ \quad \displaystyle   
u(W) \,\,=\,\, {{1}\over{\sqrt{a \, b}}} \,\,  {\rm arctg} \biggl(
\sqrt{{{b}\over{a}}} \, {{\zeta}\over{\theta}} \biggr) \,$ 

\smallskip \noindent   (7.10)   $ \quad \displaystyle  
\Pi(\xi,\, \beta) \,\,=\,\, {{1}\over{b}} \,{{\sigma^*(A,\,B)} \over{A}}
\,\,,\qquad {\rm with} \,\, (A,\, B) 
\,\,=\,\, {\rm d}\sigma(\xi,\, \beta) \, $

\smallskip \noindent   (7.11)   $ \quad \displaystyle   
\eta(\theta,\, \zeta ,\, \psi  ) \,=\, \sigma \biggl( \sqrt{\theta^2 + {{b \,
\zeta^2}\over{a}}} \,,\,\psi  \, \biggr) \,.\,$

\smallskip \noindent {\bf (iii) \quad  Nilpotent Galileo}.

\noindent
For this  third case,  we suppose

\smallskip \noindent   (7.12)   $ \quad \displaystyle    
\forall \, \alpha > 0 \,,\quad \forall \, \beta \in \R \,,\qquad \,\,
{{\partial \sigma}\over{\partial \beta}}(\alpha,\, \beta) < 0 \, .\, $
 
\smallskip \noindent
The  nilpotent Galileo  system of conservation laws takes the algebraic form

\smallskip \noindent   (7.13)   $ \quad \displaystyle   
{{\partial}\over{\partial t}} \pmatrix{\theta \cr \zeta \cr \psi \cr } \,+\,
{{\partial}\over{\partial x}} \, \Biggl\{ \, u(W) \,  \pmatrix{\theta \cr
\zeta \cr
\psi \cr} \,+\,\Pi\Bigl( \, \theta \,,\, \psi - {{b}\over{2\,a}} \,
{{\zeta^2}\over{\theta}} \,  \Bigr) \,\pmatrix{ 0 \cr 1 \cr
\displaystyle {{b}\over{a}} \, {{\zeta}\over{\theta}} \cr  } \, \Biggr\}
\,\,= \,\, 0 \,$ 

\smallskip \noindent
with a velocity field $\,\, u({\scriptstyle \bullet} ) \,\,$  given by the
relation 

\smallskip \noindent   (7.14)   $ \quad \displaystyle   
u(W) \,\,=\,\, {{\zeta}\over{a \, \theta}} \, $

\smallskip \noindent
and a mechanical pressure  $\,\, \Pi({\scriptstyle \bullet})\,\,$ satisfying

\smallskip \noindent   (7.15)   $ \quad \displaystyle    
\Pi(\alpha,\, \beta) \,\,=\,\, -{{1}\over{b}} \,{{\sigma^*(A,\,B)} \over{B}}
\,\,,\qquad {\rm with} \,\, (A,\, B) \,\,=\,\, {\rm d}\sigma(\alpha,\,
\beta) \,\, $
 
\smallskip \noindent
or in an equivalent way 

\smallskip \noindent   (7.16)   $ \quad \displaystyle   
\Pi(\alpha,\, \beta) \,\,=\,\, -{{1}\over{b}} \,{{\sigma^* \, \Bigl( \,
\displaystyle
{{\partial \sigma}\over{\partial \alpha}}(\alpha,\, \beta) \,,\, {{\partial
\sigma}\over{\partial \beta}}(\alpha,\, \beta) \, \Bigr) }\over{ \displaystyle
{{\partial \sigma}\over{\partial \beta}}(\alpha,\, \beta)  }}\,.\,$
 
\smallskip \noindent
Moreover, the function  $\,\, \eta({\scriptstyle \bullet})\,\,$ defined by

\smallskip \noindent   (7.17)   $ \quad \displaystyle 
\eta(\theta,\, \zeta ,\, \psi ) \,=\, \sigma \Bigl( \, \theta \,,\, \psi -
{{b}\over{2\,a}} \, {{\zeta^2}\over{\theta}} \, \Bigr) \,\,$

\smallskip \noindent
is a mathematical entropy associated with the hyperbolic system (7.13).

\bigskip  \noindent  {\bf Remark 4. \quad Pressure and duality.}

\noindent
The case of gas dynamics corresponds to $\,\, \theta= \rho ,\,$ $ \,\psi =
\rho \, E
,\,$ $ a = b = 1 $ for the nilpotent Galileo system of conservation laws.
The entropy
$\,\, \sigma ({\scriptstyle \bullet}) \,\,$ at null velocity is defined from the
classical thermostatic entropy function $\, \Sigma  ({\scriptstyle
\bullet})\,$ which
is concave and homogeneous of degree one relatively to the  extensive physical
 variables of mass $\, M ,\,$ volume $\, V \,$ and internal energy $\,
{\cal E} \,$
[Du90]. Introducing the intensive thermostatic variables of  temperature
$\, T ,\,$
thermodynamic pressure $\, p \,$ and massic chemical potential $\, \mu ,\,$
we have
the classical fundamental relation of thermostatics (see  {\it e.g.} Callen
[Ca85]),

\smallskip \noindent   (7.18)   $ \quad \displaystyle 
{\rm d} {\cal E} \,\,= \,\, T \,\, {\rm d} \Sigma(M,\,V,\,{\cal E}) \,-\,
p\, {\rm d}V
\,+\, \mu {\rm d}M \,\, $

\smallskip \noindent
and taking into account the Euler relation for homogeneity of degree one

\smallskip \noindent   (7.19)   $ \quad \displaystyle  
{\cal E} \,\,\equiv \,\, T \, \Sigma \,- \, p \, V \,+\, \mu \, M \,, \, $
 
\smallskip \noindent
we have necessarily~: 

\smallskip \noindent   (7.20)   $ \quad \displaystyle 
\eta \bigl( \, \rho \,,\, 0 \,,\, \psi \,\bigr) \,\,= \,\, - \Sigma \bigl(
\, \rho \,,\, 1  \,,\, \psi \,\bigr) 
\,\,\equiv \,\, \sigma ( \rho ,\, \psi) \,.\, $
 
\smallskip \noindent
We take $\, V= 1 ,\,$ $\, M = \rho \,$ and $\, {\cal E} = \psi \,;\,$ we
deduce from
the relations (7.19) and (7.20) the identity :
 
\smallskip \noindent   (7.21)   $ \quad \displaystyle 
\sigma \,\,= \,\, {{\mu}\over{T}} \, \rho \,-\,{{1}\over{T}} \,\psi \,- \,
{{p}\over{T}} \,\,  $ 
 
\smallskip \noindent
and by application of the relation (7.18) in the particular case $\, V
\equiv 1 \,, $ we deduce~: 
 
\smallskip \noindent   (7.22)   $ \quad \displaystyle 
{\rm d} \sigma \,\,= \,\, {{\mu}\over{T}} \, {\rm d}\rho \,-\,{{1}\over{T}} \,
{\rm d} \psi\,.\, $ 

\smallskip \noindent
Both identities (7.21) and (7.22)  establish that we have
 
\smallskip \noindent   (7.23)   $ \quad \displaystyle  
\sigma^* \, \Bigl( \, {{\mu}\over{T}} \,,\, -{{1}\over{T}} \, \Bigr) \,\,= \,\,
{{p}\over{T}} \,. \, $

\smallskip \noindent
In this context, the relation (7.12) can be written as
$ \,\,\,  {{1}\over{T}} > 0 \, \,\, $
and the following calculus, issued from the relation (7.16) :
 
\smallskip \noindent  $  \displaystyle   
\Pi(\rho,\, \psi) \,\,=\,\, -{{\sigma^*}\over{(-1/T)}} \,\,=\,\, {{p / T}
\over{(1 /
T)}} \,\,=\,\, p \Bigl( \, {{\mu}\over{T}} \,,\, -{{1}\over{T}} \, \Bigr) $
 
\smallskip \noindent
gives an {\bf intrinsic} definition of the mechanical  pressure 
$\,\, \Pi ({\scriptstyle \bullet})  \,\,$ 
as  identical to the thermodynamic pressure 
$\,\,  p ({\scriptstyle \bullet})  \,\,$ 
with the help of the relation  (7.23) in terms of the
{\bf dual} of the thermostatic specific entropy. We remark also that, in some
sense, the theorem 2  establishes theoretically the Euler equations of gas
dynamics.

\bigskip 
\noindent {\bf Proof of Theorem 2.}

\smallskip \noindent $\bullet \quad$
We derive the relation (1.14) relatively to the velocity $\, v \,:$

\smallskip \noindent  $  \displaystyle   
{\rm d}Y(v) \, {\scriptstyle \bullet} \, R  \, {\scriptstyle \bullet}
\,Y(v) \,+\, Y(v)  \, {\scriptstyle \bullet} 
\, R  \, {\scriptstyle \bullet} \,{\rm d}Y(v) \,\,=\,\, 0 \,   $  

\smallskip \noindent
and we consider the particular case $\, v = 0.\,$ It comes :

\smallskip \noindent   (7.24)   $ \quad \displaystyle  
{\rm d}Y(0) \, {\scriptstyle \bullet} \, R  \,+\, R  \, {\scriptstyle
\bullet} \,{\rm d}Y(0) \,\,=\,\, 0 \,.\,  $

\smallskip \noindent
Then the range of the linear space $\, \Lambda\ib{1} \,$ by  the operator
$\, {\rm
d}Y(0) \,$ is included inside the eigenspace  $\, \Lambda\ib{-1} \,$ and
the range
of the linear space $\, \Lambda\ib{-1} \,$ by  the same mapping is included
inside the
eigenspace  $\, \Lambda\ib{-1} .\,$ Moreover, the kernel of  $\, {\rm
d}Y(0) \,$ is stable under the action of the symmetry operator $\, R \,: \,$ 

\smallskip \noindent   (7.25)   $ \quad \displaystyle 
{\rm d}Y(0) \, ( \Lambda\ib{1} ) \,\,\,\subset \,\,\, \Lambda\ib{-1} \, $

\smallskip \noindent   (7.26)   $ \quad \displaystyle 
{\rm d}Y(0) \,(\Lambda\ib{-1} )\,\,\, \subset \,\,\, \Lambda\ib{1} \, $

\smallskip \noindent   (7.27)   $ \quad \displaystyle  
R \bigl( \, {\rm ker} \,  {\rm d}Y(0)  \, \bigr) \,\,\, \subset \,\,\,{\rm ker}
\,  {\rm d}Y(0)  \,. \, $

\smallskip \noindent
We deduce from the relation (4.6) $\,( {\rm dim} \,\,T\ib{W}\Omega_0 = 2
\,$ in our case) and from the relation 
 (4.14) $\,(\Omega_0 \subset \Lambda\ib{1})\,$
that we have necessarily :

\smallskip \noindent   (7.28)   $ \quad \displaystyle   
{\rm dim} \, \Lambda\ib{1}\,\ge \, 2 \,,\qquad T\ib{W}\Omega_0 \, \subset \,
\Lambda\ib{1} \,.\, $

\smallskip \noindent $\bullet \quad$
The case where $\,\, {\rm dim} \, \Lambda\ib{1}\,= \, 3 \,\,$ is not
possible. Indeed we would have 
$\,\, \Lambda\ib{-1} = \{0\} \,\,$ 
and the relation (4.13) establishes
that in this case the thermodynamic flux $\, j({\scriptstyle \bullet})
\,$ is identically null on the manifold $\, \Omega_0 \,$ then on the entire
space $\, \Omega .\,$ 
Moreover, the relation (3.10) joined with the relation 
$\, R = {\rm Id} \,$ shows that $\, u(W) \,=\, 0 \,$ 
for each $\,\, W \in \Omega .\,$ This fact
contradicts the relation (4.6) that claims that  $\, {\rm dim}
\,\,T\ib{W}\Omega_0 = 2 .\,$ 
Taking into account the relation (7.28), we have established that we have
necessarily 

\smallskip \noindent   (7.29)   $ \quad \displaystyle  
{\rm dim} \, \Lambda\ib{1}\,= \, 2 \,,\qquad  {\rm dim} \,
\Lambda\ib{-1}\,= \, 1 \,.\,  $
 
\smallskip \noindent
If $\,\, {\rm dim} \,\,{\rm ker} \, {\rm d}Y(0) \,= \, 3 \,,\,\,$ then
$\,\, Y(v)
\equiv {\rm Id} \,\, $ for each real parameter $\, v \,$ and the Galileo
group does
not operate anymore on the cone $\, \Omega. \,$ In particular, we have a
contradiction with the property (3.9) : $\, u(Y(v)\,{\scriptstyle \bullet})
\,W)=\,
u(W)-v .\,$

\smallskip \noindent $\bullet \quad$
If $\,\, {\rm dim} \,\,{\rm ker} \, {\rm d}Y(0) \,= \, 2 \,,\,\,$
due to the stability property (7.27),  we are necessarily in one of the two
following cases :

\smallskip \noindent   (7.30)   $ \quad \displaystyle   
{\rm ker} \,\, {\rm d}Y(0) \,\,=\,\, \Lambda\ib{1} \, $

\smallskip \noindent   (7.31)   $ \quad \displaystyle    
\exists \,  r_0 \in \Lambda\ib{1} , \, r_0 \ne 0 ,  \, 
\exists \, r_- \in \Lambda\ib{1} \,,\,  r_- \ne 0 ,\,  \, 
{\rm ker} \,  {\rm d}Y(0) \,=\, {\rm span} \,  <   r_0 \,,\, r_-  > . \, $
 
\smallskip \noindent
If the relation (7.30) holds, there exists a basis $\,\, (r_+ \,,\, r_0)
\,\,$ of the linear subspace 
$\, \Lambda\ib{1} \,$ with $\,\, {\rm d}Y(0) \, {\scriptstyle \bullet} 
\, r_+ \,=\, {\rm d}Y(0) \, {\scriptstyle \bullet} \, r_0 \,=\, 0
\,,\,$ a non null vector $\,\, r_- \in \Lambda\ib{-1} \,$ 
and a scalar $\,\, a > 0 \,\,$
such that inside the basis $\,\,( r_+ \,,\, r_- \,,\, r_0) \,,\,$ the
operator $\, {\rm d} Y(0) \,$ admits the following expression :

\smallskip \noindent   (7.32)   $ \quad \displaystyle     
{\rm d}Y(0) \,\,= \,\, \pmatrix{ 0 & -a & 0 \cr 0 & 0 & 0 \cr 0 & 0 & 0
\cr} \quad $
in the basis $\,\, ( r_+ \,,\, r_- \,,\, r_0) \,.\, $

\smallskip \noindent
Then \quad    $       \displaystyle      
Y(v) =    \pmatrix { 1 & -a \, v & 0 \cr 0 & 1 & 0 \cr 0 & 0 & 1 }   \, $
and for  $\displaystyle     W_0  =  \pmatrix { \theta_0 \cr 0 \cr \psi_0 } 
\in \Omega_0 , \,$ 

\smallskip \noindent
we have  $ \,\, \displaystyle    Y(v) \, {\scriptstyle \bullet} \, W_0 = 
  \pmatrix { \theta_0 \cr 0 \cr \psi_0 } \, $

\smallskip \noindent
that belongs always in the null manifold $\, \Omega_0 \, $   
 and we have a contradiction
exactly as in the preceding point. If the relation (7.31) is active, there
exists a
scalar $\, a > 0 \,$ such that inside a basis $\,\, ( r_+ \,,\, r_- \,,\,
r_0) \,\,$
consructed as previously, we have, taking into account the relation (7.25)~:

\smallskip \noindent   (7.33)   $ \quad \displaystyle     
{\rm d}Y(0) \,\,= \,\, \pmatrix{ 0 & 0 & 0 \cr -a & 0 & 0 \cr 0 & 0 & 0
\cr} \quad $  in the basis $\,\, ( r_+ \,,\, r_- \,,\, r_0) \,.\, $

\smallskip \noindent
Then \quad    $       \displaystyle   
Y(v) \,=\, \pmatrix { 1 & 0  & 0 \cr -a \, v  & 1 & 0 \cr 0 & 0 & 1 } \,\, \, $
and for $\displaystyle   \,\,\, W_0 \,=\, \pmatrix { \theta_0 \cr 0 \cr
\psi_0 } \in  \Omega_0 \,,   $ 

\smallskip \noindent   $\displaystyle   
 Y(v) \, {\scriptstyle \bullet} \, W_0
\,=\, \pmatrix { \theta_0 \cr  -a \, v \, \theta_0  \cr \psi_0 } \,\in
\Omega \,. \, $

\smallskip \noindent
We deduce from the relation (3.9) that
that  $\,\, u(\theta ,\, \zeta ,\, \psi) \,\equiv \, \zeta / (a \, \theta) \,\,$
and for $\,\, W \,=\, (\theta ,\, \zeta ,\, \psi)^{\rm t} \, \in \Omega
,\,\,$ we have
necessarily $\,\, \eta(\theta ,\, \zeta ,\, \psi) \, \equiv \, \sigma
(\theta,\, \psi)
\,\,$ where $\, \sigma ({\scriptstyle \bullet})  \,$ is the restriction of
the entropy
$\, \eta ({\scriptstyle \bullet})  \,$ to the null manifold $\, \Omega_0
.\,$ Then
the mathematical entropy $\, \eta({\scriptstyle \bullet})  \,$ cannot be a {\bf
strictly}  convex function and this case has to be excluded. In consequence
we have
necessarily $\,\, {\rm dim} \, ({\rm ker} \, {\rm d}Y(0)) \leq 1 \,.\,$

\smallskip \noindent $\bullet \quad$
We observe now that we have also necessarily $\,\,  {\rm dim} \,{\rm ker} \, {\rm
d}Y(0) \ge 1 \,\,$ because taking into account the relation (7.25), the operator
$\, {\rm d}Y(0) \,$ can be considered as a linear mapping between the
linear space $\, \Lambda\ib{1} \,$  
of dimension 2 and the linear space $\,  \Lambda\ib{-1} \,$  of
dimension 1, therefore with a necessarily non degerated kernel. According to the
previous point,  we are inconsequence in the particular case : 

\smallskip \noindent   (7.34)   $ \quad \displaystyle     
{\rm dim} \,\,\bigl( {\rm ker} \,\, {\rm d}Y(0) \bigr)  \,\,=\,\,1
\,,\qquad {\rm ker}
\,\, {\rm d}Y(0) \,\, \subset \,\, \Lambda\ib{1} \,. \,  $

\smallskip \noindent
With analogous notations as in the previous subsection, we introduce a non null
vector $\,\, r_0 \,\,$ ($r_0 \in \Lambda\ib{1})$ that generates the linear
space $\, \,
{\rm ker} \,\, {\rm d}Y(0) \,.\,$ We complete the basis of $\,
\Lambda\ib{1} \,$ by
some non null vector $\, r_+ \,$ and due to the inclusion (7.25),  its
image $\, r_-
\,$ by $\, {\rm d}Y(0) \,$ is necessarily a non null vector of the
eigenspace $\, \Lambda\ib{-1} \,:\,$ 

\smallskip \noindent   (7.35)   $ \quad \displaystyle     
\exists \, a > 0 \,,\qquad  {\rm d}Y(0) \, {\scriptstyle \bullet} \,r_+
\,\,= \,\, -a \, r_- \,.\, $

\smallskip \noindent
We write now the vector $\,\,  {\rm d}Y(0) \, {\scriptstyle \bullet} \,r_- \,\in
\Lambda\ib{1} \, \,$ inside the basis $\,(r_+ \,,\, r_0 ) \,:\,$

\smallskip \noindent   (7.36)   $ \quad \displaystyle    
{\rm d}Y(0) \, {\scriptstyle \bullet} \,r_- \,\,=\,\, c \, r_+ \,-\, d \,
r_0 \,.\, $

\smallskip \noindent
If $\,\, c \ne 0 ,\,$ we set $\,\, \widetilde{r}_+ \,=\, r_+ - d \, r_0 / c
\,\,$ and we have 
$\,\,  {\rm d}Y(0) \, {\scriptstyle \bullet} \,\widetilde{r}_ +
\,=\, \, -a \, r_- \,; \,\,$ 
$\,\,  {\rm d}Y(0) \, {\scriptstyle \bullet} \,r_- \,=\, c \,
\widetilde{r}_+ \,.$ Inside the basis $\,\, (\widetilde{r}_+ \,,\, r_-
\,,\, r_0) \,\,\,$ 
the matrix of the operator $\,{\rm d}Y(0) \,$ takes the form : 

\smallskip \noindent   (7.37)   $ \quad \displaystyle    
{\rm d}Y(0) \,\,=\,\, \pmatrix {0 & c & 0 \cr -a & 0 & 0 \cr 0 & 0 & 0 \cr
}\,,\qquad a > 0 \,,\quad c \ne 0 \,.\, $

\smallskip \noindent
We change our notations and replace the vector $\, r_+ \,$ initially
introduced by
the new vector $\, \widetilde{r}_+ \,.\,$ When the scalar $\, c \,$
introduced at the
relation (7.36)  is null, we have necessarily $\, d \ne 0 \,$ and after an
eventual
change of the sign of the vector $\, r_0 \,,\,$ the matrix of the operator
$\, {\rm
d}Y(0) \,$ inside the basis $\,\,(r_+ \,,\, r_- \,,\, r_0 ) \,\,$ can be
expressed as

\smallskip \noindent   (7.38)   $ \quad \displaystyle  
\exists \,  a > 0 \,,  \,\,  b > 0 \,,\,\,  
{\rm d}Y(0)  \,=\,  \pmatrix {0 & 0 & 0 \cr -a & 0 & 0 \cr 0 & -b & 0 \cr
}\,\,\,$ in the basis $\,\, ( r_+ \,,\, r_- \,,\, r_0) \,.\,$

\smallskip \noindent $\bullet \quad$ {\bf Case (i).} \quad
We develop now the two particular cases (7.37) and (7.38). If the relation
(7.37)
holds, we have in a first opportunity :

\smallskip \noindent   (7.39)   $ \quad \displaystyle  
\exists \, a > 0 \,,\,\, b > 0 \,,\quad
{\rm d}Y(0) \,=  \, \pmatrix {0 & -b & 0 \cr -a & 0 & 0 \cr 0 & 0 & 0 \cr
}\,\,\,$ in the basis $\,\, ( r_+ \,,\, r_- \,,\, r_0) \,\, $

\smallskip \noindent
and we are exactly in the case (i) of an ``hyperbolic Galileo'' system of
conservation laws. The proof follows what have been done at the theorem 1. By
exponentiation of the relation (7.39), we have :

\smallskip \noindent   (7.40)   $ \quad \displaystyle  
Y(v) \,\,=\,\, \pmatrix { {\rm ch} \bigl( v \, \sqrt{ a \, b \, }\,
\bigr) &  -\sqrt{ b / a \,}\, \, {\rm sh} \bigl( v \, \sqrt{a \, b }
\,\bigr) & 0 \cr
-\sqrt{ a / b  \,}\, \, {\rm sh} \bigl( v \, \sqrt{a \, b } \,\bigr) & {\rm ch}
\bigl( v \, \sqrt{a \, b \, }\, \bigr) & 0 \cr 0 & 0 & 1 } \,.\, $
 
\smallskip \noindent
For an arbitrary state $\, W \,=\, (\theta,\, \zeta ,\, \psi)^{\rm t}  \in
\Omega \,\,$ we have :

\smallskip \noindent   $ \displaystyle   
Y(v) \,{\scriptstyle \bullet} \, \pmatrix {\theta \cr \zeta \cr \psi} \,\,=\,\,
\pmatrix {\theta \,\, {\rm ch} \bigl( v \, \sqrt{ a \, b \, }\, \bigr)
\,-\, \zeta \,\, \sqrt{ b / a \,}\, \, {\rm sh} \bigl( v \, \sqrt{a \, b }
\,\bigr) \cr  -\sqrt{ a / b  \,}\, \, \theta 
\,  {\rm sh} \bigl( v \, \sqrt{a \, b } \,\bigr)
\,+\, \zeta \,\,  {\rm ch} \bigl( v \, \sqrt{ a \, b \, }\, \bigr) \cr \psi }
\,. \,   $

\medskip \noindent
Because $\,\, u \bigl( Y(u(W))  \,{\scriptstyle \bullet} \,W \bigr) \equiv
0 \,\,$
and $\,\, \Omega_0 \, \subset \, \Lambda\ib{1} \,=\, \R \times^{\rm \! t} \{ 0\}
\times^{\rm \! t} \R \,,\,$ we deduce from the previous equality  the
expression (7.4)
of the velocity field. We then have $\,\,\, \theta \, {\rm ch} \bigl( v \,
\sqrt{ a \,b \, }\, \bigr) \,-\, \zeta \,\, \sqrt{ b / a \,}\, \, {\rm sh} \bigl( v \,
\sqrt{a \, b } \,\bigr) 
\,=\, \sqrt{\theta^2 - b \, \zeta^2 / a \,} \,\,$ and the necessary
condition $\,\, \eta(W) \,=\, \eta \bigl( Y(u(W))  \,{\scriptstyle \bullet}
\,W \bigr)
\,\,$ shows that the relation (7.6) holds.

\smallskip \noindent $\bullet \quad$
We must look now to the precise conditions that makes the function $\, \eta
({\scriptstyle \bullet})\,$ defined in (7.6) a strictly convex function when the
property is satisfied for the two-variables function $\, \sigma (\alpha ,\,
\beta)
.\,$ We set as in the proof of the theorem 1 :

\smallskip \noindent   (7.41)   $ \quad \displaystyle  
\xi \,\,= \,\, \sqrt{ \theta^2 \,-\, {{b \, \zeta^2}\over{a}} \,} \,,\qquad
p \,=\, {{b}\over{a}}  \,$

\smallskip \noindent
and we have :

\smallskip \noindent   $  \displaystyle  
\xi \, {\rm d} \xi \,\,= \,\, \theta \,  {\rm d}\theta \,-\, {{b}\over{a}}
\, \zeta \,
{\rm d}\zeta \, \,,\quad {{\partial \eta}\over{\partial \theta}} \,=\,
{{\theta}\over{\xi}} \, {{\partial \sigma}\over{\partial \alpha}}
\,,\qquad   {{\partial \eta}\over{\partial \zeta}} \,=\, -{{b}\over{a}} \,
{{\zeta}\over{\xi}} \,  {{\partial \sigma}\over{\partial \alpha}} \,,\quad
 {{\partial \eta}\over{\partial \psi}} \,=\,  {{\partial
\sigma}\over{\partial \beta}}   \,,   $

\smallskip \noindent   $  \displaystyle   
 {{\partial^2 \eta}\over{\partial \theta^2}} \,=\, -p \,
{{\zeta^2}\over{\xi^3}}
\,{{\partial \sigma}\over{\partial \alpha}}  \,+\,   {{\theta^2}\over{\xi^2}}
\,\,{{\partial^2 \sigma}\over{\partial \alpha^2}}
\,,\qquad  {{\partial^2 \eta}\over{\partial  \theta \, \partial \zeta}}
\,=\,p \,   {{\theta \, \zeta}\over{\xi^3}} \,{{\partial \sigma}\over{\partial
\alpha}}  \,-\, p \,   {{\theta \, \zeta}\over{\xi^2}} \, {{\partial^2
\sigma}\over{\partial \alpha^2}} \,    $

\smallskip \noindent   $  \displaystyle   
{{\partial^2 \eta}\over{\partial  \theta \, \partial \psi}} \,=\,
{{\theta}\over{\xi}} \,  {{\partial^2 \sigma}\over{\partial  \alpha \,
\partial \beta}}
\,,\qquad  {{\partial^2 \eta}\over{\partial \zeta^2}} \,=\, -p \,
{{\theta^2}\over{\xi^3}} \,  {{\partial \sigma}\over{\partial \alpha}}
\,+\,p^2 \,
{{\zeta^2}\over{\xi^2}} \, {{\partial^2 \sigma}\over{\partial \alpha^2}} \, $ 

\smallskip \noindent   $  \displaystyle    
{{\partial^2 \eta}\over{\partial  \zeta \, \partial \psi}} \,=\, -p \,
{{\zeta}\over{\xi}} \,  {{\partial^2 \sigma}\over{\partial  \alpha \,
\partial \beta}}
\,,\qquad   {{\partial^2 \eta}\over{\partial \psi^2}} \,=\,  {{\partial^2
\sigma}\over{\partial \beta^2}} \,.    $ \qquad

\smallskip \noindent  Then

\smallskip \noindent   (7.42)   $ \,\,  \displaystyle   
{\rm d}^2 \eta \,= \, \pmatrix { \displaystyle
-p \,  {{\zeta^2}\over{\xi^3}} \,{{\partial \sigma}\over{\partial \alpha}}
\,+\, {{\theta^2}\over{\xi^2}} 
\,{{\partial^2 \sigma}\over{\partial \alpha^2}}
& \displaystyle
p \,   {{\theta \, \zeta}\over{\xi^3}} \,{{\partial \sigma}\over{\partial
\alpha}}  -  p \,   {{\theta \, \zeta}\over{\xi^2}} \, {{\partial^2
\sigma}\over{\partial \alpha^2}}
& \displaystyle {{\theta}\over{\xi}} \,
  {{\partial^2 \sigma}\over{\partial  \alpha \, \partial \beta}}
\cr \displaystyle
p \,   {{\theta \, \zeta}\over{\xi^3}} \,{{\partial \sigma}\over{\partial
\alpha}}  - p \,   {{\theta \, \zeta}\over{\xi^2}} \, {{\partial^2
\sigma}\over{\partial \alpha^2}}
& \displaystyle
-p \, {{\theta^2}\over{\xi^3}} \,  {{\partial \sigma}\over{\partial \alpha}}  
+  p^2 \, {{\zeta^2}\over{\xi^2}} \, {{\partial^2 \sigma}\over{\partial \alpha^2}}
& \displaystyle
-p \, {{\zeta}\over{\xi}} \,  {{\partial^2 \sigma}\over{\partial  \alpha \,
\partial \beta}}
\cr \displaystyle
{{\theta}\over{\xi}} \,  {{\partial^2 \sigma}\over{\partial  \alpha \,
\partial \beta}}
& \displaystyle  -p \, {{\zeta}\over{\xi}} 
\,  {{\partial^2 \sigma}\over{\partial  \alpha \,
\partial \beta}}
& \displaystyle
{{\partial^2 \sigma}\over{\partial \beta^2}} \cr  } \! .\, $ 

\smallskip \noindent
The two by two sub-matrix at the top and the left of the matrix of the
expression
(7.42) admits the following determinant :

\smallskip \noindent   $  \displaystyle     
{\det} \,\,  \Bigl( {\rm d}^2 \eta \Bigr)\ib{1 \le i \le 2 \,,\,\,1 \le j
\le 2} \,= \,{{1}\over{\xi^5}} \,    {{\partial \sigma}\over{\partial \alpha}} \, 
{{\partial^2 \sigma}\over{\partial \alpha^2}} \,  \Bigl( \, -p \, \theta^4
\,-\, p^3
\, \zeta^4 \,+\, 2 \, p^2 \, \theta^2 \, \zeta^2 \, \Bigr) $ 

\smallskip \noindent   $  \displaystyle  \qquad \qquad  \qquad \qquad \qquad \,\,=\,\,
-{{b}\over{a}} \,\, {{1}\over{\xi}} \,   {{\partial \sigma}\over{\partial
\alpha}} \,\, {{\partial^2 \sigma}\over{\partial \alpha^2}} \,    $

\smallskip \noindent
which is necessarily strictly positive in order to get the strict convexity
of the
function $\, \eta .\,$ We know that the second derivative $\,\, {{\partial^2
\sigma}\over{\partial \alpha^2}} \,\,$ is strictly positive and that $\, a
\,$ $\, b
\,$ and $\, \xi \,$ have the same property. Then we have $\,\, - {{\partial
\sigma}
\over {\partial \alpha}} \,> 0 \, \,$ and the relation (7.1) is
established. We develop now
the determinant of $\,\,  {\rm d}^2 \eta \,\,$ by using the third column.
It comes :

\smallskip \noindent $\displaystyle
{\det} \,\,  \Bigl( {\rm d}^2 \eta \Bigr) \,\,=\,\, -{{p}\over{\xi}} \,
{{\partial
\sigma} \over {\partial \alpha}} \,  {{\partial^2 \sigma} \over {\partial
\alpha^2}}
\,  {{\partial^2 \sigma} \over {\partial \beta^2}} \, \,\,+\,  $

\smallskip \noindent $\displaystyle \qquad  + \,\,\,\,
{{\theta}\over{\xi}} \,   {{\partial^2 \sigma} \over {\partial \alpha \,
\partial
\beta}} \,\,\, \left| \matrix { \displaystyle p \,   {{\theta \,
\zeta}\over{\xi^3}}
\,{{\partial \sigma}\over{\partial \alpha}}  \,-\, p \,   {{\theta \,
\zeta}\over{\xi^2}} \, {{\partial^2 \sigma}\over{\partial \alpha^2}}
& \displaystyle
-p \, {{\theta^2}\over{\xi^3}} \,  {{\partial \sigma}\over{\partial
\alpha}}  \,+\,p^2
\, {{\zeta^2}\over{\xi^2}} \, {{\partial^2 \sigma}\over{\partial \alpha^2}}
\cr \displaystyle
{{\theta}\over{\xi}} \,  {{\partial^2 \sigma}\over{\partial  \alpha \, \partial
\beta}} & \displaystyle
-p \, {{\zeta}\over{\xi}} \,  {{\partial^2 \sigma}\over{\partial  \alpha \,
\partial
\beta}} \cr} \right| \,  $

\smallskip \noindent $\displaystyle \qquad  + \,\,\,\,
p \, {{\zeta}\over{\xi}} \,   {{\partial^2 \sigma} \over {\partial \alpha
\, \partial
\beta}}  \,\, \left| \matrix { \displaystyle
-p \,  {{\zeta^2}\over{\xi^3}} \,{{\partial \sigma}\over{\partial \alpha}}
\,+\,
{{\theta^2}\over{\xi^2}} \,\,{{\partial^2 \sigma}\over{\partial \alpha^2}}
& \displaystyle
p \,   {{\theta \, \zeta}\over{\xi^3}} \,{{\partial \sigma}\over{\partial
\alpha}}  \,-\, p \,   {{\theta \, \zeta}\over{\xi^2}} \, {{\partial^2
\sigma}\over{\partial \alpha^2}}   \cr \displaystyle {{\theta}\over{\xi}} \,
{{\partial^2 \sigma}\over{\partial  \alpha \, \partial \beta}}
& \displaystyle
-p \, {{\zeta}\over{\xi}} \,  {{\partial^2 \sigma}\over{\partial  \alpha \,
\partial \beta}} \cr} \right| \,$

\smallskip \noindent $\displaystyle  \quad   = \,\, 
-{{p}\over{\xi}} \, {{\partial  \sigma} \over {\partial \alpha}} \,
{{\partial^2
\sigma} \over {\partial \alpha^2}} \,  {{\partial^2 \sigma} \over {\partial
\beta^2}} \,$ 

\smallskip \noindent $\displaystyle \qquad  + \,\,\,\,
 {{\theta}\over{\xi^4}} \, \Bigl({{\partial^2 \sigma}\over{\partial
\alpha \, \partial \beta}}\Bigr)^2 \,\, \left| \matrix{ \displaystyle
p \, {{\theta \,\zeta}\over{\xi}} \,  {{\partial  \sigma} \over {\partial \alpha}} \,
-\, p \, \theta \, \zeta \,  {{\partial^2 \sigma} \over {\partial \alpha^2}} &
\displaystyle -p \, {{\theta^2}\over{\xi}} \,  {{\partial  \sigma} \over
{\partial
\alpha}} \,+\, p^2 \, \zeta^2 \,   {{\partial^2 \sigma} \over {\partial
\alpha^2}}
\cr    \theta & -p \, \zeta \cr } \right|  \,  \,$

\smallskip \noindent $\displaystyle \qquad  + \,\,\,\, 
p \, {{\zeta}\over{\xi^4}} \, \, \Bigl({{\partial^2 \sigma}\over{\partial
\alpha \, \partial \beta}}\Bigr)^2 \,\, \left| \matrix{ \displaystyle
-p \, {{\zeta^2}\over{\xi}} \,  {{\partial  \sigma} \over {\partial
\alpha}} 
\, + \,
\theta^2 \,  {{\partial^2 \sigma} \over {\partial \alpha^2}} & \displaystyle
p \, {{\theta \, \zeta}\over{\xi}} \, {{\partial  \sigma} \over {\partial
\alpha}} \,
- \, p \, \theta \, \zeta \,  {{\partial^2 \sigma} \over {\partial
\alpha^2}} \cr
\theta & -p \, \zeta \cr } \right| \,$

\smallskip \noindent $\displaystyle  \quad   = \,\,  
-{{p}\over{\xi}} \, {{\partial  \sigma} \over {\partial \alpha}} \,
{{\partial^2
\sigma} \over {\partial \alpha^2}} \,  {{\partial^2 \sigma} \over {\partial
\beta^2}} \,$ 

\smallskip \noindent $\displaystyle \qquad  + \,\,\,\, 
 {{1}\over{\xi^4}} \, \Bigl({{\partial^2 \sigma}\over{\partial
\alpha \, \partial \beta}}\Bigr)^2 \,\,{{\partial  \sigma} 
\over {\partial \alpha}} \,
\Bigl[ \, {{\theta^2}\over{\xi}} \bigl( -p^2 \, \zeta^2 \,+\, p \, \theta^2
\bigr)
\,+\, {{p \, \zeta^2}\over{\xi}} \,  \bigl( p^2 \, \zeta^2 \,-\, p \,
\theta^2 \bigr)
\, \Bigr] \, $
\smallskip \noindent $ \displaystyle  \quad = \,\,\,\,
-{{p}\over{\xi}} \, {{\partial  \sigma} \over {\partial \alpha}} \,
{{\partial^2
\sigma} \over {\partial \alpha^2}} \,  {{\partial^2 \sigma} \over {\partial
\beta^2}}
\, \,\,+\,\, {{p}\over{\xi^5}} \, \Bigl({{\partial^2 \sigma}\over{\partial
\alpha \, \partial \beta}}\Bigr)^2 \,\,{{\partial  \sigma} \over {\partial
\alpha}} \,
\bigl( \theta^2 \,-\, p \, \zeta^2 \bigr) \,\,\bigl( \theta^2 \,-\, p \, \zeta^2
\bigr)  \,$
\smallskip \noindent $ \displaystyle  \quad = \,\,\,\,
-{{p}\over{\xi}} \,\,  {{\partial  \sigma} \over {\partial \alpha}} \,
\Big[  \,
{{\partial^2 \sigma} \over {\partial \alpha^2}} \,  {{\partial^2 \sigma} \over
{\partial \beta^2}} \,-\  \Bigl({{\partial^2 \sigma}\over{\partial \alpha
\, \partial
\beta}}\Bigr)^2 \, \Bigr] \, \hfill $ due to (7.41)

\smallskip \noindent
and

\smallskip \noindent   (7.43)   $ \quad \displaystyle  
{\det} \,\,  \bigl( {\rm d}^2 \eta \bigr) \,\,=\,\, -{{b}\over{a\,\sqrt{
\theta^2 - b \, \zeta^2 / a \,}}} \, 
 {{\partial  \sigma} \over {\partial \alpha}} \, \,
\Big[  \, {{\partial^2 \sigma} \over {\partial \alpha^2}} \, 
 {{\partial^2 \sigma} \over {\partial \beta^2}} 
\,-\  \Bigl({{\partial^2 \sigma}\over{\partial \alpha
\, \partial \beta}}\Bigr)^2 \, \Bigr] \,.\,  $ 

\smallskip \noindent
The matrix $\,\, {\rm d}^2 \eta  \,\,$ is definite positive  if it is the
case for the
matrix $\,\, {\rm d}^2 \sigma  \,\,$ and if $\,\,  {{\partial  \sigma}
\over {\partial
\alpha}} < 0 \,, \,$ {\it i.e.} when the condition (7.1) is satisfied.

\smallskip \noindent $\bullet \quad$
In order to obtain the algebraic expression of the thermodynamic flux $\,
j({\scriptstyle \bullet}) \,,\,$ we first  evaluate the entropy variables 

\smallskip \noindent   (7.44)   $ \quad \displaystyle  
\varphi \,\, =\,\, \Bigl( \,  {{\partial  \eta} \over {\partial \theta}} \,,\,
{{\partial  \eta} \over {\partial \zeta}} \,,\,   {{\partial  \eta} \over
{\partial
\psi}} \, \Bigr) \,\,=\,\,  \Bigl( \,  {{\theta}\over{\xi}} \,  {{\partial
\sigma}
\over {\partial \alpha}} \,,\, -{{b \, \zeta}\over{a \, \xi}} \,
{{\partial  \sigma}
\over {\partial \alpha}} \,,\, {{\partial  \sigma} \over {\partial \beta}}
\, \Bigr) \,  $ 

\smallskip \noindent
on the null-velocity manifold $\, \Omega_0 \,$ that corresponds to $\,\, (\theta
\,,\, \zeta \,,\, \psi) \,\,=\,\, ( \theta_0 \,,\, 0 \,,\, \psi_0 ) \,. \,$
Then
$\displaystyle  \,\, \varphi_0 \,= \, \Bigl( \,
{{\partial  \sigma}
\over {\partial \alpha}}(\theta_0,\, \psi_0) \,\,,\,\, 0 \,\,,\,\,
{{\partial  \sigma}
\over {\partial \beta}}(\theta_0,\, \psi_0) \, \Bigr) \,.\,$ Taking into account the
relation (7.39), it comes :

\smallskip \noindent   (7.45)   $ \quad \displaystyle   
\varphi_0 \, {\scriptstyle \bullet} \, {\rm d}Y(0) \,\,=\,\, \Bigl( \, 0
\,\,,\,\, -b  \,  {{\partial  \sigma} \over {\partial \alpha}} 
\,\,,\,\, 0 \, \Bigr) \,.\, $

\smallskip \noindent
We introduce the mechanical pressure $\,\, \Pi({\scriptstyle \bullet})\,\,$ as a
notation~: 

\smallskip \noindent   (7.46)   $ \quad \displaystyle   
\forall \,\, W_0 \,=\, (\theta_0,\, 0 ,\, \psi_0)^{\rm t} \, \in \Omega_0
\,,\qquad j(W_0) \,\,=\,\, \bigl( \, 0 \,,\, \Pi(\theta_0,\,\psi_0)\,,\, 0
\,\bigr)^{\rm t}\, $
 
\smallskip \noindent
and the relation $\,\, \varphi \,{\scriptstyle \bullet}\, {\rm d}j \,+\,
\eta^* \,
{\rm d}u \,\equiv\, 0 \,\,$ applied against the vector $\,\,{\rm d}Y(0)
\,{\scriptstyle
\bullet}\,W_0 \,\,\,$ gives

\smallskip \noindent   $  \displaystyle   
b \,  {{\partial  \sigma} \over {\partial \alpha}} (\theta_0,\, \psi_0) \,\,
\Pi(\theta_0,\,\psi_0) \,\,+\,\, \sigma^*(A,\,B) \,=\, 0 $

\smallskip \noindent 
 with  
$ \quad A ={{\partial  \sigma} \over {\partial \alpha}}(\theta_0,\, \psi_0) \,,\,\,\, B =
{{\partial  \sigma} \over {\partial \beta}}(\theta_0,\, \psi_0) ,\, \, $  
and this property is exactly the relation (7.5). The general expression of the
thermodynamic flux $\, j(W) \,$ is obtained thanks to the relations (3.12) and
(7.40)~:

\smallskip \noindent   $  \displaystyle    
j(W) \,\,= \,\, Y(-u(W)) \,{\scriptstyle \bullet}\, j \bigl(
Y(u(W))\,{\scriptstyle
\bullet}\, W\bigr) $

\smallskip \noindent   $  \displaystyle    \quad \,\, = \, 
\pmatrix { {\rm ch} \bigl( u \, \sqrt{a \, b \, }\, \bigr) &  \sqrt{ b /
a \,}\, \, {\rm sh} \bigl( u \, \sqrt{a \, b } \,\bigr) & 0 \cr  \sqrt{ a /
b  \,}\, \,
{\rm sh} \bigl( u \, \sqrt{a \, b } \,\bigr) & {\rm ch} \bigl( u \, \sqrt{a
\, b \, }\,
\bigr) & 0 \cr 0 & 0 & 1 } \,{\scriptstyle \bullet}\, \pmatrix {0 \cr \Pi\bigl(
\sqrt{\theta^2 - b \, \zeta^2 / a } \,,\,\psi \bigr) \cr 0 } \,$

\smallskip \noindent   (7.47)   $ \quad \displaystyle   
\forall \, W \,=\, \pmatrix {\theta \cr \zeta \cr \psi \cr} \, \in \Omega
\,,\qquad
j(W) \,\,= \,\, {{ \Pi\bigl( \sqrt{\theta^2 - b \, \zeta^2 / a } \,,\,\psi
\bigr) }\over{\sqrt{\theta^2 - b \, \zeta^2 / a }}} \,\, 
\pmatrix{b \, \zeta / a \cr \theta  \cr 0 } \,.\, $

\smallskip  \noindent
Joined with the expression (7.4) of the velocity field, the relation (7.47)
establishes the expression (7.3) of the hyperbolic Galileo system of
conservation
laws.

\smallskip \noindent $\bullet \quad$
We verify now that the function $\,\, \eta({\scriptstyle \bullet}) \,\,$
defined at
the relation (7.6) is effectively a mathematical entropy associated with
the flux
$\,\, u(W) \, \eta(W) \,;\,$  in other words  we have the additional
conservation law

\smallskip \noindent   (7.48)   $ \quad \displaystyle    
{{\partial \eta(W)}\over{\partial t}} \,+\,
{{\partial \eta(W)}\over{\partial x}} \bigl( u(W) \, \eta(W) \bigr)
\,\,=\,\,0 \,\, $
 
\smallskip \noindent
if $\,\, W(x,\,t) \,\,$ is a regular  solution of the conservation law (7.3). By
differentiation of the relation (7.4), we have :

\smallskip \noindent   (7.49)   $ \quad \displaystyle   
a \, {\rm d} u \,\,=\,\, {{1}\over{\theta^2 \,- \, b \, \zeta^2 / a}} \, \bigl(
\theta \, {\rm d}\zeta \,-\, \zeta \, {\rm d}\theta \bigr) \,.\,  $
 
\smallskip \noindent
We have now do develop some algebraic calculus :

\smallskip \noindent  $ \displaystyle     
{{\partial \eta}\over{\partial t}} \,+\,  {{\partial}\over{\partial x}}
\bigl( \eta \, u \bigr) \,\,= \,\, \Bigl(  {{\partial}\over{\partial t}} + u \,
{{\partial}\over{\partial x}} \Bigr) \eta(W) \,+\, \eta \,  {{\partial
u}\over{\partial x}} \,\,=\, $

\smallskip  \noindent $\displaystyle \qquad = \,\,
{{\partial \sigma}\over{\partial \alpha}} \, \, \Bigl[ \,
{{\partial \xi}\over{\partial \theta}} \, \Bigl(  {{\partial}\over{\partial
t}} + u \,  {{\partial}\over{\partial x}} \Bigr) \theta \,+\,  {{\partial
\xi}\over{\partial \zeta}} \, \Bigl(  {{\partial}\over{\partial t}} + u \,
{{\partial}\over{\partial x}}
\Bigr) \zeta \,\Bigr] \,+\,  {{\partial \sigma}\over{\partial \beta}} \, \Bigl(
{{\partial}\over{\partial t}} + u \,  {{\partial}\over{\partial x}} \Bigr)
\psi \,+\, \eta \,  {{\partial u}\over{\partial x}} \,$

\smallskip  \noindent $\displaystyle \qquad = \,\,
{{\partial \sigma}\over{\partial \alpha}} \, \, \Bigl[ \,
{{\theta}\over{\xi}} \,
\Bigl(  {{\partial}\over{\partial t}} + u \,  {{\partial}\over{\partial x}}
\Bigr) \theta \,-\,  {{b}\over{a}} \, {{\zeta}\over{\xi}} \, \Bigl(  {{\partial}\over
{\partial t}} + u \,  {{\partial}\over{\partial x}} \Bigr) \zeta \,\Bigr] \,+\,
{{\partial \sigma}\over{\partial \beta}} \, \Bigl(
{{\partial}\over{\partial t}} + u
\,  {{\partial}\over{\partial x}} \Bigr) \psi \,+\, \eta \,  {{\partial
u}\over{\partial x}}  \,$

$ \hfill $ due to (7.44)

\smallskip  \noindent $\displaystyle \qquad = \,\,
{{\theta}\over{\xi}} \, {{\partial \sigma}\over{\partial \alpha}} \, \Bigl[
\, -\theta
\, {{\partial u}\over{\partial x}} \,-\, {{\partial}\over{\partial x}} \, \Bigl(
\Pi(\xi,\,\psi) \, {{b \, \zeta}\over{a \, \xi}} \Bigr) \, \Bigr]  $ 

\smallskip  \noindent $\displaystyle \qquad \qquad \qquad  - \,
{{b}\over{a}} \, {{\zeta}\over{\xi}} \,  {{\partial \sigma}\over{\partial
\alpha}} \, \Bigl[ \,-\zeta \, {{\partial u}\over{\partial x}}  \, 
\,-\,  {{\partial}\over {\partial x}} \, 
\Bigl( \Pi(\xi,\,\psi) \, {{\theta}\over{\xi}} \, \Bigr)  \, \Bigr]  \,$

\smallskip  \noindent $\displaystyle \qquad \qquad \qquad +\,\,\,
{{\partial \sigma}\over{\partial \beta}} \,\bigl( -\psi \, {{\partial
u}\over{\partial
x}} \bigr) \,+\, \eta \,  {{\partial u}\over{\partial x}} \, \hfill $
according to (7.3)

\smallskip  \noindent $\displaystyle \qquad = \,\,
{{\partial \sigma}\over{\partial \alpha}} \, {{\partial u}\over{\partial x}} \,
\Bigl[ -{{\theta^2}\over{\xi}} \,+\, {{b}\over{a}} \, {{\zeta^2}\over{\xi}}
\Bigr]
\,-\, {{\partial \sigma}\over{\partial \beta}} \, {{\partial
u}\over{\partial x}} \,
\psi \,+\, \eta \,  {{\partial u}\over{\partial x}}  \,$

\smallskip  \noindent $\displaystyle \qquad \qquad \qquad  +\, {{1}\over{\xi}} \,
{{\partial \sigma}\over{\partial \alpha}} \,\Pi(\xi,\,\psi)
\,\Bigl[-{{\theta}\over{\xi}} \, {{b}\over{a}} \, {{\partial
\zeta}\over{\partial x}}
\,+\,{{b}\over{a}} \, {{\zeta}\over{\xi}} \, {{\partial
\theta}\over{\partial x}}
\Bigr] \,$

\smallskip  \noindent $\displaystyle \qquad = \,\,
-{{\partial u}\over{\partial x}} \,\Bigl[ \, \xi \,  {{\partial
\sigma}\over{\partial
\alpha}}(\xi,\,\psi) \,+\, \psi \, {{\partial \sigma}\over {\partial
\beta}}(\xi,\,\psi) \,-\, \sigma (\xi,\,\psi) \, \Bigr] \,$

\smallskip  \noindent $\displaystyle \qquad \qquad \qquad  -
\, {{b}\over{a \, \xi^2}}
\,  {{\partial \sigma}\over{\partial \alpha}} \, \Pi(\xi,\,\psi) \, \Bigl[
\theta \,{{\partial \zeta}\over{\partial x}} 
\,-\, \zeta \,  {{\partial  \theta}\over{\partial  x}} \Bigr] \, $

\smallskip  \noindent $\displaystyle \qquad = \,\,
-{{\partial u}\over{\partial x}} \,\sigma^* \,-\, {{b}\over{a}} \,  {{\partial
\sigma}\over{\partial \alpha}}\, \Bigl( -{{1}\over{b}} \,
{{\sigma^*}\over{\partial
\sigma \,/\, \partial \alpha}} \Bigr) \,\, \Bigl( a \, {{\partial
u}\over{\partial x}} \Bigr) \, \hfill $ due to (7.5) and (7.49)

\smallskip  \noindent $\displaystyle \qquad = \,\,0 \,$
\smallskip \noindent
and the property (7.48) is established. The study of the first case is over.

\smallskip \noindent $\bullet \quad$ {\bf Case (ii).} \quad
We consider again the expression (7.37) of the jacobian matrix $\, {\rm
d}Y(0) \,$
now with $\, c > 0 \,$ and the relation (7.37) can be rewritten as :

\smallskip \noindent   (7.50)   $ \quad \displaystyle    
\exists \,   a > 0  ,\,\,  b > 0 , \quad
{\rm d}Y(0) \,=\, \pmatrix {0 & b & 0 \cr -a & 0 & 0 \cr 0 & 0 & 0 \cr
} \quad   \,$ in the basis $\,\, ( r_+ \,,\, r_- \,,\, r_0) \,. \, $

\smallskip \noindent
The relation (6.30) established during the proof of the theorem 1 can be
reproduced
without any modification and we have in consequence :

\smallskip \noindent   (7.51)   $ \quad \displaystyle     
Y(v) \,\,=\,\, \pmatrix { {\rm cos} \bigl( v \, \sqrt{ a \, b \, }\, \bigr)
&  \sqrt{ b
/ a \,}\, \, {\rm sin} \bigl( v \, \sqrt{a \, b } \,\bigr) & 0 \cr  -\sqrt{ a / b
\,}\, \, {\rm sin} 
\bigl( v \, \sqrt{a \, b } \,\bigr) & {\rm cos} \bigl( v \, \sqrt{ a
\, b \, }\, \bigr) & 0 \cr 0 & 0 & 1 } \,;\, $

\smallskip \noindent
then for $\,\,(\theta,\, \zeta,\, \psi)^{\rm t} \in \Omega \,,\,$ we have

\smallskip \noindent  $\displaystyle      
Y(v) \,{\scriptstyle \bullet} \, \pmatrix {\theta \cr  \zeta \cr \psi \cr}
\,\,=\,\,  \pmatrix { \theta \,\, 
 {\rm cos} \bigl( v \, \sqrt{ a \, b \, }\, \bigr) \,\,+\,\,
\sqrt{ b / a \,}\, \, \zeta \,\,  {\rm sin} \bigl( v \, \sqrt{a \, b }
\,\bigr)  \cr  -\sqrt{ a / b  \,}\, \, \theta \,\, {\rm sin} \bigl( v \,
\sqrt{a \, b } \,\bigr) \,+\, \zeta \,\, 
 {\rm cos} \bigl( v \, \sqrt{ a \, b \,  }\, \bigr)  \cr \psi} \,.\, $

\smallskip \noindent
We know that the reference basis $\,\, (r_+ \,,\, r_- \,,\, r_0 ) \,\,$
belongs to
the product of spaces $\,\, \Lambda\ib{1} \times  \Lambda\ib{-1} \times
\Lambda\ib{1}
\,\,$ and $\,\, \Omega_0 \, \subset \Lambda\ib{1} \,. \,$ We deduce that
the second
component of the state $\,\,  Y(v) \,{\scriptstyle \bullet} \, W \,\,$ that
belongs
to the null velocity manifold is necessarily null and we have

\smallskip \noindent   (7.52)   $ \quad \displaystyle      
{\rm tg} \, \bigl( u\sqrt{a \, b} \,\bigr) \,\,= \,\, \sqrt{{{b}\over{a}}} \,
{{\zeta}\over{\theta}} \, $

\smallskip \noindent
and the relation (7.9) is established. We deduce naturally

\smallskip \noindent   (7.53)   $ \quad \displaystyle       
{\rm cos} (u\sqrt{a \, b} \, ) \,\,= \,\, {{\theta}\over{\xi}} \,,\qquad
{\rm sin} (u\sqrt{a \, b} \, ) \,\,= \,\,  \sqrt{{{b}\over{a}}} \,
{{\zeta}\over{\xi}} \,,\qquad \xi \,=\, \sqrt{\theta^2 \,+\, b \, \zeta^2 / a\,} $

\smallskip \noindent   (7.54)   $ \quad \displaystyle        
Y(u(W)) \,{\scriptstyle \bullet} \, W \,\,=\,\, ( \, \xi \,,\, 0 \,,\, \psi
\, )^{\rm t} \,,
\qquad \xi \,=\, \sqrt{\theta^2 \,+\, b \, \zeta^2 / a\,}\,  $
 
\smallskip \noindent
and the relation (7.11) is a consequence of the invariance (3.2)  of the
mathematical
entropy for the transformation $\, Y(v) .\, $

\smallskip \noindent $\bullet \quad$
We consider now as given the strictly convex  function $\, \sigma({\scriptstyle
\bullet})\,$ which is the restriction of the mathematical entropy $\,
\eta({\scriptstyle \bullet})\,$ to the null manifold $\, \Omega_0 .\, $ We must
verify that the mathematical entropy is also a strictly convex function of the
triplet $\,\, (\theta ,\, \zeta,\, \psi) \,.\,$ We set as above $\,\, p = b
/ a \,\,$
and we have

\smallskip \noindent   (7.55)   $ \quad \displaystyle     
\varphi \,\,=\,\, \Bigl( \, {{\partial \eta}\over{\partial \theta}} \,,\,
{{\partial \eta}\over{\partial \zeta}} \,,\,
 {{\partial \eta}\over{\partial \psi}} \,\Bigr)
\,\,= \,\, \Bigl( \, {{\theta}\over{\xi}} \, {{\partial
\sigma}\over{\partial \alpha}}
\,,\, p \, {{\zeta}\over{\xi}} \, {{\partial \sigma}\over{\partial \alpha}}
\,,\,{{\partial \sigma}\over{\partial \beta}} \, \Bigr)  \,.\,  $

\noindent   Then

\smallskip \noindent  $  \displaystyle   
 {{\partial^2 \eta}\over{\partial \theta^2}} \,=\, p \,  {{\zeta^2}\over{\xi^3}}
\,{{\partial \sigma}\over{\partial \alpha}}  \,+\,   {{\theta^2}\over{\xi^2}}
\,\,{{\partial^2 \sigma}\over{\partial \alpha^2}}
\,,\qquad  {{\partial^2 \eta}\over{\partial  \theta \, \partial \zeta}}
\,=\,-p \,   {{\theta \, \zeta}\over{\xi^3}} \,{{\partial \sigma}\over{\partial
\alpha}}  \,+\, p \,   {{\theta \, \zeta}\over{\xi^2}} \, {{\partial^2
\sigma}\over{\partial \alpha^2}} \,,\qquad $

\smallskip \noindent  $  \displaystyle   
{{\partial^2 \eta}\over{\partial  \theta \, \partial \psi}} \,=\,
{{\theta}\over{\xi}} \,  {{\partial^2 \sigma}\over{\partial  \alpha \,
\partial \beta}} \,,\qquad   
{{\partial^2 \eta}\over{\partial \zeta^2}} \,=\, p \,
{{\theta^2}\over{\xi^3}} \,  {{\partial \sigma}\over{\partial \alpha}}
\,+\,p^2 \,
{{\zeta^2}\over{\xi^2}} \, {{\partial^2 \sigma}\over{\partial \alpha^2}} \,, $ 

\smallskip \noindent  $  \displaystyle    
{{\partial^2 \eta}\over{\partial  \zeta \, \partial \psi}} \,=\, p \,
{{\zeta}\over{\xi}} \,  {{\partial^2 \sigma}\over{\partial  \alpha \, \partial \beta}}
\,,\qquad   {{\partial^2 \eta}\over{\partial \psi^2}} \,=\,  {{\partial^2
\sigma}\over{\partial \beta^2}} \,   $

\noindent   and
 
\smallskip \noindent   (7.56)   $ \,  \displaystyle     
{\rm d}^2 \eta  \,= \,  \pmatrix { \displaystyle
 \!\! p \,  {{\zeta^2}\over{\xi^3}} \,{{\partial \sigma}\over{\partial \alpha}}
+  {{\theta^2}\over{\xi^2}} \, \,{{\partial^2 \sigma}\over{\partial \alpha^2}}
& \displaystyle
  -p \,   {{\theta \, \zeta}\over{\xi^3}} \,{{\partial \sigma}\over{\partial
\alpha}} +  p \,   {{\theta \, \zeta}\over{\xi^2}} \, {{\partial^2
\sigma}\over{\partial \alpha^2}}
& \displaystyle
{{\theta}\over{\xi}} \,  {{\partial^2 \sigma}\over{\partial  \alpha \,
\partial \beta}}
\cr \displaystyle
 \!\!  -p \,   {{\theta \, \zeta}\over{\xi^3}} \,{{\partial \sigma}\over{\partial
\alpha}}  +  p \,   {{\theta \, \zeta}\over{\xi^2}} \, {{\partial^2
\sigma}\over{\partial \alpha^2}}
& \displaystyle
p \, {{\theta^2}\over{\xi^3}} \,  {{\partial \sigma}\over{\partial \alpha}}
+ p^2 \, {{\zeta^2}\over{\xi^2}} \, {{\partial^2 \sigma}\over{\partial \alpha^2}}
& \displaystyle
p \, {{\zeta}\over{\xi}} \,  {{\partial^2 \sigma}\over{\partial  \alpha \,
\partial  \beta}}
\cr \displaystyle
{{\theta}\over{\xi}} \,  {{\partial^2 \sigma}\over{\partial  \alpha \,
\partial \beta}}
& \displaystyle
p \, {{\zeta}\over{\xi}} \,  {{\partial^2 \sigma}\over{\partial  \alpha \,
\partial  \beta}}
& \displaystyle
{{\partial^2 \sigma}\over{\partial \beta^2}} \cr  }  . $

\medskip \noindent
This matrix is identical to the one presented at the relation (7.42),
except that the
variable $\, p \,$ must be changed into $\,-p .\,$ But this change of sign
is exactly what is necessary to modify the 
definition of the variable $\, \xi \,$ from  (7.41) to
(7.53). We observe that the two by two minor determinant composed by the
two left  lines and the two first columns of the right hand 
side of the  relation (7.56) is equal
to $\,\,{{p}\over{\xi}} \,  {{\partial \sigma}\over{\partial \alpha}}  \,
{{\partial^2  \sigma}\over{\partial \alpha^2}}  \,\,$ 
and is strictly positive when $\,\, {{\partial
\sigma}\over{\partial \alpha}}  \,\,$ is strictly positive. In consequence the
hypothesis (7.7) is clearly established. The computation of $\,\, {\det} \,
({\rm d}^2 \eta) \,\,$ 
done during the study of the hyperbolic case conducts to the
relation (7.43) and the change of the variable $\, p \,$ 
into $\, -p \,$ establishes  that

\smallskip \noindent   (7.57)   $ \quad \displaystyle   
{\det} \,\,  \bigl( {\rm d}^2 \eta \bigr) \,\,=\,\, {{p}\over{\xi}} \,
{{\partial
\sigma} \over {\partial \alpha}} \, \,  \Big[  \,  {{\partial^2 \sigma} \over
{\partial \alpha^2}} \,  {{\partial^2 \sigma} \over {\partial \beta^2}} \,-\
\Bigl({{\partial^2 \sigma}\over{\partial \alpha \, \partial
\beta}}\Bigr)^2 \, \Bigr] \,\, $

\smallskip \noindent
which is strictly positive when $\, \sigma({\scriptstyle \bullet}) \,$ is
strictly convex and when the relation (7.7) is satisfied.

\smallskip \noindent $\bullet \quad$
The constitution of the thermodynamic flux $\,\, j({\scriptstyle
\bullet})\,\,$ is obtained exactly as the hyperbolic case. 
Taking into account the relations  (7.50) and 
(7.55), we have, for $\,\, W_0 \,=\, (\theta_0,\, 0,\, \psi)^{\rm t} $ $\in
\Omega_0  \,:\,$

\smallskip \noindent   $ \displaystyle    
\varphi(\theta_0,\, 0,\, \psi) \, {\scriptstyle \bullet} \, {\rm d}Y(0)
\,=\, \Bigl( {{\partial
\sigma} \over {\partial \alpha}} \,,\, 0 \,,\, {{\partial  \sigma} \over
{\partial
\beta}} \Bigr) \, \pmatrix{0 & b & 0 \cr -a & 0 & 0 \cr 0 & 0 & 0 } \,=\,
\Bigl( \, 0
\,,\, b \,  {{\partial  \sigma} \over {\partial \beta}} \,,\, 0 \, \Bigr) \, $  

\smallskip  \noindent
and if we introduce the mechanical pressure  $\,\, \Pi( {\scriptstyle
\bullet}) \,\,$
with the relation (7.46), we have from the relation (2.9) : $\,\,\, \varphi_0 \,
{\scriptstyle \bullet} \, {\rm d}Y(0) \,{\scriptstyle \bullet} \,j(W_0)
\,+\, \sigma^*
\,(-1) \,=\,0 \,,\,\,\,$ and this calculus proves exactly the relation
(7.10). The
general expression of the thermodynamic flux is given by somes lines of algebra that
are consequence of the relations (3.12)  and (7.51)~:

\smallskip \noindent   $ \displaystyle     
j(W) \,\,= \,\, Y(-u(W)) \,{\scriptstyle \bullet}\, j \bigl(
Y(u(W))\,{\scriptstyle  \bullet}\, W\bigr) $

\smallskip \noindent $ \displaystyle   \quad   =   \,
\pmatrix { \!\! {\rm cos} \bigl( u \, \sqrt{ a \, b \, }\, \bigr) 
&   \!\!   -\sqrt{ b / a \,}\, \,
{\rm sin} \bigl( u \, \sqrt{a \, b } \,\bigr) & 0 \cr 
  \!\! \sqrt{ a / b  \,} \, {\rm sin}
\bigl( u \, \sqrt{a \, b } \,\bigr) & {\rm cos} \bigl( u \, \sqrt{ a \, b
\, }\, \bigr) & 0 \cr 0 & 0 & 1 }
\,{\scriptstyle \bullet}\, \pmatrix {0 \cr \Pi\bigl( \sqrt{\theta^2 + b \,
\zeta^2 / a } \,,\,\psi \bigr) \cr 0 } \,$

\smallskip \noindent $ \displaystyle   \quad   =   \, 
\Pi (\xi,\, \psi) \,\, \pmatrix{ \displaystyle  -\sqrt{{{b}\over{a}}} \,
\sqrt{{{b}\over{a}}} \, {{\zeta}\over{\xi}} \cr \theta / \xi \cr 0 }
\,\hfill $ taking  into account the relation (7.53)

\smallskip \noindent $ \displaystyle   \quad   =   \,  
{{1}\over{\xi}} \,\,  \Pi (\xi,\, \psi) \,\, \pmatrix{ \displaystyle
-{{b\, \zeta }\over{a}} \cr \theta \cr 0 \cr } \,.\,$

\smallskip  \noindent
Then the algebraic expression (7.8) of the elliptic Galileo system of
conservation  laws is established.

\smallskip \noindent $\bullet \quad$
We still have to show that any regular solution of the system (7.8) satisfy the
conservation (7.48) of the mathematical  entropy, with a velocity field
given by the
relation (7.9) that satisfies in consequence :

\smallskip \noindent   (7.58)   $ \quad \displaystyle    
a \, {\rm d} u \,\,=\,\,{{1}\over{\theta^2 + b \, \zeta / a }} \, ( \theta
\, {\rm  d}\zeta - \zeta \, {\rm d} \theta) \,.\, \, $

\smallskip \noindent
We have very simply

\smallskip \noindent  $  \displaystyle   
  {{\partial \eta}\over{\partial t}} \,+\,  {{\partial}\over{\partial x}}
\bigl( \eta \, u \bigr) \,\,= \,\, \Bigl(  {{\partial}\over{\partial t}} + u \,
{{\partial}\over{\partial x}} \Bigr) \eta(W) \,+\, \eta \,  {{\partial u}\over
{\partial x}} \,\,=\,  $

\smallskip  \noindent $\displaystyle \qquad = \,\,
{{\partial \sigma}\over{\partial \alpha}} \, \, \Bigl[ \,
{{\partial \xi}\over{\partial \theta}} \, \Bigl(  {{\partial}\over{\partial
t}} + u \,  {{\partial}\over{\partial x}} \Bigr) \theta 
\,+\,  {{\partial  \xi}\over{\partial
\zeta}} \, \Bigl(  {{\partial}\over{\partial t}} + u \,
{{\partial}\over{\partial x}} \Bigr) \zeta \,\Bigr] 
\,+\,  {{\partial \sigma}\over{\partial \beta}} \, \Bigl(
{{\partial}\over{\partial t}} + u \,  {{\partial}\over{\partial x}} \Bigr)
\psi \,+\,  \eta \,  {{\partial u}\over{\partial x}} \,$

\smallskip  \noindent $\displaystyle \qquad = \,\,
{{\partial \sigma}\over{\partial \alpha}} \, \, \Bigl[ \,
{{\theta}\over{\xi}} \,
\Bigl(  {{\partial}\over{\partial t}} + u \,  {{\partial}\over{\partial x}}
\Bigr)
\theta \,+\,  {{b}\over{a}} \, {{\zeta}\over{\xi}} \, \Bigl(  {{\partial}\over
{\partial t}} + u \,  {{\partial}\over{\partial x}} \Bigr) \zeta \,\Bigr] \,+\,
{{\partial \sigma}\over{\partial \beta}} \, \Bigl(
{{\partial}\over{\partial t}} + u
\,  {{\partial}\over{\partial x}} \Bigr) \psi \,+\, \eta \,  {{\partial
u}\over{\partial x}}  \,$

$ \hfill $ due to (7.55)

\smallskip  \noindent $\displaystyle \qquad = \,\,
{{\theta}\over{\xi}} \, {{\partial \sigma}\over{\partial \alpha}} \, \Bigl[
\, -\theta
\, {{\partial u}\over{\partial x}} \,+\, {{\partial}\over{\partial x}} \, \Bigl(
\Pi(\xi,\,\psi) \, {{b \, \zeta}\over{a \, \xi}} \Bigr) \, \Bigr]  \,$

\smallskip  \noindent $\displaystyle \qquad \qquad \qquad + \,
{{b}\over{a}} \, {{\zeta}\over{\xi}} \,  {{\partial \sigma}\over{\partial
\alpha}} \,  \Bigl[ \,-\zeta \, {{\partial u}\over{\partial x}}
 \,-\,  {{\partial}\over {\partial x}} \, \Bigl( \Pi(\xi,\,\psi) 
\, {{\theta}\over{\xi}} \, \Bigr)  \, \Bigr]  \,$

\smallskip  \noindent $\displaystyle \qquad \qquad \qquad +\,\,\,
{{\partial \sigma}\over{\partial \beta}} \,\bigl( -\psi \, {{\partial
u}\over{\partial
x}} \bigr) \,+\, \eta \,  {{\partial u}\over{\partial x}} \, \hfill $
applying the equation (7.8)

\smallskip  \noindent $\displaystyle \qquad = \,\,
- {{\partial u}\over{\partial x}} \,\Bigl[ \, {{\theta^2}\over{\xi}} \,
{{\partial
\sigma}\over{\partial \alpha}} \, +\, {{b}\over{a}} \,
{{\zeta^2}\over{\xi}} \, {{\partial
\sigma}\over{\partial \alpha}} \, +\, \psi \,  {{\partial
\sigma}\over{\partial \beta}}
\,-\, \sigma \, \Bigr] \,+\,  {{\partial }\over{\partial x}} \Bigl(
{{\Pi}\over{\xi}}
\, \Bigr) \, \Bigl[  {{\theta}\over{\xi}} \,  {{\partial
\sigma}\over{\partial \alpha}}
\,  {{b\, \zeta }\over{a}} \,- \,  {{b }\over{a}} \,  {{\zeta}\over{\xi}} \,
{{\partial \sigma}\over{\partial \alpha}} \, \theta \Bigr] \,$

\smallskip  \noindent $\displaystyle \qquad \qquad \qquad +\,\,\,
{{b }\over{a}} \, {{\partial \sigma}\over{\partial \alpha}} \,
{{\Pi}\over{\xi^2}}
\,\, \Bigl( \theta \,  {{\partial \zeta}\over{\partial x}} \,-\, \zeta \,
{{\partial
\theta}\over{\partial x}} \, \Bigr) \,$

\smallskip  \noindent $\displaystyle \qquad = \,\,
- \sigma^* \Big(  {{\partial  \sigma}\over{\partial \alpha}}  (\xi,\,\psi) ,\,
 {{\partial  \sigma}\over{\partial \beta}}  (\xi,\,\psi) \Big)  
\, {{\partial u}\over{\partial x}} \,+\, b \,  {{\partial
\sigma}\over{\partial \alpha}} (\xi,\,\psi) \,\, \Pi (\xi,\,\psi) \, {{\partial
u}\over{\partial x}} \,, $


\smallskip  \noindent $\displaystyle \qquad = \,\,0 \,\hfill $ 
taking into account the relation (7.15).

\smallskip \noindent
This property establishes the structure of the  elliptic Galileo system.

\smallskip \noindent $\bullet \quad$ {\bf Case (iii).} \quad
We detail now the third case, when the matrix $\,\, {\rm d}Y(0) \,\,$ is
given by
the relation (7.38) with $\, a \! > \! 0 \,$ and $\, b \! > \! 0 \,.\,$ We first
remark that the matrix  $\,\, {\rm d}Y(0) \,\,$ is nilpotent because

\smallskip \noindent   (7.59)   $ \quad \displaystyle   
{\rm d}Y(0)^2 \,\,=\,\, \pmatrix{ 0 & 0 & 0 \cr 0 & 0 & 0 \cr a \, b & 0 &
0 \cr}  \,,\qquad {\rm d}Y(0)^3 \,\,=\,\,0 \,. \,  $  

\smallskip \noindent
This remark justifies the name ``nilpotent'' given for this third Galileo group
preserving system of conservation laws. The exponentiation of the matrix
$\,\,Y(v)  \,\,$ is easy :

\smallskip \noindent   (7.60)   $ \quad \displaystyle    
Y(v) \,\,= \,\,  \pmatrix{ 1 & 0 & 0 \cr -a \, v  & 1 & 0 \cr  a\, b \,
{{v^2}\over{2}} & -bv & 1 \cr} \,,\qquad v \in \R \,,\qquad a > 0 \,,\qquad
b > 0 \,. $  

\smallskip \noindent
The determiation of the velocity field $\,\,u( {\scriptstyle \bullet})
\,\,$  is a
direct consequence of the evaluation of the product $\,\, Y(v) \,  {\scriptstyle
\bullet} \, W \,\,$ and of the remark that $\,\, u\bigl( Y(u(W)) \,
{\scriptstyle
\bullet} \, W \bigr) \,=\, 0 \,.\, $ We have

\smallskip \noindent  $  \displaystyle     
\forall \, W \,=\, \pmatrix { \theta \cr \zeta \cr \psi \cr } \, \in \Omega
\,,\qquad  Y(v) \, \pmatrix { \theta \cr \zeta \cr \psi \cr } 
\,\,=\,\, \pmatrix { \theta \cr -a \, v \, \theta 
\,+\, \zeta \cr a \, b \, {{v^2}\over{2}} \, \theta \,-\, b
\, v \,  \zeta \,+\, \psi \cr } \,  $

\smallskip \noindent
and we deduce that we have necessarily $\,\,u(W) \,=\, {{\zeta}\over{a \,
\theta}} .
\,\,$ This expression is exactly the relation (7.14). We deduce :

\smallskip \noindent   (7.61)   $ \quad \displaystyle     
\forall \, W \,=\, (\theta,\, \zeta,\, \psi)^{\rm t} \, \in \Omega \,,\qquad
Y(u(W) \,  {\scriptstyle \bullet}  W \bigr) \,\,=\,\, \Bigl( \, \theta \,,\, 0
\,,\, \psi \,-\, {{b}\over{2 \, a}} \, {{\zeta^2}\over{\theta}} \,\Bigr)^{\rm t}
\,.\, $ 

\smallskip \noindent $\bullet \quad$
The computation of the entropy variables is a simple consequence of the
relations (7.17) and (7.61). It comes

\smallskip \noindent   (7.62)   $ \quad \displaystyle     
\varphi \,\,= \,\, \Bigl( \, {{\partial \sigma}\over{\partial \alpha}} \,+\,
{{b}\over{2 \, a}} \, {{\zeta^2}\over{\theta^2}} \, {{\partial
\sigma}\over{\partial
\beta}} \,,\, -{{b}\over{a}} \, {{\zeta}\over{\theta}} \, {{\partial
\sigma} \over  {\partial \beta}} \,,\, 
{{\partial \sigma}\over{\partial \beta }} \, \Bigr) \,.\, $
 
\smallskip \noindent
With the notation $\,\, p \,=\, b / a \,,\,$ the exact expression of the
Hessian of the mathematical entropy only needs some care~:

\smallskip \noindent  $  \displaystyle    
{{\partial^2 \eta}\over{\partial \theta^2}} \,\,=\,\, {{\partial^2 \sigma} \over
{\partial \alpha^2}} \,-\, p \, {{\zeta^2} \over {\theta^3}} \, {{\partial
\sigma}\over{\partial \beta}} \,+\, {{1}\over{4}} \, p^2 \, {{\zeta^4}\over
{\theta^4}} \,  {{\partial^2 \sigma} \over {\partial \beta^2}} \,, \qquad
{{\partial^2 \eta}\over{\partial \theta \, \partial \zeta }} \,\,=\,\, p \,
{{\zeta} \over {\theta^2}} \,  {{\partial \sigma}\over{\partial \beta }} 
\,-\, {{p^2} \over{2}} \, {{\zeta^3} \over{\theta^3}} \, 
 {{\partial^2 \sigma}\over{\partial \beta^2 }}  \,,\, $

\smallskip \noindent  $  \displaystyle  
{{\partial^2 \eta}\over{\partial \theta \, \partial \psi }} \,\,=\,\,{{p}
\over{2}} \, {{\zeta^2} \over{\theta^2}} 
\,   {{\partial^2 \sigma}\over{\partial
\beta^2 }} \,,  \qquad
{{\partial^2 \eta}\over{\partial \zeta^2}} \,\,= \,\, -{{p} \over{\theta}} \,
{{\partial \sigma}\over{\partial \beta }} \,+\, p^2 \, {{\zeta^2}\over
{\theta^2}} \,
{{\partial^2 \sigma}\over{\partial \beta^2 }} \,, $

\smallskip \noindent  $  \displaystyle   
{{\partial^2 \eta}\over{\partial \zeta \, \partial \psi}}\,\,= \,\, -p \,
{{\zeta} \over{\theta}} \, {{\partial^2 \sigma}\over{\partial \beta^2 }}  
\,, \qquad  {{\partial^2 \eta}\over{\partial \psi^2}} 
\,\,= \,\, {{\partial^2 \sigma}\over{\partial \beta^2 }} \,,\,    $

\smallskip  \noindent
and with the complementary notation $\,\, q \,\equiv\, \zeta / \theta \,$
we get 

\setbox21=\hbox{$\displaystyle  {\rm d}^2 \eta \,\,=\,\,   $}
\setbox22=\hbox{$\displaystyle 
\pmatrix { \displaystyle \!  {{\partial^2 \sigma} \over
{\partial \alpha^2}}- p \,{{q^2} \over{\theta}} \, {{\partial
\sigma}\over{\partial \beta }} + {{1}\over{4}} \, p^2 \, q^4 \,
{{\partial^2 \sigma} \over {\partial \beta^2}}   & \displaystyle
{{p \, q} \over{\theta}} \, {{\partial \sigma}\over{\partial \beta }} 
-   {{1} \over{2}} \, p^2  q^3   {{\partial^2 \sigma} \over {\partial \beta^2}}
& \displaystyle {{1} \over{2}} \, p   q^2 
\, {{\partial^2 \sigma} \over {\partial \beta^2}}
\cr   \displaystyle \!    {{p \, q} \over{\theta}} 
\, {{\partial \sigma}\over{\partial \beta }}   
-  {{1} \over{2}} \, p^2  q^3  {{\partial^2 \sigma} \over {\partial \beta^2}}
& \displaystyle -{{p} \over{\theta}} 
\, {{\partial \sigma}\over{\partial \beta }} \, +\,  p^2 \, q^2
\, {{\partial^2 \sigma} \over {\partial \beta^2}} &  \displaystyle
-p \, q \,  {{\partial^2 \sigma} \over {\partial \beta^2}}    \cr
\displaystyle \!   {{1} \over{2}} \, p   q^2 
  {{\partial^2 \sigma} \over {\partial \beta^2}}
& \displaystyle  -p \, q \,  {{\partial^2 \sigma} \over {\partial
\beta^2}}  & \displaystyle  
{{\partial^2 \sigma} \over {\partial \beta^2}} \cr  } .  $ }
\setbox30= \vbox {\halign{#\cr \box21 \cr  \cr   \box22 \cr     }}
\setbox31= \hbox{ $\vcenter {\box30} $}
\setbox44=\hbox{\noindent  (7.63) $\displaystyle     \left\{ \box31 \right. $}  
\smallskip \noindent $ \box44 $

\smallskip  \noindent
The two by two minor determinant at the bottom and the right of $\,\,  {\rm
d}^2 \eta
\,\,$ is equal to $\,\, -{{b}\over{a \, \theta}} \,  {{\partial \sigma}
\over {\partial
\beta}} \,  {{\partial^2 \sigma} \over {\partial \beta^2}} \,,\,\,$ which
establishes the necessary condition (7.12) when we have supposed that  the
domain $ \,
\Omega \, $ is composed by states $\, W \,=\, (\theta ,\, \zeta ,\,
\psi)^{\rm t} \,$
that satisfy the condition

\smallskip \noindent   (7.64)   $ \quad \displaystyle     
\theta > 0 \,.\, $

\smallskip  \noindent The evaluation of the determinant of the relation
(7.63) is  easy. We multiply the last line by $\, p \, q \,$ 
and we add it to the second line.  We  obtain :

\smallskip \noindent  $ \displaystyle    
{\rm det} \, ({\rm d}^2 \eta ) \,\,=\,\,  $ 

\smallskip \noindent  $ \displaystyle   \quad  = \, 
 \left|  \matrix { \displaystyle  {{\partial^2 \sigma} \over
{\partial \alpha^2}} \,-\, p \,{{q^2} \over{\theta}} \, {{\partial
\sigma}\over{\partial \beta }} \,+\, {{1}\over{4}} \, p^2 \,  q^4 \,
{{\partial^2 \sigma} \over {\partial \beta^2}}
& \displaystyle {{p \, q} \over{\theta}} 
\, {{\partial \sigma}\over{\partial \beta }} \, -
\, {{1} \over{2}} \, p^2 \, q^3 \, 
{{\partial^2 \sigma} \over {\partial \beta^2}}
& \displaystyle {{1} \over{2}} \, p \, q^2 \, 
{{\partial^2 \sigma} \over {\partial \beta^2}}
\cr   \displaystyle {{p \, q} \over{\theta}} \, 
{{\partial \sigma}\over{\partial \beta }}
& \displaystyle -{{p} \over{\theta}} \, 
{{\partial \sigma}\over{\partial \beta }}   & 0
   \cr \displaystyle {{1} \over{2}} \, p \, q^2 \, {{\partial^2 \sigma}
\over {\partial \beta^2}}  
& \displaystyle   -p \, q \,  {{\partial^2 \sigma}
\over {\partial \beta^2}}  
 & \displaystyle  {{\partial^2 \sigma} \over  {\partial
\beta^2}} \cr  } \right| \,;\, $

\smallskip  \noindent
then we multiply the third column of the previous expression by $\,
{{1}\over{2}} \,p
\, q^2 \,$ and we substract  it from the first column ; in an analogous way, we
multiply the last column by $ \, p \, q \,$ and we add it to the second
column of the
determinant. We find~:

\smallskip \noindent  $ \displaystyle     
{\rm det} \, ({\rm d}^2 \eta )\,\,=\,\, \left|  \matrix { \displaystyle
{{\partial^2
\sigma} \over {\partial \alpha^2}} \,-\, p \,{{q^2} \over{\theta}} \, {{\partial
\sigma}\over{\partial \beta }}               & \displaystyle
{{p \, q} \over{\theta}} \, {{\partial \sigma}\over{\partial \beta }} &
\displaystyle
{{1} \over{2}} \, p \, q^2 \, {{\partial^2 \sigma} \over {\partial \beta^2}}
\cr   \displaystyle
{{p \, q} \over{\theta}} \, {{\partial \sigma}\over{\partial \beta }}
& \displaystyle
-{{p} \over{\theta}} \, {{\partial \sigma}\over{\partial \beta }}   & 0
\cr  0 & 0 &  \displaystyle  {{\partial^2 \sigma} \over {\partial \beta^2}}
\cr  }
\right| \,\,=\,\, \displaystyle  {{\partial^2 \sigma} \over {\partial
\beta^2}} \,
\Bigl( -{{p}\over{\theta}}  \Bigr) \, {{\partial \sigma} \over {\partial
\beta}} \, \,
{{\partial^2 \sigma} \over {\partial \alpha^2}} \,.\,  $
 
\smallskip  \noindent
The sign of $\,\,  {\rm det} \, ({\rm d}^2 \eta )\,\,$ is the one of $\,\,
-{{\partial \sigma} \over {\partial \beta}} \,$ when the condition (7.64)
is satisfied then is coherent with the inequality (7.12).
 Then the function 
$\,\, \eta({\scriptstyle \bullet}) \,\,$ 
defined with the relation (7.17) is strictly convex.

\smallskip \noindent $\bullet \quad$
The  thermodynamic flux $\,\, j({\scriptstyle \bullet}) \,\,$ is
constructed exactly as
in the two previous cases. We use the definition (7.46) of the function $\,\,
\Pi({\scriptstyle \bullet}) \,;\,$ then the expression (7.62)  of the entropy
variables shows :

\smallskip \noindent  $ \displaystyle     
\varphi \, {\scriptstyle \bullet} \, {\rm d}Y(0) \, {\scriptstyle \bullet}
\,j(W_0)  \,=\, \Bigl(  {{\partial \sigma} \over {\partial \alpha}} \,,\,0 \,,\,
{{\partial \sigma} \over {\partial \beta}} \, \Bigr)  
\pmatrix {0 & 0 & 0 \cr -a & 0 & 0 \cr
0 & -b & 0 \cr }  \pmatrix {0 \cr \Pi \cr 0 \cr } \,=\, -b \, \Pi \, {{\partial
\sigma} \over {\partial \beta}} \,. \, $
 
\smallskip \noindent
The relation (7.16) is then a consequence of the remark that 
$\,\, {\rm d}j(W) \, {\scriptstyle \bullet} \, 
 {\rm d}Y(0) \, {\scriptstyle \bullet} \, W $ 
$ \,=\,  {\rm d}Y(0) \, {\scriptstyle \bullet}   \, j(W) \, $
joined with
the relation (4.7). The general expression for the function $\,\, j(W) \,$ is a
consequence of the relation (3.12) and the expresison (7.61) of a state
inside the null velocity manifold :

\smallskip \noindent  $ \displaystyle      
j(W) \,\,= \,\, Y\bigl(-u(W)\bigr) \, {\scriptstyle \bullet} \, j\Bigl(
\theta \,,\,
0 \,,\, \psi - {{b} \over{2 \, a}} \, {{\zeta^2} \over{\theta}} \, \Bigr) $

\smallskip  \noindent $\displaystyle \qquad \,\,\,\, \, = \,\,\,  
\,\pmatrix { 1 & 0 & 0 \cr a \, u & 1 & 0 \cr a \, b \, {{u^2}\over{2}} & b
\, u & 1 } \, \pmatrix {0 \cr  \displaystyle \Pi\Bigl( \theta \,,\, \psi 
- {{b} \over{2 \, a}} \, {{\zeta^2} \over{\theta}} \, \Bigr) \cr 0 } $

\smallskip  \noindent $\displaystyle \qquad \,\,\,\, \, = \,\,\,
\Pi\Bigl( \theta \,,\, \psi - {{b} \over{2 \, a}} \, {{\zeta^2}
\over{\theta}} \,
\Bigr) \,\, \pmatrix { 0 \cr 1 \cr \displaystyle {{b}\over{a}}\, {{\zeta}
\over{\theta}} \cr} \,\hfill $ due to the relation (7.14).

\smallskip  \noindent
The algebraic expression (7.13) of the nilpotent Galileo system of conservation laws is
an immediate consequence of what have been done at  the previous line.

\smallskip \noindent $\bullet \quad$
As in the two preceding  cases, we verify that the candidate (7.17) for beeing a
mathematical entropy satisfies the relation (7.48) for  regular solutions of the
conservation law (7.13). We have

\smallskip \noindent  $ \displaystyle     
{{\partial \eta}\over{\partial t}} \,+\,  {{\partial}\over{\partial x}}
\bigl( \eta \, u \bigr) \,\,= \,\, \Bigl(  {{\partial}\over{\partial t}} + u \,
{{\partial}\over{\partial x}} \Bigr) \eta(W) \,+\, \eta \,  {{\partial u}\over
{\partial x}} $

\smallskip  \noindent $\displaystyle \,\,\, = \,\,
\Bigl(  {{\partial \sigma}\over{\partial \alpha}} \,+\, {{a \,b}\over{2}}
\, u^2 \,
{{\partial \sigma}\over{\partial \beta}}\, \Bigr) \,  \Bigl(
{{\partial}\over{\partial t}} + u
\,  {{\partial}\over{\partial x}} \Bigr) \theta \,-\, b \, u \,  {{\partial
\sigma}\over{\partial \beta}} \, \,  \Bigl(  {{\partial}\over{\partial t}} + u
\,  {{\partial}\over{\partial x}} \Bigr) \zeta  $

\smallskip  \noindent $\displaystyle   \qquad + \, 
{{\partial \sigma}\over{\partial
\beta}} \,  \Bigl(  {{\partial}\over{\partial t}} + u \,
{{\partial}\over{\partial
x}} \Bigr) \psi \,+\, \eta \,  {{\partial u} \over {\partial x}} \,  $
\hfill   due to the expression (7.62)

\smallskip  \noindent $\displaystyle \,\,\, = \,\,
\Bigl(  {{\partial \sigma}\over{\partial \alpha}} \,+\, {{a \,b}\over{2}}
\, u^2 \,
{{\partial \sigma}\over{\partial \beta}}\, \Bigr) \,  \Bigl(-\theta \,
{{\partial u}
\over {\partial x}} \,\Bigr) \,-\, b \, u \,  {{\partial
\sigma}\over{\partial \beta}} \, \,  \Bigl( -\zeta \,  {{\partial u}
\over {\partial x}} \,-\,  {{\partial \Pi} \over {\partial x}} \, \Bigr) \,$
\smallskip  \noindent $\displaystyle \qquad  \quad + \,\,
 {{\partial \sigma}\over{\partial \beta}} \, \Bigl( -\psi \,   {{\partial u}
\over {\partial x}} \,- \,   {{\partial } \over {\partial x}} \,\bigl(  \,
\Pi \, u
\, b \bigr) \, \Bigr) \,+\, \sigma \,   {{\partial u} \over {\partial x}} $
 
 \hfill taking into account the equation (7.13)

\smallskip  \noindent $\displaystyle \,\,\, = \,\,
-  {{\partial u} \over {\partial x}} \, \Bigl[ \,  \theta \, {{\partial
\sigma}\over{\partial \alpha}} \,+\, \Bigl( \psi - {{b}\over {2 \, a }} \,
{{\zeta^2} \over{\theta}} \,  \Bigr) 
\, {{\partial \sigma}\over{\partial \beta}} \,-\,  \sigma \,
\Bigr] \,-\, b \, \Pi \, {{\partial \sigma}\over{\partial \beta}} \,
{{\partial u} \over {\partial x}}  \,\hfill $ due to (7.14)

\smallskip  \noindent $\displaystyle \,\,\, = \,\,
-  {{\partial u} \over {\partial x}} \, \, \Bigl[ \,  \theta \, \, {{\partial
\sigma}\over{\partial \alpha}} \Bigl( \theta ,\, \psi - {{b}\over {2 \, a }} \,
{{\zeta^2} \over{\theta}} \,  \Bigr) \,\,+\,\, \Bigl( \psi - {{b}\over {2
\, a }} \, {{\zeta^2} \over{\theta}} \,  \Bigr) \,\, 
 {{\partial \sigma}\over{\partial   \beta}}
\Bigl( \theta ,\, \psi - {{b}\over {2 \, a }} \, {{\zeta^2} \over{\theta}}
\,  \Bigr)  \,$

\smallskip  \noindent $\displaystyle   \qquad - \,  
\sigma \Bigl( \theta ,\, \psi - {{b}\over {2 \, a }} \, {{\zeta^2}
\over{\theta}} \,  \Bigr) \,\,-\,\, \sigma^*  \Bigl({{\partial
\sigma}\over{\partial
\alpha}} ,\, {{\partial \sigma}\over{\partial \beta}} \, \Bigr) \, \Bigr] $

\smallskip  \noindent $\displaystyle \,\,\, = \,\, 0 \,$
\qquad  
and the theorem 2 is proven. $ \hfill \square \kern0.1mm    $

\bigskip \bigskip \centerline{ \smcap 8. \quad The Cemracs System} \smallskip

This section describes the funny hyperbolic  system of conservation laws
that we have
 derived during the Cemracs 99. It corresponds to the elliptic Galileo system of
conservation laws with $\, a = b = 1 \,$ and the thermodynamics of  the
polytropic
perfect gas.

\smallskip \noindent $\bullet \quad$
Let $\,\, \gamma \! > \! 1 \,\,$ be a fixed real number, {\it e.g.} $\,
\gamma = 7 / 5  \,$  for the air (di-atomic gas)  
at usual conditions of tempeature and pressure, see  [Ca85]. We set

\smallskip \noindent   (8.1)   $ \quad \displaystyle      
\sigma(\theta,\, \psi) \,\,= \,\, - \theta \,\, {\rm log} \, \Bigl[ {{(\gamma-1) \,
\psi} \over {\theta^{\gamma}}} \, \Bigr] \,,\qquad \theta > 0 \,,\quad \psi
> 0 \, $ 

\smallskip  \noindent
and as considered in the remark 4, we can introduce the density $\, \rho
\,$ and the  internal energy $\, e \,$ according to

\smallskip \noindent   (8.2)   $ \quad \displaystyle      
 \theta \,\,=\,\, \rho \,,\qquad \psi \,\,=\,\, \rho \, e \,. \,  $

\smallskip \noindent 
We have  $ \displaystyle \quad 
{\rm d} \sigma \,\,=\,\, Q(\theta,\, \psi) \, {\rm d}\theta \,+\, \chi(\theta,\,
\psi)  \, {\rm d} \psi \, \,\quad  $  with

\smallskip \noindent   (8.3)   $ \quad \displaystyle      
Q(\theta,\, \psi) \,\,=\,\, \gamma \,+\, {{\sigma(\theta,\, \psi)}\over
{\theta}} \, $

\smallskip \noindent   (8.4)   $ \quad \displaystyle     
\chi(\theta,\, \psi) \,\,=\,\, -{{\theta}\over{\psi}} \quad > 0 \,.\, $

\smallskip  \noindent
The determination of the dual function of the entropy is easy. We have to
solve, for  a given pair $\,\, (A,\, B) \,,\,$ the system

\smallskip \noindent   (8.5)   $ \quad \displaystyle     
\left\{ \matrix { \displaystyle \gamma \,+\, {{\sigma(\theta,\, \psi)}\over
{\theta}}
\,=\, A \cr \displaystyle  -{{\theta}\over{\psi}} \,=\, B \,. \cr } \right. $ 

\smallskip  \noindent
Then, following the relation (7.10), we evaluate the ``mechanical pressure''
$\, \Pi({\scriptstyle \bullet}) \,: \,$

\smallskip \noindent  $  \displaystyle     
\Pi(\theta,\, \psi) \,=\, {{\sigma^*(A,\,B)}\over {A}} \,= \, {{1}\over
{A}} \, \bigl( \, \theta \, A \,+ \, \psi \, B 
\,- \, \sigma(\theta,\, \psi) \, \bigr) \, $

\smallskip \noindent  $  \displaystyle     \qquad  \quad  \,\,\, \,=\,
\theta \,+\, {{1}\over {\gamma + \sigma / \theta}} \, (-\theta - \sigma)
\,=\, {{(\gamma-1) \, \theta } \over {\gamma + \sigma / \theta}} \,,  $

\smallskip \noindent   (8.6)   $ \quad \displaystyle      
\Pi(\theta,\, \psi) \,\,=\,\,  {{(\gamma-1) \, \theta } \over
{\displaystyle  \gamma \,-\, {\rm log} 
\, \Bigl[ {{(\gamma-1) \, \psi} \over {\theta^{\gamma}}} \,
\Bigr] }} \, \,.  \, $

\smallskip  \noindent
Taking into account the relation (7.9), we set

\smallskip \noindent   (8.7)   $ \quad \displaystyle       
u(W) \,\,=\,\, {\rm arctg} \, {{\zeta}\over{\theta}} \,,\qquad W \,=\,
(\theta,\,  \zeta,\, \psi)^{\rm t } \in \Omega \,\, $

\smallskip  \noindent
and the Cemracs system is a simple re-writing of the relation (7.8) :

\setbox21=\hbox{$\displaystyle  
{{\partial}\over{\partial t}} \pmatrix{\theta \cr \zeta \cr \psi \cr }\,+\,
{{\partial}\over{\partial x}} \, \Biggl\{ \, u(W) \,  \pmatrix{\theta \cr
\zeta \cr \psi \cr } 
\,+\, {{\Pi\bigl( \, \sqrt{\theta^2 + \zeta^2  \,}\,,\,\psi
\,  \bigr)} \over {\sqrt{\theta^2 +  \zeta^2 \,}}} \, 
 \pmatrix{ -\zeta  \cr \theta \cr  0 \cr } \, \Biggr\} \,= \, 0 \,,   $}
\setbox22=\hbox{$\displaystyle 
\qquad \qquad \qquad \qquad  \qquad \qquad  \qquad \qquad  \qquad   
W \,=\, \pmatrix{ \theta \cr \zeta \cr  \psi \cr } \,.\,  $ }
\setbox30= \vbox {\halign{#\cr \box21 \cr   \box22 \cr     }}
\setbox31= \hbox{ $\vcenter {\box30} $}
\setbox44=\hbox{\noindent  (8.8) $\displaystyle \, \,   \left\{ \box31 \right. $}  
\smallskip \noindent $ \box44 $ 

\bigskip \noindent  {\bf Proposition 9. \quad  The Cemracs system is
hyperbolic.}

\noindent
Under the hypotheses

\smallskip \noindent   (8.9)   $ \quad \displaystyle       
\theta > 0 \,,\qquad \psi > 0 \,,\qquad \Pi\bigl( \, \sqrt{\theta^2 + \zeta^2
\,}\,,\,\psi \,  \bigr) > 0 \,,\, $

\smallskip  \noindent
the system (8.6) (8.7) (8.8) of conservation laws is hyperbolic.

\bigskip \noindent  {\bf Proof of proposition 9.}
\smallskip \noindent $\bullet \quad$
It is some kind of exercice for the studients. We do a direct proof of this
proposition  to improve the preceding assertions. The simplest way to do it
is to
consider the following new set of variables :

\smallskip \noindent   (8.10)   $ \quad \displaystyle 
V \,\,\equiv \,\, \Bigl( \, \theta \,,\, \xi \equiv {{\zeta}\over{\theta}} \,,\,
\varphi \equiv {{1}\over{\sqrt{1 + \xi^2}}} \, \Pi (\theta \, \sqrt{1 +
\xi^2} ,\,  \psi) \, \Bigr)^{\rm t}  \,.\, \, $

\smallskip  \noindent
We have

\smallskip \noindent   (8.11)   $ \quad \displaystyle      
(1 + \xi^2) \,  {\rm d}u  \,\,=\,\, {\rm d}\xi \,\,=\,\, {\rm d}
\Bigl({{\zeta}\over{\theta}} \, \Bigr) \,\,=\,\, -{{\zeta}\over {\theta^2}} \,
{\rm d} \theta \,+\, {{1}\over{\theta}} \,{\rm d}\zeta \,$

\noindent and

\smallskip \noindent   (8.12)   $ \quad \displaystyle   
{{ \Pi (\theta \, \sqrt{1 + \xi^2} ,\, \psi)}\over {\sqrt{1 + \xi^2}}} \,\,=\,\,
{{(\gamma - 1) \, \theta} \over {\displaystyle \gamma \,-\, {\rm log} \Bigl[
{{(\gamma-1) \, \psi}\over {\theta^{\gamma} \, (1+\xi^2)^{\gamma / 2 }}} \,
\Bigr] }} \,\, \equiv \,\, \varphi \,. \,  $

\noindent
Then  $\displaystyle \quad 
{{{\rm d}\varphi} \over {\varphi}} \,\,=\,\, {{{\rm d}\theta} \over
{\theta}} \,-\, {{\varphi}\over {(\gamma-1) \, \theta}} \, 
\Bigl[ \, - {{{\rm d}\psi} \over  {\psi}}
\,+\, \gamma \, {{{\rm d}\theta} \over {\theta}} \,+\, {{\gamma}\over {1 +
\xi^2}} \,  \xi \, {\rm d} \xi \, \Bigr] \, $

\smallskip \noindent   (8.13)   $ \quad \displaystyle   
{\rm d}\varphi \,\,= \,\, \Bigl( \, 1 \,- \, {{\gamma}\over{\gamma -1}} \,
{{\varphi}\over{\theta}} \,  \Bigr) \, {{\varphi}\over{\theta}} \, {\rm d}\theta
\,\,-\,\, {{\gamma}\over{\gamma -1}} \,\, {{\varphi^2}\over{\theta}} \,\,
{{\xi}\over {1 + \xi^2}} \, {\rm d}\xi 
\,\,+\,\, {{\varphi^2} \over{(\gamma-1)\,  \theta}} \,
{{{\rm d}\psi} \over {\psi}}\,. \, $

\smallskip \noindent $\bullet \quad$
We write again the two first equations of the system (8.8)~:

\smallskip \noindent   (8.14)   $ \quad \displaystyle    
{{\partial \theta}\over{\partial t}} 
\,+\, u \,  {{\partial \theta}\over{\partial x}}
\,+\, \theta \,  {{\partial u}\over{\partial x}} \,-\,  {{\partial}\over
{\partial x}} \bigl( \varphi \, \xi \bigr)  \,\,= \,\, 0  \, $

\smallskip \noindent   (8.15)   $ \quad \displaystyle    
{{\partial \zeta}\over{\partial t}} \,+\, u \,  {{\partial
\zeta}\over{\partial x}}  \,+\, \zeta \,  {{\partial u}\over{\partial x}} 
\,+\, {{\partial \varphi }\over {\partial x}} \,\,= \,\, 0  \,.\,  $

\smallskip \noindent
We multiply the equation (8.14) by $\,\,\, -\zeta / \theta^2 \,=\,  -\xi /
\theta
\,\, \,$ and the equation (8.15) by $\,\, 1 / \theta \,.\,$ 
Thanks to the relation (8.11), we have the following  evolution 
in time of the variable $\, \, \xi  \,:\,$

\smallskip \noindent   (8.16)   $ \quad \displaystyle     
{{\partial \xi}\over{\partial t}} \,+\, u \,  {{\partial \xi}\over{\partial x}}
\,+\, {{\varphi \, \xi}\over{\theta}} \,  {{\partial \xi}\over{\partial x}}
\,+\, {{1+\xi^2}\over{ \theta}} 
\, {{\partial \varphi }\over {\partial x}} \,\,= \,\, 0 \,.\, $

\smallskip \noindent
The equation for the variable $\,\, \varphi \,\,$ is composed from the relations
(8.14), (8.16), the third equation of the system (8.8) that is~:

\smallskip \noindent   (8.17)   $ \quad \displaystyle   
{{\partial \psi}\over{\partial t}} 
\,+\, u \,  {{\partial \psi}\over{\partial x}}
\,+\, \psi \,  {{\partial u}\over{\partial x}} \,\,= \,\, 0  \,     $

\smallskip \noindent
and from the relation (8.13).  We set

\smallskip \noindent   (8.18)   $ \quad \displaystyle   
y \,\, \equiv \,\, {{\varphi}\over{\theta}} \,;\, $

\smallskip \noindent
we multiply the equation (8.14) by the coefficient
$\,\, \displaystyle  \bigl(1-{{\gamma}\over {\gamma-1}} \, {{\varphi}\over
{\theta}}\bigr) \, {{\varphi}\over {\theta}} \,,\,$ the equation (8.16) by $\,\,
\displaystyle  -{{\gamma}\over {\gamma-1}} \, {{\varphi^2 }\over {\theta }} \,
{{\xi}\over {(1 + \xi^2)}} \,,\,$  the equation (8.17) by $\,\,\displaystyle
{{\varphi^2}\over {(\gamma-1) \, \theta \, \psi}} \,\,$ and we add these three
equations. The  $\,\,\displaystyle   {{\partial \theta}\over {\partial x}}
\,\,$ term is absent, the term associated  to 
$\,\,\displaystyle  {{\partial \xi}\over  {\partial x}} \,\,$ is equal to

\smallskip \noindent  $  \displaystyle   
\theta \, \Bigl[ \, \Bigl({{1}\over{1+\xi^2}} - y \Bigr) \, \Bigl( 1 -
{{\gamma}\over {\gamma-1}} \, y \Bigr) \, y \,- \, {{\gamma}\over {\gamma
-1}} \, y^3 \, {{\xi^2}\over {1 + \xi^2}} 
\, +\, {{1}\over{ \gamma-1}} \, {{y^2} \over{1+\xi^2}}  \, \Bigr] \, $

\smallskip \noindent  $  \displaystyle   \qquad    = \,\,
 {{\theta}\over {1+\xi^2}} \,\, \Bigl( \, y \,-\, (2+\xi^2) \, y^2 \,+\,
{{\gamma}\over {\gamma - 1}} \, y^3 \, \Bigr) \, $  
 
\smallskip  \noindent
and the coefficient of the  term relatively to $\,\, \displaystyle  {{\partial
\varphi}\over {\partial x}} \,\, $ is :

\smallskip \noindent  $  \displaystyle    
-\xi \, y \, \Bigl( 1 \,-\, {{\gamma}\over {\gamma - 1}} \,y \Bigr) \,-\,
{{\gamma}\over {\gamma - 1}} \xi \, y^2 \,\,=\,\, -\xi \,y \,.\, $  

\smallskip  \noindent
We deduce from this calculus that with the variables $\, V \,$ introduced
in (8.10),  the system (8.8) takes the form

\smallskip \noindent   (8.19)   $ \quad \displaystyle     
{{\partial V}\over {\partial t}} \,+\, u \, {{\partial V}\over {\partial
x}} \, +\,  B(V) \,  {{\partial V}\over {\partial x}} \,\, = \,\, 0 \,$ 

\noindent with

\smallskip \noindent   (8.20)   $ \quad \displaystyle      
B(V) \,\,= \,\, \pmatrix { 0 &  \displaystyle  {{\theta}\over{1 + \xi^2}} \,-\,
\varphi & -\xi \cr 0 &  \xi \, y &  \displaystyle {{1+\xi^2}\over {\theta}} \cr
0 & \displaystyle  {{\theta}\over {1+\xi^2}} \,\, \Bigl( \, y \,-\,  (2+\xi^2) \, y^2 
\,+\, {{\gamma}\over {\gamma - 1}} \, y^3 \, \Bigr) & -\xi \, y \cr } \,.\, $

\medskip \noindent $\bullet \quad$
We see in an clear  way that the real number zero is an eigenvalue of the
matrix $\,\, B(V) \,\,$ of the relation (8.20). 
The two other eigenvalues $\,\, \lambda \,\,$  satisfy  the equation~:

\smallskip \noindent   (8.21)   $ \quad \displaystyle      
\lambda^2 \,+\, \Bigl[ \, -y \,+\, 2 \, y^2 \,-\,  {{\gamma}\over {\gamma -
1}} \,y^3 \, \Bigr] \,\,=\,\, 0 \,.\,$

\smallskip  \noindent
The equation (8.21) has two opposite  real eigenvalues when $\, y > 0 \,\,$ as
proposed in our  study [Du99]. We have effectively in this case :

\smallskip \noindent   $ \displaystyle      
-y \,+\, 2 \, y^2 \,-\, {{\gamma}\over {\gamma - 1}} \, y^3 \,\,=\,\, -y \,
(1-y)^2  \,-\, {{1}\over {\gamma - 1}} \, y^3 \,\, < \,\, 0 \, $

\smallskip  \noindent
and the proposition 9 is established.  $ \hfill \square \kern0.1mm    $

\bigskip  \noindent  {\bf Remark 5. \quad Curious Physics.}

\noindent
The condition $\,\, y >  0 \,\,$ {\it i.e.} $\,\,  \varphi / \theta  > 0 \,\,$
or in an equivalent way $\,\, \Pi > 0 \,\,$ because $\,\, \theta > 0 \,\,$ by
convention. It corresponds to

\smallskip \noindent   (8.22)   $ \quad \displaystyle    
\Pi \,\, \equiv \,\, {{p}\over{\mu}} \,> \,0 \, \, $

\smallskip  \noindent
because as claimed in the relation (7.22) : $\,\,  {\rm d}
\sigma \,\,=  \,\, {{\mu}\over{T}} \, {\rm d}\theta 
\,-\,{{1}\over{T}} \, {\rm d} \psi . \,\, \,$ Then\br
  $  \Pi \,=\, {{1}\over{\mu / T}} \, {{p}\over {T}} \,\,=
\,\,  {{p}\over{\mu}} \,\, \,$ and the mechanical pressure $\,\,
\Pi({\scriptstyle
\bullet}) \,\,$ is the quotient of the physical pressure $\,\, p({\scriptstyle
\bullet}) \,\,$ divided by the chemical potential $\,\, \mu \,\,$  which is
{\bf not} natural. It is easier to interpret the Cemracs 
system (8.8) as an elliptic  Galieo
system of conservation laws whose associated thermostatics is obtained by
{\bf exchanging} the roles of the variables $\,
\theta \,$ and $\, \psi \equiv \rho \, e \,$ inside the relation (8.2). This
corresponds to :

\smallskip \noindent   (8.23)   $ \quad \displaystyle     
\theta \,=\, \rho \, e \,,\qquad \psi \,=\, \rho \,.  $
 
\smallskip  \noindent
Then the relation $\,\,\,\displaystyle  {\rm d} \sigma \,\,=
-\,{{1}\over{T}} \, \theta
\,+\,{{\mu}\over{T}}  \, {\rm d}\psi \,\,\,$ and the condition (7.7) can be
written

\smallskip \noindent   (8.24)   $ \quad \displaystyle     
{{\partial \sigma}\over {\partial \theta}} \,\,= \,\, -{{1}\over {T}} \,\,
> \,\, 0 \, $

\smallskip  \noindent
corresponding to a thermostatics system with a negative temperature and
associated with a thermostatic pressure $ \,\,p \,\,$  
equal to the mechanical  pressure $ \,\, \Pi \,\,$ and given by  the formula

\smallskip \noindent   (8.25)   $ \quad \displaystyle      
\Pi \, (\rho \, e,\, \rho) \,\,=\,\,  {{(\gamma-1) \, \rho \, e } \over
{\displaystyle \gamma \,-\, {\rm log} \, 
\Bigl[ {{(\gamma-1) \, \rho} \over {(\rho \, e)^{\gamma}}} \, \Bigr] }} \, \,. $

\bigskip \bigskip

\centerline{  \smcap   9  \quad Acknowledgements} \smallskip

We thank Bruno Despr\'es  and Pierre-Arnaud Raviart  for preliminary discussions
(june 1998)  that have motivated this work, Denis Serre for judicious remarks
concerning a preliminary version of this report (november 1998)  and
Fr\'ed\'eric
Coquel for his kind encouragements to develop with paper,   ink and {\TeX}
software
an idea initially  proposed with chalk and blackboard  during a three days
stay at
Cemracs 99. Last but not least,  we apologize to T. Ruggeri for having not
been aware
of his work [Ru89].

$\hfill$ FD, december 8, 2000.

\bigskip \bigskip
 
\centerline{  \smcap  10  \quad  References} \smallskip

\smallskip \hangindent=11mm \hangafter=1 \noindent   
[Ca85]   H.B. Callen. 
{ \it Thermodynamics and an introduction to thermostatics, 
second edition}, John Wiley \& sons, New York, 1985.  

\smallskip \hangindent=11mm \hangafter=1 \noindent   
 [De98] B. Despr\'es. Lagrangian systems of
conservation laws,  {\it Numerische Mathematik}, vol.~89, p.~99-134, 2001.
 
\smallskip \hangindent=11mm \hangafter=1 \noindent  
 [Du90]  F. Dubois. Concavity of
thermostatic entropy and convexity of  Lax's mathematical entropy, 
{\it La Recherche A\'erospatiale},  n$^{\rm o}$1990-3, p.77-80,\br 
 may 1990. 

\smallskip \hangindent=11mm \hangafter=1 \noindent 
 [Du99]  F. Dubois.  Systems of conservation laws
invariant for  Galileo group and space reflection, 
{\it CNAM-IAT Research Report} n$^{\rm o}$318, february 1999.  

\smallskip \hangindent=11mm \hangafter=1 \noindent 
[FL71] K.O. Friedrichs, P.D. Lax. Systems of
conservation equations with a convex extension, 
{\it Proc. Nat. Acad. Sci. U.S.A.}, vol$.\,$68, n$^{\rm o}$8,  p. 1686-1688,  1971.

\smallskip \hangindent=11mm \hangafter=1 \noindent 
Go61] S.K. Godunov. On an interresting class of
quasilinear systems,  {\it Dok. Akad. Nauk. SSSR.}, vol$.\,$139, n$^{\rm o}$3,
p$.\,$521-523, 1961, and {\it Soviet Math.}, vol$.\,$2, p$.\,$947-949, 1961. 

\smallskip \hangindent=11mm \hangafter=1 \noindent 
[Go71] C. Godbillon. {\it El\'ements de topologie
alg\'ebrique}, Hermann, Paris,  1971.  

\smallskip \hangindent=11mm \hangafter=1 \noindent 
[GR96]  E. Godlewski, P.A. Raviart. {\it
Numerical Approximation of Hyperbolic Systems of Conservation Laws}, 
Applied Mathematical Sciences, vol$.\,$118, Springer, New York,  1996. 

\smallskip \hangindent=11mm \hangafter=1 \noindent 
[Mo66] J.J. Moreau. Fonctionnelles convexes,
{\it S\'eminaire Leray}, Coll\`ege de France, Paris, 1966.

\smallskip \hangindent=11mm \hangafter=1 \noindent 
[Ru89] T. Ruggeri. Galilean invariance  and entropy
principle for systems of balance laws, {\it Continuum Mechanics and
Thermodynamics},
vol$.\,$1, p$.\,$3-20, 1989.  

\smallskip \hangindent=11mm \hangafter=1 \noindent 
[Se82]  D. Serre.  Invariants des \'equations de
la M\'ecanique. Calcul des varia\-tions~: formes quadratiques positives sur
un c\^one. 
Remarques sur les \'equations de Navier-Stokes dans le cas stationnaire.
{\it Th\`ese d'Etat}, Paris~6 University, june~1982.   

\bye